\documentclass[12pt,preprint2]{aastex}
\usepackage{emulateapj5,psfig}
\usepackage{onecolfloat5}

\newcommand{\NH}{[\ion{N}{2}]$\lambda 6584/$H$\alpha$}
\newcommand{\NHA}{[\ion{N}{2}]/H$\alpha$}
\newcommand{\HA}{H$\alpha$}
\newcommand{\HB}{H$\beta$}

\newcommand{\OFIVE}{[\ion{O}{3}]$\lambda 5007$}

\newcommand{\OTHREE}{[\ion{O}{2}]$\lambda 3727$}

\newcommand{\MET}{[Fe$/$H]}
\newcommand{\meanfe}{$<\!{\rm Fe}\!>$}
\newcommand{\msa}{mag/$\Box$\arcsec}
\newcommand{\HII}{\ion{H}{2}}
\newcommand{\HI}{\ion{H}{1}}

\slugcomment{Accepted for publication in the Astronomical Journal}

\begin{document}
\twocolumn[


\title{Spectroscopy of Low Surface Brightness Galaxies with the Hobby-Eberly Telescope}


\author{Marcel Bergmann}
\affil{Astronomy Department, University of Texas, Austin, TX 78712}
\email{marcel@astro.as.utexas.edu}

\and

\author{Inger J{\o}rgensen}
\affil{Gemini Observatory, Hilo, HI 96720}
\email{ijorgensen@gemini.edu}

\and

\author{Gary J. Hill}
\affil{McDonald Observatory, University of Texas, Austin, TX 78712}
\email{hill@astro.as.utexas.edu}


\begin{abstract}
We have obtained low resolution spectra of nineteen 
red and blue low surface brightness galaxies, using the Marcario Low 
Resolution Spectrograph on the 9.2m Hobby-Eberly Telescope.  
These galaxies form a very heterogeneous class, whose spectra qualitatively resemble those of 
high surface brightness galaxies covering the full range of spectra seen in galaxies of Hubble types from E 
to Irr.  
We use a 
combination of emission line (EW(H$\alpha$), \NH) and absorption line (Mgb, 
H$\beta$, \meanfe) based diagnostics to investigate the star-formation and
chemical enrichment histories of these galaxies.  These are diverse, with
some galaxies having low metallicity and very young mean stellar ages, and other
galaxies showing old, super-solar metallicity stellar populations. 
In contrast with some previous
studies which found a strong trend of decreasing metallicity with decreasing
central surface brightness, we find a population of galaxies with low surface
brightness and near-solar metallicity.  Correlations between several of the
gas phase and stellar population age and metallicity indicators are used to
place contraints on plausible evolutionary scenarios for LSB galaxies.
The redshift range spanned by these galaxies is broad, with radial 
velocities from 3400 km/s to more than 65000 km/s.  A subset of the sample 
galaxies have published \HI\ redshifts and gas masses based on observations with
the Arecibo 305m single-dish radio telescope, which place these galaxies far off
of the mean Tully-Fisher relation.  Our new optical redshifts do not agree with 
the published \HI\ redshifts for these galaxies.  Most of the
discrepancies can be explained by beam confusion in the Arecibo observations, 
causing erroneous \HI\ detections for some of the galaxies.  
\end{abstract}



] 

\section{Introduction}

The class of low surface brightness (LSB) galaxies is
a diverse one.  LSB galaxies exhibit a range of absolute 
magnitudes, colors, and morphologies 
comparable to those of high surface brightness (HSB) galaxies.  The defining 
characteristic of LSB galaxies is, of course, their faint central surface 
brightness, variously defined as being at least one magnitude fainter than 
the cannonical Freeman (1970) value of 21.65 B mag/$\square\arcsec$, or fainter
than 23 B mag/$\square\arcsec$;
the precise limit varies from investigator to investigator.  However, whether
the low surface brightness results from a common evolutionary history for these
galaxies remains unknown.

Studies of LSB galaxies fall into two broad classes.  
The first class comprises the surveys to find LSB galaxies, and attempts 
to quantify what fraction of the total galaxy number density, 
luminosity density, and mass density 
are made up by them.  The second class comprises studies of the 
internal structure of LSB galaxies, including investigations into their gas
content, star formation history, and evolutionary processes.

The large area surveys which were initially mined for LSB galaxies were
photographic surveys, including the Upsala Galaxy Catalog \citep[UGC;][]{nil73}, 
the POSS-2 survey \citep{sch88,sch92}, and the APM survey \citep{imp96}.  In the past decade, large area
CCD surveys have also found a significant population of LSB galaxies. Notable
among these are the drift scan survey of \citet{dal97}, who use well 
quantified automated detection methods, and the multicolor survey of 
\citet*[hereafter OBC97]{one97a} and 
\citet[hereafter OBSCI97]{one97b}, 
which turned up a previously unseen population of 
red LSB galaxies.

These surveys proved the existence of LSB galaxies as more than the occasional
oddity.  \citet{dal97} make a conservative estimate of the number density
of LSB galaxies of ${\it N}=0.01^{+0.006}_{-0.005}$\ galaxies $h^3_{50}{\rm Mpc}^{-3}$, equal to the number density of HSB galaxies.
However, the \citeauthor{dal97} survey suffers from small number statistics, 
with only 7 LSBs detected.  \citet{mcg95} have a larger sample, and come
to a similar conclusion about the LSB galaxy number density. 
More recently,  \citet{dej00} have used a 
photographic survey of $\simeq$ 1000 Sb-Sdm galaxies to calculate
the bivariate luminosity and number density of galaxies as a function of 
both surface brightness and scale length.  They find that while the luminosity 
density is dominated by HSB galaxies (peaked at $\mu_B(0)\simeq21.0$ \msa), 
the number density of galaxies has an essentially flat distribution at all
surface brightnesses fainter than $\mu_B(0)\simeq21.0$ \msa, and therefore
LSB galaxies make up a significant fraction of the total number of galaxies
(See also the surveys of \citet{one00c} and \citet{cro02}).

The best studied LSB disk galaxies are those in the thesis sample of \citet{deb97b},
originally selected from the photographic survey catalogs of \citet{sch92} 
and the UGC catalog.  Observations of these $\sim 20$ galaxies include
multicolor optical surface photometry \citep{deb95}, VLA mapping of the \HI\ distribution
and kinematics \citep*{deb96a,deb97}, and spectroscopy of the \HII\ regions \citep{deb98}.
The general picture obtained from these observations is that LSB galaxies are
metal poor and gas rich, with a range of sizes (scale lengths) comparable to 
HSB galaxies, but with much lower surface densities. 
However, note that de Blok's selection criteria included an existing \HI\ 
detection.  The 
kinematics, and therefore the gravitational potentials, are dominated by dark matter at all
radii \citep{deb97,mcg01}, and the low global star formation rates (SFR) are the combined result
of both the low surface density, and the low metallicity \citep{ger99}.  

\citet{spr95} obtained photometry, low-resolution optical spectroscopy, and HI spectroscopy for
a sample of 8 giant LSB galaxies.  Their optical spectroscopic data pertain primarily to the bulges
of these galaxies, and they find no difference between the bulges of large LSB and HSB galaxies.

The wide field CCD survey of OBC97 found a large number
of LSB galaxies, covering a wider range of colors than those found by 
previous photographic surveys.  In particular, red LSB galaxies with (B$-$V)$> 0.9$, 
colors consistent with purely old stellar populations, were catalogued for the first 
time.  The follow-up single dish \HI\ observations using the Arecibo 305m
single-dish radio 
telescope \citep*{one00} suggested that many of these red LSBs (and some of the blue LSBs) were quite 
peculiar indeed,
having gas mass-to-light ratios (M/L) in excess of 8, with a few higher than 30.
Further, the rotational line widths were far larger than would be expected from their luminosities given 
the standard Tully-Fisher (1977) relation for spiral galaxies.

Larger scatter in the Tully-Fisher relation among LSB galaxies was also measured
by \citet*{bur01}.  Their survey of over 250 HSB and LSB galaxies selected from the APM survey \citep{imp96}
 included photometry, HI spectroscopy, and low resolution optical spectroscopy \citep*{imp01,bur01}.   
They used emission line ratios to measure the metallicity for 49 HSB and 17 LSB galaxies 
(where they define $\mu_{\rm B}$ = 22\msa as the dividing line)
and found there to be a roughly 50\% overlap in the metallicity range of the two groups.

\citet{bel00b} combined optical and near-infrared surface photometry
to study the stellar populations in subsets of both the de Blok sample and the 
red galaxies from the OBSCI97 survey.  Broadband optical studies of stellar
populations suffer from a degeneracy in color space between age and metallicity;
red colors could indicate either older stars or more
metal-rich stars, while bluer colors indicate either younger stars or metal-poor
stars.  The inclusion of near-infrared colors partially breaks this
degeneracy, essentially giving a handle on the fractional light contribution
from old stars which dominate the light in the near-infrared, and younger stars which
will dominate at bluer wavelengths.  \citeauthor{bel00b} found that the bluer galaxies
from the de Blok sample were younger than similar
HSB late-type galaxies. Further, the five red LSB galaxies in their study
were a mixture of two types: genuinely red galaxies with old
stellar populations, but central surface brightnesses on the bright end of
the LSB range ($\sim 22.5$\msa\ in B), and a group of objects with lower 
surface brightness whose true colors, after correcting for galactic 
extinction, were in fact as blue as the de Blok sample galaxies and had similar
stellar mean ages.  Interestingly, the three red galaxies with old stellar 
populations were not detected in the \citet{one00} \HI\ survey, while the two galaxies
with blue extinction corrected colors were detected.  However, none of the red
galaxies with extreme M$_{\rm HI}$/L ratios were studied by \citeauthor{bel00b}

We have obtained spectra for a sample of 19 red and blue LSB galaxies chosen 
from the catalog of OBC97, with the intent of studying the star 
formation histories of these galaxies.  Using a combination of emission and
absorption line measures, we compare the chemical enrichment of the stars and
gas, and we use the stellar population models of 
\citeauthor{vaz99} \citetext{\citeyear{vaz99}, \citeyear{vaz00}\footnote{\url{ 
http://www.iac.es/galeria/vazdekis/MODELS\_2000/out\_li\_BI}}, 
hereafter V1999 and V2000, respectively}
to interpret these data quantitatively.  In particular, we attempt to determine
whether a single LSB galaxy formation model can be reconciled with the various
observational attributes of these galaxies.

This paper is organized as follows.  In Section \ref{obsdataredux} we present the
sample, the spectroscopic and photometric observations, and the data reduction.
Sections \ref{datapresentation} through \ref{globcomp} presents the LSB spectra, analysis of the
emission line and absorption index strengths, and comparison to HSB galaxies.
We discuss these results in the context of LSB formation models in Section 
\ref{discussion}. Finally, we summarize our conclusions in Section \ref{summary}.


\begin{deluxetable}{lrrccccccccc}
\tabletypesize{\scriptsize}
\setlength{\tabcolsep}{0.05in}
\tablecaption{Literature data for galaxies in spectroscopic sample\tablenotemark{a}
   \label{littable}}
\tablewidth{0pt}
\tablehead{
\colhead{galaxy} & \colhead{RA}   & \colhead{Dec}   &
\colhead{cz (HI)} & \colhead{r$_e$} & \colhead{B(5")\tablenotemark{b}} & 
\colhead{I$_{\rm c}$(5")\tablenotemark{b}} &
\colhead{B$_{\rm tot}$\tablenotemark{b}} & \colhead{I$_{\rm c tot}$\tablenotemark{b}} & \colhead{$\mu_{\rm B}(0)_{\rm cor}$\tablenotemark{b}} & \colhead{\it i\tablenotemark{~c}} &\colhead{E(B-V)}\\
 & \colhead{(J2000)} & \colhead{(J2000)} & \colhead{(km s$^{-1}$)} & 
\colhead{('')} & \colhead{(mag)} & \colhead{(mag)} & \colhead{(mag)}&\colhead{(mag)} & \colhead{(\msa)} & \colhead{(deg)} &  \\
\colhead{(1)} & \colhead{(2)} &\colhead{(3)} & \colhead{(4)} &\colhead{(5)} &
\colhead{(6)} & \colhead{(7)} &\colhead{(8)} & \colhead{(9)} &\colhead{(10)} &
\colhead{(11)} &\colhead{(12)}
}
\startdata
  {[}OBC97]P1-3  &  23:21:18 & 08:14:28   &      ~3746  &  ~7.1   &   18.28  &           16.70 &     17.15&           15.81& 22.2 & 45 & 0.071\\
  {[}OBC97]P6-1  &  23:23:33 & 08:37:25   &     10882   &  18.9   &   19.55  &           16.79 &     17.23&           15.04& 24.2 & 66 & 0.061\\
  {[}OBC97]P9-4  &  23:18:40 & 07:30:44   &      ~4205  &  10.9   &   20.84  &           18.23 &     18.86&           16.29& 25.0 & 67 & 0.144\\
  {[}OBC97]C1-2  &  08:19:55 & 20:51:43   &      ~8531  &  ~5.2   &   18.99  &           16.91 &     18.10&           15.84& 22.5 & 21 & 0.049\\
  {[}OBC97]C1-4  &  08:19:24 & 21:00:12   &     \nodata &  12.6   &   18.09  &           15.80 &     16.55&           14.48& 22.3 & 63 & 0.050\\
  {[}OBC97]C3-2  &  08:22:36 & 20:59:46   &     \nodata &   ~6.0\tablenotemark{d}&18.17& 16.06$\pm0.06$ &     16.34&           14.14& 22.8 & 47 & 0.038\\
  {[}OBC97]C4-1  &  08:24:33 & 21:27:04   &      ~7905  &   11.0\tablenotemark{d}&18.97& \nodata&    17.83&       \nodata  & 23.6 & 39 & 0.044\\
  {[}OBC97]C4-5  &  08:25:48 & 21:52:45   &     \nodata &  ~5.6   &   18.96  &         \nodata &     18.23&       \nodata  & 23.0 & 57 & 0.046\\
  {[}OBC97]C5-3  &  08:27:44 & 22:28:52   &     12942   &  ~8.6   &   20.12  &           18.31$\pm0.1$ &     18.88&           16.37& 24.0 & 20 & 0.041\\ 
  {[}OBC97]C5-5  &  08:27:12 & 22:53:39   &      ~5524  &  23.0   &   19.90$\pm0.2$  &           18.60$\pm0.2$ &     18.08$\pm0.2$&           16.19$\pm0.2$& 24.3 & 45 & 0.038\\
  {[}OBC97]C6-1  &  08:27:31 & 21:30:03   &      ~4322  &  ~6.5   &   18.57  &           16.22$\pm0.06$ &     17.53&           15.27& 22.5 & 44 & 0.041\\
  {[}OBC97]C8-3  &  08:17:17 & 21:39:12   &      ~3570  &  ~4.8   &   19.22  &        \nodata  &     17.69&       \nodata  & 24.3 & 72 & 0.039\\
  {[}OBC97]N3-1  &  12:32:59 & 07:29:26   &     \nodata &  ~8.6   &   19.05  &        \nodata  &     18.02&       \nodata  & 23.5 & 64 & 0.023\\
  {[}OBC97]N9-1  &  10:20:18 & 27:45:10   &     \nodata &  ~7.6   &   18.41  &           19.18$\pm1$ &     17.29&           17.60$\pm1$& 22.3 & 41 & 0.036\\
  {[}OBC97]N9-2  &  10:20:22 & 28:07:56   &      ~7746  &  ~6.4   &   19.80  &           20.59$\pm0.06$ &     18.89&           17.49& 24.0 & 47 & 0.037\\
  {[}OBC97]N10-2 &  11:58:42 & 20:34:43   &     \nodata &  10.8   &   18.03  &           16.22$\pm0.06$ &     16.57&           14.85& 22.8 & 52 & 0.023\\
  {[}OBC97]N10-4 &  11:58:52 & 20:58:28   &      ~7478  &  ~9.6   &   19.94  &        \nodata  &     18.90&       \nodata  & 24.2 & 31 & 0.027\\
  {[}OBC97]U1-4  &  11:38:25 & 17:05:14   &      ~3450  & 32.5    &   19.00  &           17.26$\pm0.06$ &     13.08&           12.21& 23.3 & 55 & 0.025\\
  {[}OBC97]U1-8  &  11:39:23 & 16:50:31   &     \nodata &  10.3   &   19.30  &           16.55$\pm0.06$ &     18.15&           15.65& 24.1 & 70 & 0.025\\
\enddata
\tablenotetext{a}{Columns (1)-(3) give the names and positions from \citet{one00}, and the redshifts of
their HI detections are listed in column (4).  The exponential scale lengths measured
by OBC97 have been converted to half light radii using the formula
r$_e$ = 1.68$\alpha$ \citep{spa88}, and these are listed in column (5).  
Columns (6)-(7) list the B-band and I-band magnitudes inside apertures of radius 5\arcsec, and
the corresponding total magnitudes are given in
columns (8)-(9) (OBC97, OBSCI97).  
The inclination corrected B band central surface brightnesses 
are given in column (10), and the inclinations are
in column (11), both from OBC97.
Column (12) lists the galactic extinction derived from the \citet{sch98} dust maps.}
\tablenotetext{b}{Unless otherwise noted, the typical intrinsic uncertainty on the photometry 
is reported to be $\pm$0.05 mag for the aperture magnitudes,
and $\pm$0.1 mag for the total magnitudes and $\mu_{\rm B}(0)_{\rm cor}$.  
The uncertainty in the photometric zeropoints are 0.03 mag in the B-band, and 0.04 mag in the I-band  (OBC97, OBSCI97).}
\tablenotetext{c}{The intrinsic uncertainty on the inclination angle is $\pm 5^{\rm o}$ (OBC97).}
\tablenotetext{d}{Not measured by OBC97. We use this value for spectral extraction to optimize the signal-to-noise ratio.
(see Sect. \ref{specextract}).}
\end{deluxetable}


\section{Observations \& Data Reduction \label{obsdataredux}}
\subsection{Sample selection for the spectroscopic observations} \label{sampselect}

This project began during the early science operations of the 
Hobby-Eberly Telescope \citep[HET;][]{ram98} in 1999 November 
and observations continued through 2001 January.  For the first
set of spectroscopic observations, during winter 1999-2000, we chose a group of LSB galaxies 
from the sample of OBC97
which covered a range of color (0.3 $<$ B$-$V $<$ 1.35) and central surface 
brightness (22.0 $< \mu_B(0) <$ 24.0).  
Twelve galaxies were observed during the first observing period.
A new sample of six galaxies was chosen for observation during winter 
2000-2001.  These objects were selected from the \HI\ follow-up survey of 
\citet{one00}, and stand out due to their extremely high M$_{\rm HI}$/L$_{\rm B}$, and that 
they lie more than 5 $\sigma$ from the I-band Tully-Fisher relation \citep{tul77,pie88}.  
Additionally, one galaxy ([OBC97]P06-1\footnote{The official (IAU) names for
the LSB galaxies in this work originated in OBC97, and are listed in
full in Table \ref{littable}.  For convenience, however, we will use
abbreviated galaxy names without [OBC97] for the remainder of this paper.}) 
was chosen because it is the first LSB 
(albeit a giant LSB) galaxy for which CO has been detected 
\citetext{\citealp*{one00b}; see also \citealp{mat01}}. These seven galaxies fall
into the same range of color and central surface brightness as the original
sample.  We obtained spectra for a total of 19 galaxies.
Table \ref{littable} lists the galaxies in our spectroscopic sample, along with positions and
relevant data from the literature.   

\subsection{Photometry}  \label{photometry}
\subsubsection{Observations \& Basic Reductions} \label{photoobs}
We obtained new photometry for a subset of the LSB galaxies in the spectroscopic
sample, using
the Prime Focus Camera \citep[PFC;][]{cla95} on the McDonald Observatory 0.8m telescope in
November and December of 1999.  The PFC is a wide-field imager, with a 45\arcmin 
$\times\!$ 45\arcmin \ field of view.  The wide-field images include eleven of
the nineteen galaxies in the spectroscopic sample, as well as sixteen galaxies 
from the OBC97 catalog not included in our spectroscopic observations.
The instrumental configuration is given in 
Table \ref{instrumenttable}.  Images were taken through the B, R$_c$, and I$_c$ 
filters,  centered on the OBC97 survey fields C1, C3, C4, C5, N9, 
and U1.  
Multiple images were taken in each band to aid in cosmic ray rejection.  
Additionally, we took images of 10 fields containing a total of 217 standard 
stars included in \citet{lan92} to facilitate photometric calibration of the 
data.  For each field observed, Table \ref{photobstable} lists the position
of the field center, the total integration time and number of exposures through 
each filter, and the image quality of the coadded data, described in terms of
the FWHM and ellipticity of the PSF as determined from gaussian fits to radial 
profiles of more than 100 stars per image.
\begin{deluxetable}{lll}
\tabletypesize{\scriptsize}
\tablecaption{Instrumentation\label{instrumenttable}}
\tablewidth{0pt}
\tablehead{ & Spectroscopy & Photometry
}
\startdata
Telescope       & Hobby-Eberly Telescope (9.2m) &McDonald Observatory 0.8m  \\
Instrument      & Low Resolution Spectrograph           &Prime Focus Camera  \\
Dates           & 10/99-2/01 & 11/99-12/99 \\
CCD             & Ford Aerospace $3072\times 1024$  &Loral-Fairchild $2048\times 2048$ \\
                & \phm{Ford Aero}(binned $2\times 2$) & \\
Gain            & 1.82e$^-$/ADU &1.6 e$^-$/ADU \\
RON             & 7.0 e$^-$     &5.87e$^-$ \\
Pixel size spatial      & 0.465\arcsec  &1.354\arcsec  \\ 
Wavelength Range & 4300\AA\ - 7240 \AA\ & B, R$_c$, I$_c$ \\
Slit Width $\times$ Length      & 2\arcsec $\times$ 4\arcmin & \nodata\\
Grism           & 600 l/mm      & \nodata \\
Dispersion      & 2\AA /pix     & \nodata  \\
Resolution($\sigma$)    & 3.3\AA \ (190 km s$^{-1}$ @ 5200\AA)& \nodata \\
\enddata
\end{deluxetable}



\begin{deluxetable}{ccccccc}
\tabletypesize{\scriptsize}
\tablecaption{PFC Photometry Observations\label{photobstable}}
\tablewidth{0pt}
\tablehead{\colhead{Field\tablenotemark{a}} & \colhead{Filter} & 
\colhead{Exposure Time\tablenotemark{b}} & 
\colhead{N$_{\rm obs}$\tablenotemark{c}} & 
\colhead{FWHM$_{\rm PSF}$\tablenotemark{d}} & 
\colhead{$\epsilon_{\rm PSF}$\tablenotemark{e}} & 
\colhead{N$_{\rm PSF}$\tablenotemark{f}} 
}
\startdata
C1                      & B~    & ~300  & 1     & 2.8   & 0.12  & 176 \\
(08:19:52.6 +21:00:17)  & R$_c$ & ~180  & 1     & 2.7   & 0.13  & 111 \\
                        & I$_c$ & ~120  & 1     & 2.4   & 0.12  & 101 \\
C3                      & B~    & 1500  & 3     & 3.0   & 0.08  & 229 \\
(08:22:46.6 +21:03:34)  & R$_c$ & ~780  & 3     & 2.7   & 0.06  & 176 \\
                        & I$_c$ & ~600  & 3     & 3.0   & 0.04  & 127 \\
C4                      & B~    & 2100  & 4     & 2.8   & 0.05  & 508 \\
(08:26:01.3 +21:57:06)  & R$_c$ & 1380  & 5     & 2.8   & 0.04  & 201 \\
                        & I$_c$ & ~600  & 3     & 3.0   & 0.06  & 109 \\
C5                      & B~    & 2100  & 4     & 3.1   & 0.10  & 180 \\
(08:26:54.1 +22:35:23)  & R$_c$ & 1080  & 4     & 3.0   & 0.04  & 126 \\
                        & I$_c$ & ~600  & 3     & 3.1   & 0.07  & 145 \\
N9                      & B~    & 2100  & 4     & 3.2   & 0.07  & 264 \\
(10:20:29.3 +27:55:29)  & R$_c$ & ~780  & 3     & 2.8   & 0.07  & 141 \\
                        & I$_c$ & ~600  & 3     & 3.0   & 0.05  & 118 \\
U1                      & B~    & 1200  & 2     & 3.8   & 0.04  & 147 \\
(11:39:29.1 +17:04:21)  & R$_c$ & 1200  & 4     & 3.2   & 0.03  & 101 \\
                        & I$_c$ & ~720  & 3     & 3.5   & 0.06  & 145 \\
\enddata
\tablenotetext{a}{First row of each entry is the field name from OBC97.  
Second row lists the coordinates (J2000) of the field center.}
\tablenotetext{b}{Total exposure time for all observations, in seconds.}
\tablenotetext{c}{Number of individual exposures.}
\tablenotetext{d}{Mean full-width at half-maximum of a gaussian fit to the 
point-spread-function (PSF), measured in arcsec.}
\tablenotetext{e}{Mean measured ellipticity of the PSF.}
\tablenotetext{f}{Number of stars used in determination of FWHM and $\epsilon$.}
\end{deluxetable}

Data reduction was done within the
IRAF\footnote{IRAF is distributed by the National Optical Astronomy 
Observatories, which are operated by the Association of Universities for 
Research in Astronomy, Inc., under cooperative agreement with the National
Science Foundation.} environment.  The images were overscan and bias subtracted,
then a 
shutter correction was applied.  
Next, the images
were divided by a normalized dome flatfield, as well as a normalized 
illumination correction.  
The illumination correction is a highly clipped 
average of all science frames observed at low airmass, in the absence of 
clouds or moonlight, and corrects large scale gradients not reproduced by the 
dome flatfield image.  After this step, the science images were individually 
examined
for residual real gradients caused by scattered moonlight, clouds, or sky 
illumination \citep[see][]{chr96}.  Any residual gradients larger than 1\% across
the full image were fit using {\tt imsurfit} and subtracted off, with the mean 
sky level then added back in as a constant to maintain counting statistics.

Prior to coaddition, a badpixel mask was generated for each image by taking
the standard mask file for the CCD, and manually adding all the pixels affected
by satellite trails or by ghost images caused by reflections within the PFC
optics.  The coaddition
process was done using the task {\tt imcoadd} (J{\o}rgensen, in prep.) 
in the Gemini IRAF 
package.  This task determines spatial offsets between all the input images,
shifts them accordingly,  generates cosmic ray masks, and averages all the
good pixel values to produce a coadded output image.  The
photometric zeropoint offsets between all the input images and the output 
image were determined.  As long as at least one input image was taken during photometric
conditions, the coadded output can then be easily calibrated onto a standard system.
Finally, we used the HST guide star catalog reference frame to establish the
world coordinate system (WCS) for the coadded images.

\subsubsection{Object Extraction \& Standard Photometry \label{stdphot}}
Object extraction was done using SExtractor \citetext{version 2.1.6;
\citealt{ber96}}, and
the output catalogs were matched using the WCS positions to produce a single
table with photometry in all three bands (B, R$_c$, I$_c$) for each object.  

To calibrate the photometry onto a standard system, we used observations
of the \citet{lan92} selected area fields SA92, SA95, SA98, SA101, SA104, and SA113, 
as well as the fields near the standards PG0918+029, PG0231+051, 
PG1633+091, and PG0233+051.  
Observations were made each photometric night, interspersing science and standard fields. 
In total, we have 380 stellar observations in each of the three bands, covering 217 different stars.
The standard fields were observed over a range of airmasses
spanning from 1 to 3, to better constrain the extinction coefficients. In
contrast, the science fields were all observed in the airmass range 1 to 1.25. 
Average extinction coefficients covering all three photometric nights 
were used in the standard calibration.  The rms scatter
of the final standard calibration of the photometric zeropoints 
was 0.06 mag (B), 0.05 mag (R$_c$), and 0.05 mag(I$_c$).

There were no observations of field U1 taken during photometric 
conditions.  To calibrate this field, we used the
$\sim 1000$ stars in our observation in common with the USNO-A2.0 catalog
\citep{usno}.  We derived relations between the SExtractor ``best'' magnitudes
and the USNO magnitudes for both the U1 and N9 fields, and used the zeropoint
offset between these two relations to calibrate the photometry for the U1 field to
the standard system.  The USNO magnitudes do not have to be on a standard system
to make this technique work,  merely be internally consistent. 
The extra uncertainty contributed by this calibration is 0.02 mag, 
smaller than our zeropoint uncertainty as derived using SExtractor.  

We applied galactic extinction corrections to our photometry based on 
reddening values derived from the \citet{sch98} dust maps, and the
coefficients derived by them for extinction in each band versus color
excess (e.g. A$_{\rm B}/$E(B$-$V) = 4.315, A$_{\rm V}/$E(B$-$V) = 3.315, 
A$_{\rm R_c}/$E(B$-$V) = 2.673, A$_{\rm I_c}/$E(B$-$V) = 1.940).
Table \ref{photdatatable} lists the extinction corrected magnitudes 
for the 27 galaxies in the OBC97 sample that
are included in our observations with the PFC.  Also listed are the E(B-V) 
values used to determine the extinction correction in each band.

\begin{deluxetable}{cccccccc}
\tabletypesize{\scriptsize}
\tablecaption{New Photometry from PFC Observations\tablenotemark{a}\label{photdatatable}}
\tablewidth{0pt}
\tablehead{\colhead{Name} & \colhead{B(5\arcsec)} & 
\colhead{R$_{\rm c}$(5\arcsec)} & \colhead{I$_{\rm c}$(5\arcsec)} & 
\colhead{B$_{\rm tot}$} & \colhead{R$_{\rm c,tot}$} &
\colhead{I$_{\rm c,tot}$} & \colhead{E(B-V)} 
}
\startdata
C1-1 & 17.95$\pm$0.03 & 16.31$\pm$0.01 & 15.70$\pm$0.01 & 17.05$\pm$0.05 & 15.59$\pm$0.02 & 15.12$\pm$0.02 & 0.051 \\
C1-2 & 18.66$\pm$0.06 & 17.26$\pm$0.03 & 16.80$\pm$0.04 & 17.79$\pm$0.08 & 16.57$\pm$0.04 & 15.97$\pm$0.06 & 0.049 \\
C1-4 & 17.45$\pm$0.02 & 16.01$\pm$0.01 & 15.44$\pm$0.01 & 16.35$\pm$0.03 & 14.89$\pm$0.01 & 14.32$\pm$0.02 & 0.050 \\
C1-6 & 17.53$\pm$0.02 & 16.48$\pm$0.01 & 15.99$\pm$0.02 & 16.57$\pm$0.03 & 15.49$\pm$0.02 & 15.07$\pm$0.03 & 0.051 \\
C3-1 & 18.26$\pm$0.01 & 16.63$\pm$0.01 & 16.02$\pm$0.01 & 17.21$\pm$0.02 & 15.84$\pm$0.01 & 15.30$\pm$0.01 & 0.034 \\
C3-2 & 17.60$\pm$0.01 & 16.24$\pm$0.00 & 15.65$\pm$0.00 & 16.38$\pm$0.01 & 15.02$\pm$0.01 & 14.39$\pm$0.01 & 0.038 \\
C3-6 & 17.81$\pm$0.01 & 16.60$\pm$0.01 & 16.06$\pm$0.01 & 16.62$\pm$0.01 & 15.42$\pm$0.01 & 14.88$\pm$0.01 & 0.049 \\
C4-1 & 18.58$\pm$0.03 & 18.00$\pm$0.03 & 17.67$\pm$0.03 & 17.71$\pm$0.05 & 17.08$\pm$0.04 & 16.61$\pm$0.05 & 0.044 \\
C4-2 & 19.05$\pm$0.05 & 18.32$\pm$0.04 & 17.84$\pm$0.04 & 17.50$\pm$0.06 & 16.62$\pm$0.04 & 16.09$\pm$0.04 & 0.047 \\
C4-3 & 18.46$\pm$0.03 & 17.67$\pm$0.02 & 17.22$\pm$0.02 & 17.33$\pm$0.04 & 16.55$\pm$0.03 & 16.07$\pm$0.03 & 0.048 \\
C4-5 & 18.62$\pm$0.03 & 17.50$\pm$0.01 & 16.99$\pm$0.04 & 18.05$\pm$0.03 & 16.85$\pm$0.02 & 16.36$\pm$0.06 & 0.046 \\
C4-6 & 17.65$\pm$0.01 & 16.23$\pm$0.00 & 15.97$\pm$0.02 & 17.47$\pm$0.01 & 15.41$\pm$0.00 & 15.92$\pm$0.01 & 0.050 \\
C4-7 & 19.63$\pm$0.06 & 18.73$\pm$0.03 & 18.40$\pm$0.16 & 19.13$\pm$0.06 & 18.08$\pm$0.04 & 17.85$\pm$0.15 & 0.044 \\
C4-8 & 19.11$\pm$0.05 & 17.92$\pm$0.03 & 17.45$\pm$0.03 & 18.29$\pm$0.07 & 16.98$\pm$0.05 & 16.58$\pm$0.04 & 0.049 \\
C5-1 & 17.95$\pm$0.01 & 17.12$\pm$0.01 & 16.72$\pm$0.01 & 17.62$\pm$0.01 & 16.82$\pm$0.01 & 16.43$\pm$0.02 & 0.043 \\
C5-2 & 18.91$\pm$0.02 & 17.68$\pm$0.01 & 17.13$\pm$0.02 & 18.26$\pm$0.02 & 17.09$\pm$0.02 & 16.56$\pm$0.02 & 0.045 \\
C5-3 & 19.63$\pm$0.04 & 18.94$\pm$0.04 & 18.53$\pm$0.06 & 19.01$\pm$0.05 & 18.43$\pm$0.04 & 18.02$\pm$0.06 & 0.041 \\
C5-4 & 18.14$\pm$0.01 & 16.76$\pm$0.01 & 16.22$\pm$0.01 & 17.23$\pm$0.01 & 16.01$\pm$0.01 & 15.57$\pm$0.01 & 0.040 \\
C5-5 & 19.28$\pm$0.03 & 18.56$\pm$0.03 & 18.24$\pm$0.05 & 17.91$\pm$0.03 & 17.12$\pm$0.03 & 16.70$\pm$0.05 & 0.038 \\
N9-1 & 18.01$\pm$0.01 & 17.14$\pm$0.01 & 16.76$\pm$0.01 & 17.19$\pm$0.01 & 16.40$\pm$0.01 & 16.09$\pm$0.02 & 0.036 \\
N9-2 & 20.32$\pm$0.07 & 19.13$\pm$0.06 & 18.46$\pm$0.06 & 20.13$\pm$0.11 & 19.04$\pm$0.07 & 18.38$\pm$0.10 & 0.037 \\
U1-1\tablenotemark{b} & 18.17$\pm$0.02 & 16.96$\pm$0.01 & 16.42$\pm$0.01 & 17.09$\pm$0.02 & 15.91$\pm$0.01 & 15.38$\pm$0.01 & 0.023 \\
U1-2\tablenotemark{b} & 19.19$\pm$0.04 & 17.85$\pm$0.01 & 17.25$\pm$0.02 & 18.66$\pm$0.05 & 17.47$\pm$0.02 & 16.89$\pm$0.03 & 0.025 \\
U1-3\tablenotemark{b} & 17.96$\pm$0.01 & 16.08$\pm$0.00 & 15.42$\pm$0.00 & 16.90$\pm$0.02 & 14.96$\pm$0.01 & 14.33$\pm$0.01 & 0.026 \\
U1-4\tablenotemark{b} & 18.05$\pm$0.02 & 17.09$\pm$0.01 & 16.64$\pm$0.01 & 16.64$\pm$0.01 & 15.81$\pm$0.01 & 15.40$\pm$0.01 & 0.025 \\
U1-6\tablenotemark{b} & 19.32$\pm$0.04 & 18.37$\pm$0.02 & 17.95$\pm$0.03 & 18.06$\pm$0.05 & 17.08$\pm$0.03 & 16.64$\pm$0.05 & 0.027 \\
U1-8\tablenotemark{b} & 18.73$\pm$0.02 & 17.05$\pm$0.01 & 16.41$\pm$0.01 & 17.91$\pm$0.04 & 16.47$\pm$0.01 & 15.83$\pm$0.01 & 0.025 \\
\enddata 
\tablenotetext{a}{The uncertainties included in the table are the intrinsic
uncertainties associated with each measurment.  Additionally, there is a 
systematic zeropoint uncertainty of 0.06 mag in the B-band, 0.05 mag in the 
R$_{\rm c}$-band, and 0.05 mag in the I$_{\rm c}$-band.}
\tablenotetext{b}{Photometric zeropoints for field U1 were determined relative 
to field N9.  The zeropoint uncertainty for this field is 0.02 mag larger in
each band than the other fields. See Section \ref{stdphot} for details.} 
\end{deluxetable}


\subsubsection{Comparison to literature data}
\begin{figure*}[t!]
  \centerline{\psfig{figure=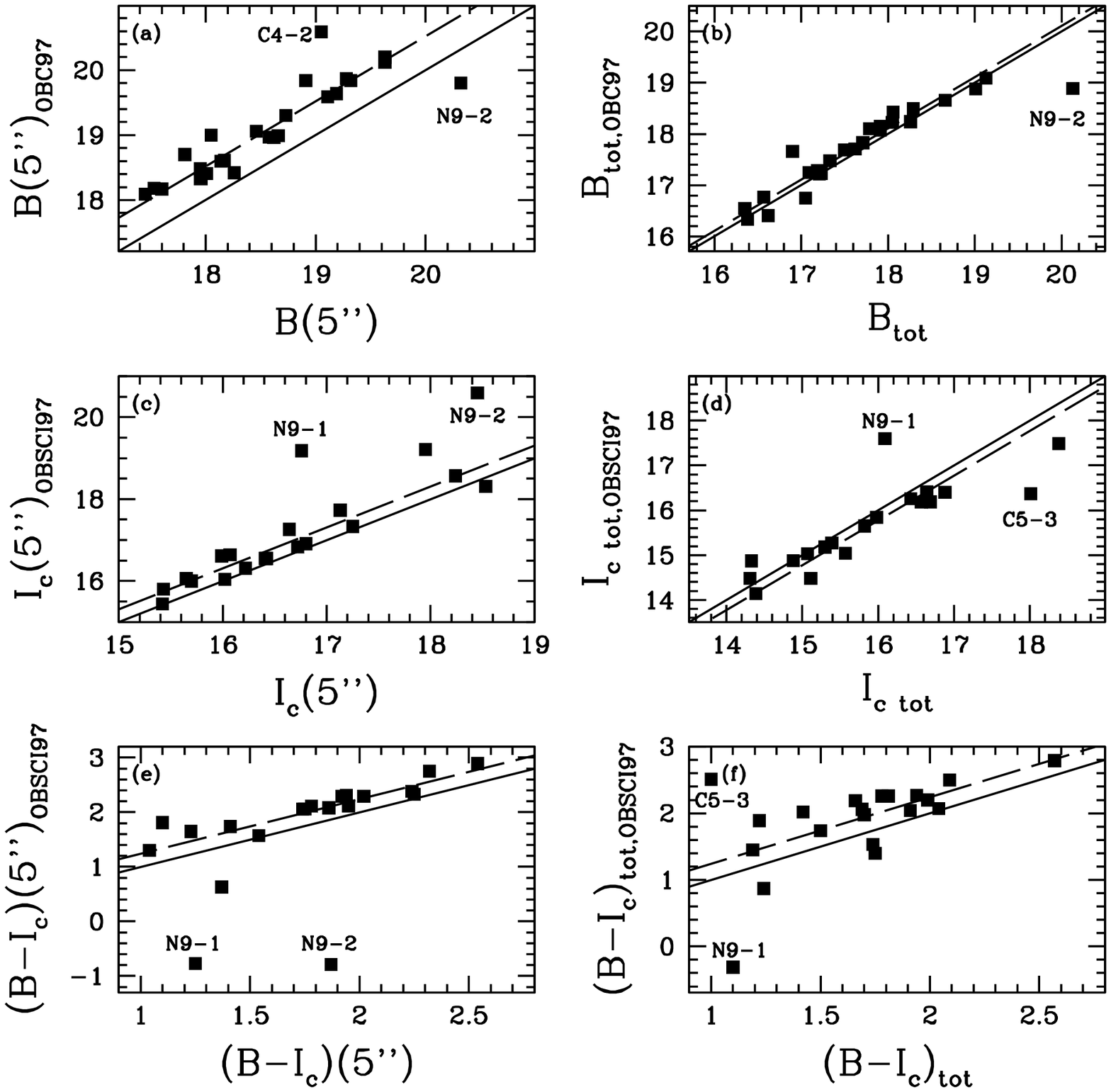,width=6.0in}}
  \caption{ Comparison between our photometry and that of 
 OBC97 and OBSCI97.  Both datasets have been corrected for galactic extinction.
 Solid lines denote lines of
 equality; Dashed lines show the best fit to the zeropoint offset. 
 Typical intrinsic uncertainties for all of the data are less than 0.05 mag,
 comparable to the size of the boxes.  
 (a) Comparison of B-band magnitudes measured inside apertures of radius 5''.  
 The fit to the zeropoint offset gives $\Delta m_{zero} = -0.52\pm0.04$ mag 
 (rms=0.19 mag).  The two furthest outliers are labelled and were not included
 in the fit.
 (b) Comparison of B-band total magnitudes.  The zeropoint offset is 
 $\Delta m_{zero} = -0.11\pm0.04$ mag (rms=0.21 mag), excluding the two 
 outliers N9-2 and U1-4 (off the plot).  The complex surface brightness profiles
 of these two galaxies were not well fit by OBC97 using single
 exponential disk fits.
 (c) Comparison of I-band 5'' radius aperture magnitudes.  Excluding N9-1 and
 N9-2, the mean zeropoint offset is $\Delta m_{zero} = -0.31\pm0.08$ mag 
 (rms=0.34 mag).  (d) The I-band total magnitude comparison.  Excluding N9-1, 
 C5-3, and U1-4 (off the plot) gives a best-fit zeropoint offset of 
 $\Delta m_{zero} = 0.23\pm0.08$ mag (rms = 0.33 mag).   
 (e) Comparison of (B$-$I$_{\rm c}$) aperture colors.  We derive a zeropoint 
 offset of $\Delta$(B$-$I$_{\rm c})(5'') = -0.24\pm0.07$ mag (rms=0.29 mag).   
 The two labelled outliers were not included in the fit.  (f) Comparison of
 the total (B$-$I$_{\rm c}$) colors.  The zeropoint offset is the same as for
 the aperture colors: $\Delta$(B$-$I$_{\rm c})(5'') = -0.24\pm0.07$ mag 
 (rms=0.30 mag), excluding the two labelled points.} 
 \label{photcompplots}
\end{figure*}

We have compared our photometry with the measurements of OBC97 
and OBSCI97 (which were also obtained with the PFC). 
Figure \ref{photcompplots} shows the comparison.

Figures \ref{photcompplots}a and \ref{photcompplots}c show the comparison 
of the aperture magnitudes in the B-band and I-band, respectively.
In the B-band, we find an offset of $-0.52\pm0.04$ mag with
an rms scatter of 0.19 mag (excluding the two most deviant measurements, N9-2 
and C4-2). Offsets are our magnitudes minus OBC97 magnitudes. In the I-band, we 
find an offset of $-0.31\pm0.08$ mag with an rms scatter
of 0.34 mag (excluding N9-1 and N9-2).  The offsets and the large
scatter may be due to the sensitivity of the small apertures used (5\arcsec) to 
changes in seeing or image quality.  In the interim between when OBC97 obtained
their data and the time when we made our observations, the PFC optics were anti-reflection coated 
and re-aligned, which improved the delivered image quality.  In our 
coadded images, the stellar PSF has a typical FHWM of 2.2 pixels (2.9\arcsec),
with an ellipticity of 0.07.  O'Neil has provided some of her images to the 
\anchor{http://ned.ipac.caltech.edu}{NED}\footnote[1]{\url{http://ned.ipac.caltech.edu}} database, and the typical image 
quality we measure for these small images is $\sim$ 3 pixels fwhm (4\arcsec).
Poorer image quality will scatter more of the light outside the small 5\arcsec\ 
aperture, resulting in fainter aperture magnitude measurements, consistent
with the offsets we see.

OBC97 derive total magnitudes from exponential and King profile 
fits to the galaxy surface brightness profiles. We compare these to the 
SExtractor ``best'' magnitudes, which are also a measure of the 
total magnitude (though systematically too faint by 6\%, see \citet{ber96} for details) 
in Figures \ref{photcompplots}b and \ref{photcompplots}d. 
The agreement in the B-band is good, except for two outliers, N9-2 and U1-4. 
Examination of the surface brightness profiles in OBC97 for
these two objects shows the profiles to be more complex than could be well
fit by a single exponential.  Excluding these two points, the offset between 
the two datasets is $-0.11\pm0.04$ mag with an rms scatter of 0.21 mag.
In the I-band, the offset is $0.23\pm0.08$ mag with an rms scatter of 0.33 mag.

We have not measured central surface brightness, $\mu(0)$, from our photometry,
so we cannot 
compare to the values in OBC97.  However, since they derive total 
magnitudes based on $\mu(0)$ and the exponential scale length ($\alpha$, 
derived over an area with radius typically larger than 12\arcsec), we expect offsets in 
$\mu(0)$ to be comparable to offsets in the total magnitudes, which are 
small in the B-band.

Figures \ref{photcompplots}e \ref{photcompplots}f show the comparisons 
of the (B$-$I$_c$) aperture and total colors, respectively,  
with the results from OBSCI97.  For both the aperture and total colors, we
derive a zeropoint offset of $-0.24\pm0.07$ mag with an rms scatter of 0.30 mag,
when we exclude the two most deviant points (N9-1 and N9-2 for the 
aperture colors, N9-1 and C5-3 for the total colors). 
For N9-1 the uncertainty on (B$-$I$_{\rm c}$) from 
OBSCI97 is 1 mag.  The typical intrinsic uncertainty
for our (B$-$I$_c$) color measurements is 0.05 mag, derived from the SExtractor 
uncertainties on the individual band magnitudes, added in quadrature.  
In the following analysis we use our new photometry when available.  For the
remaining galaxies with spectroscopic observations we use photometry from 
OBC97.  We calibrate the colors from OBSCI97 to 
be consistent with our data by applying an offset of $-$0.24 mag, and adopt an 
uncertainty in this calibration of 0.3 mag.

\subsection{Spectroscopy}
\subsubsection{Spectroscopic observations} \label{specobservations}
Spectroscopy was obtained using the Marcario Low Resolution Spectrograph \citetext{LRS; \citealt{hil98}}
on the 9.2m HET.
Table \ref{instrumenttable} details the instrumental setup.
The weather ranged from photometric to some cirrus.
Typical image quality was $\sim$2.6\arcsec \ 
FWHM during the 1999-2000 season, and $\sim$2.3\arcsec \ FWHM during the 2000-2001
winter, measured on images taken before the spectrum was observed.  
Additionally, the shape of the point spread function (PSF) was much more 
regular and symmetric during the second season, as a result of image quality
improvements made to the telescope.
The only major
instrumental change between the two observing seasons was an upgrade of the CCD controller
electronics from ``version 1'' in 1999-2000 to ``version 2'' in 2000-2001.  The
version 2 controller provides a much more stable bias and overscan level, faster
readout, and lower readout noise.

The typical exposure time was 30 minutes, and multiple observations were
made for most of the galaxies to increase the signal-to-noise ratio (S/N) and aid in cosmic ray rejection.
Table \ref{specobstable} details the spectroscopic observations.

\begin{deluxetable}{ccccl}
\tabletypesize{\scriptsize}
\tablecaption{HET Spectroscopic Observations\label{specobstable}}
\tablewidth{0pt}
\tablehead{\colhead{Name} & \colhead{Exposure Time\tablenotemark{a}} & 
\colhead{N$_{\rm obs}$\tablenotemark{b}} & \colhead{PA slit\tablenotemark{c}} & 
\colhead{Dates of observations}
}
\startdata
  P1-3  & 1195 & 1 & 46 & Nov 30, 2000\\
  P6-1  & 4200 & 3 & 45 & Dec 21,22($\times 2$), 2000 \\
  P9-4  & 3600 & 2 & 41 & Nov 23,24, 2000 \\
  C1-2  & 2014 & 2 & -66 & Dec 10, 1999($\times 2$)  \\
  C1-4  & 5160 & 3 & -67 & Nov 11, 1999($\times 2$);Mar 11, 2000 \\
  C3-2  & 3300 & 2 & -67 & Nov 10, 1999; Dec 7, 1999 \\
  C4-1  & 2770 & 2 & -67, 67 & Nov 21, 2000; Jan 26, 2001 \\
  C4-5  & 3750 & 2 & -67, 68 & Nov 12, 1999; Mar 6, 2000 \\
  C5-3  & 6700 & 4 & -68, 67 & Nov 14, 1999; Dec 14,17, 1999; Mar 13, 2000 \\
  C5-5  & 1666 & 1 & -69 & Nov 30,  2000  \\
  C6-1  & 1200 & 1 & -67 & Nov 24, 2000 \\
  C8-3  & 4800 & 4 & -67, 67&Dec 22, 2000($\times 2$); Jan 27, 2001($\times 2$)\\
  N3-1  & 3300 & 2 & -49, 47 & Dec 10, 1999; Mar 12, 2000 \\
  N9-1  & 3600 & 2 & -75 & Nov 13, 1999; Dec 8, 1999 \\
  N9-2  & 5200 & 3 & -76, 76 & Dec 13, 1999; Mar 3, 2000  \\
  N10-2 & 3600 & 2 & -66 & Dec 6, 1999; Dec 7, 1999 \\
  N10-4 & 5280 & 3 & -66, 66 & Dec 11, 1999; Mar 12,13, 2000 \\
  U1-4  & 3600 & 2 & -62 & Dec 13,14, 1999 \\
  U1-8  & 1800 & 2 & -60 & Dec 8,10, 1999 \\
NGC 3872 & ~900& 1 & -59& Dec 7, 1999 \\
UGC 3844 & ~600& 1 &-82 & Dec 7, 1999 \\
NGC 1569 & ~600& 1 & 8 & Nov 10, 2000 \\
\enddata
\tablenotetext{a}{Total exposure time for all observations, in seconds}
\tablenotetext{b}{Number of individual exposures}
\tablenotetext{c}{Measured in degrees North through East; always parallactic}
\end{deluxetable}


In total we observed 19 LSB galaxies, two HSB elliptical galaxies
(NGC 3872 and UGC 3844) and one HSB post-starburst irregular galaxy (NGC 1569).
The HSB observations provide comparisons of old and young stellar populations, 
respectively.  We
also observed a K giant star (HD 12402) to use as cross-correlation template
for the old stellar populations. 

\subsubsection{Basic reductions of spectroscopic observations\label{fringecorrect}}
The spectral data reduction was done using IRAF. 
The observations spanned about 2.5 years beginning with early science operations
of the HET, continuing through the telescope and instrument shake-down
periods, and into normal operations.  Consequently, we will go into some detail
about the basic data reductions, and steps taken to find and correct systematic
effects which might be present in the data.  Readers not interested in the
detailed steps of the basic reduction may skip ahead to 
Section \ref{specextract}.  The steps of the basic data reduction
are as follows:

\noindent
\makebox[5.0in][l]{\hspace{1.0in}1) bias and overscan subtraction}\\ 
\makebox[5.0in][l]{\hspace{1.0in}2) dark subtraction}\\ 
\makebox[5.0in][l]{\hspace{1.0in}3) flatfielding} \\
\makebox[5.0in][l]{\hspace{1.0in}4) cosmic ray cleaning} \\
\makebox[5.0in][l]{\hspace{1.0in}5) wavelength calibration and rectification}\\
\makebox[5.0in][l]{\hspace{1.0in}6) sky subtraction} \\
\indent

The dispersion axis in the LRS runs along rows.
The version 1 CCD controller electronics, in use from 1999 October - 2000 March, had a 
variable overscan level, depending on the flux hitting the CCD.  If there were rows
with high flux such as the spectroscopic trace for bright objects, then the
overscan level for those rows would be depressed.  Consequently, for bright
traces, the overscan had to be subtracted row by row.  For faint traces, or
uniform illuminations such as with the flatfields, the overscan level was
a smooth function of row number, and we subtracted a fit to the overscan, rather
than doing it row by row.  The version 2 controller produces a very stable
overscan level, which is a constant for all rows regardless of the illumination.

Mean bias frames were produced each month by combining all bias (zero exposure time)
images taken on nights when science data was obtained.  These images were 
overscan subtracted, then averaged together, with sigma clipping to eliminate
cosmic rays.  The coadded bias image was then fit with a low-order surface,
and this fit was subtracted from all the science data for that month.  
The mean of the fit was about 1.4 ADU. Month to month variations were 0.1 ADU.

Flatfield correction was done using lamp flats to correct small scale 
pixel-to-pixel
variations, and using twilight spectra to correct for the slit illumination
in the spatial dimension.  Mean lamp flats were derived for each night,
and checked against each other for variations, which were negligible.  While
fringing is evident redward of $\sim$7000 \AA, it is not seen to vary in the flats
over the course of a month. There is some variation between months, caused by
shifts in the slit position with respect to the image field center, and the 
corresponding wavelength range hitting the CCD.  The
lamp flats taken for each month were combined into a single master flat.
The rms uncertainty in these master flats is about 0.1\% redward of 6000 \AA, 
increasing slowly blueward to 0.15\% at 4800 \AA.  
 
Spectra of the sky taken during morning twilight were used to determine the slit 
illumination.  The twilight flats were averaged together, then collapsed along
the spectral dimension leaving a one dimensional ``image'' of the slit.  This
spatial profile was then applied to the lamp flat to correct for the large
scale response of the instrument, along the slit.
The way the HET operates, the spherical primary mirror is not 
moved during an observation.  Instead, tracking is accomplished by moving the
entire Prime Focus Instrument Package (PFIP) relative to the primary mirror
much in the way the Arecibo 305m radio telescope is pointed.  As objects are 
tracked across the mirror, the pupil entering the instrument changes and, 
consequently, so does the slit illumination.  We found that the twilight flats 
taken with the PFIP location fixed often did not
represent the slit illumination during our science exposures.  The gross 
vignetting pattern of the telescope optics was represented, but small gradients
remained. These gradients were corrected as part of the sky subtraction 
procedure, described below.  

Cosmic ray rejection was done using an IRAF script called {\tt speccrrej} (a 
prototype of a task which is planned for the Gemini IRAF package).  This
task makes a two dimensional model of the longslit spectrum, sensitive to 
features larger than $\sim$ 30 square pixels.  It does a fairly good job of
simulating the spectrum, as well as most of the sky lines.  However, 
sharp features are not modeled, and show up in the difference image.  
Most of these
features are cosmic rays, though strong skylines and sometimes strong
emission lines or deep absorption troughs are also flagged.  The cosmic ray
masks were all checked manually to make sure that real spectral features were
not flagged.  The flagged cosmic rays were then replaced by the values from the
model spectrum.   Pixels affected by cosmic rays were added to the bad pixel
masks for each exposure.

Spectral arc lamps (Ne and HgCdZn) were observed each night with our setup.
We used
the IRAF tasks {\tt identify}, {\tt reidentify}, and {\tt fitcoords} 
to establish the spectral dispersion function in the two dimensional spectra.  
There were
34 lamp lines used over the full wavelength range, and the fifth order
polynomial fits typically
had an RMS scatter of 0.2 \AA.  There were small (less than 2 \AA) wavelength 
offsets found for the skylines in many of the science exposures, caused
by a shifting of the slit position between setups.  These shifts caused a
wavelength shift in the dispersion function, but no change in its shape.
The wavelengths of several strong skylines (OI$\lambda$5577, OI$\lambda$6300) 
were used to correct for these offsets.
The {\tt identify} task was also used to find the spectral trace along the
spatial dimension.  In cases where the trace of a galaxy was too faint to be
seen, the fit for a brighter object observed the same night was used instead.
We then rectified the two dimensional spectra.

Prior to sky subtraction, we checked the quality of the spatial illumination 
correction for each spectrum, by measuring the flux of several strong sky lines  
along the slit.  The sky level should be constant along the slit, but due to the
varying illumination of the HET image pupil, the simple twilight flatfield was
not always effective.  In these cases, we fitted any 
residual illumination gradient with a smooth low order function, and divided 
it out.  Sky regions bracketing the spectral trace were averaged, and then
subtracted from the full 2D spectrum.  This produced good results, 
except near the OI$\lambda$5577, NaD$\lambda 5895$, and OI$\lambda$6300
skylines, which were simply interpolated over.  These bad regions are marked
in the spectra presented in Section \ref{datapresentation}.

\subsubsection{Spectral extraction, coaddition, flux calibration \label{specextract}}

We extracted the spectra in apertures of length r$_e$ (thus including the
signal within r$_e$/2 of the center, rounding off to the nearest integer
pixel), by simply summing the appropriate rows.  Since the galaxies cover a
wide range of distances, the fixed slit width of 2\arcsec \ will cover a
varying fraction of the galaxy disks.  However, since no good spectral gradient
data exist for these types of galaxies, we choose to simply note this fact and
not attempt any corrections for it.  The galaxies C3-2 and C4-1 
did not have exponential scale lengths reported in OBC97.
We chose the apertures for these galaxies based
on their light profiles along the slit and to optimize the S/N; 
we use an aperture length of 6\arcsec \ for C3-2, and 11\arcsec \ for C4-1.

A special correction had to be made for the data from 1999 November.  During
the first month of science observations with the telescope, a glass membrane
pellicle was used as part of the telescope guiding system.  This membrane
caused a fringe pattern to appear in the data, with a fringe wavelength of 
300-400 pixels, and a strength and location that varied from one exposure to 
the next.  For each of the galaxies observed in 1999 November, we obtained at
least one additional spectrum during a later month.  To remove the fringe pattern
from the data, we fit the continuum, and normalized the spectrum by this fit.
We then multiplied the normalized spectrum with the average continuum fit from
the data not affected by the fringing problem.

Most of the galaxies had multiple observations, often taken several months
apart.  Before coadding the spectra, we corrected the wavelength scales to a
consistent velocity with respect to the sun.  The spectra were then averaged.

Each night of science observations, a spectrophotometric standard star 
was observed.
The relative flux response of the instrument is very stable (with the
exception of the data from 1999 November), and we 
based the relative flux calibration on all the standard stars observed 
during an observing season.  For the 1999-2000 season, we used 15 standard
star observations, and for the 2000-2001 season we averaged 7 observations.
The stars observed were: HD 84937, HD 19445, HZ 4, HZ 44, Feige 34, Feige 66, 
Feige 67, Gl 191B2B, Hiltner 600, BD+25 3941, and BD+26 2606.
The observed spectral energy distributions (SED) were scaled to account for 
grey extinction due to cirrus, 
and then fit with a sixth order polynomial, with an RMS deviation of less
than 0.1\%.  When the different observations are scaled to the same value
at 5800 \AA, the full range of measurements at both 4800 \AA \ and 7150 \AA \ 
spans 0.25\%, which we take to be the maximum expected relative uncertainty over 
large wavelength ranges.  We then applied these relative flux calibrations to
all the coadded spectra.

In parallel with the spectral reductions and extractions, variance spectra 
were produced.  These are derived by first assuming poisson noise from the sky 
and object counts, then folding in readout noise, flat-fielding uncertainty, the 
effects of rectification and extraction, and finally the co-addition of spectra.

\subsubsection{Redshifts}

Redshifts are derived in two separate ways, (1) from cross-correlation with 
absorption line templates and (2) from measurements of the emission line 
wavelengths.

We used the IRAF task {\tt xcor} to compute the cross-correlations for galaxies
showing stellar absorption features, with
an observation of HD12402 (spectral type K1III) as the stellar template.
The restframe wavelength range used for
the computation was 4300 \AA \ - 6500 \AA\ for galaxies with no significant 
H$\beta$ emission (C1-2, C1-4, U1-8, N3-1). For galaxies with H$\beta$
emission, the cross-correlation was performed over
the range 5000 \AA \ - 6000 \AA, which still includes several strong absorption 
features.  The uncertainties determined by the {\tt xcor} task were
60 km/s for the galaxies 
without H$\beta$ emission and 75 km/s for the systems with both absorption and
emission features.

Redshifts for the emission line galaxies were based on positions of the
lines H$\beta$, [\ion{O}{3}]$\lambda\lambda$4959,5007, H$\alpha$, 
[\ion{N}{2}]$\lambda$6584, and [\ion{S}{2}]$\lambda\lambda$6717,6731.  
We take the mean velocity of all the strong lines as the
systemic velocity, and the rms scatter as the uncertainty in the velocity.  
This uncertainty was typically $\sim 30$km/s for galaxies with at least four
emission lines detected.

For the galaxies where we measure redshifts from both the emission lines and
stellar absorption features, the two measurements were always
consistent with no velocity offset, to within the uncertainties.  
We transform the redshifts into the heliocentric reference frame.
The measured redshifts are listed in Table \ref{specdatatable}.  We list
the emission line redshifts when available, otherwise the absorption line
redshifts are shown.

\begin{figure*}[htb!]
  \centerline{\psfig{figure=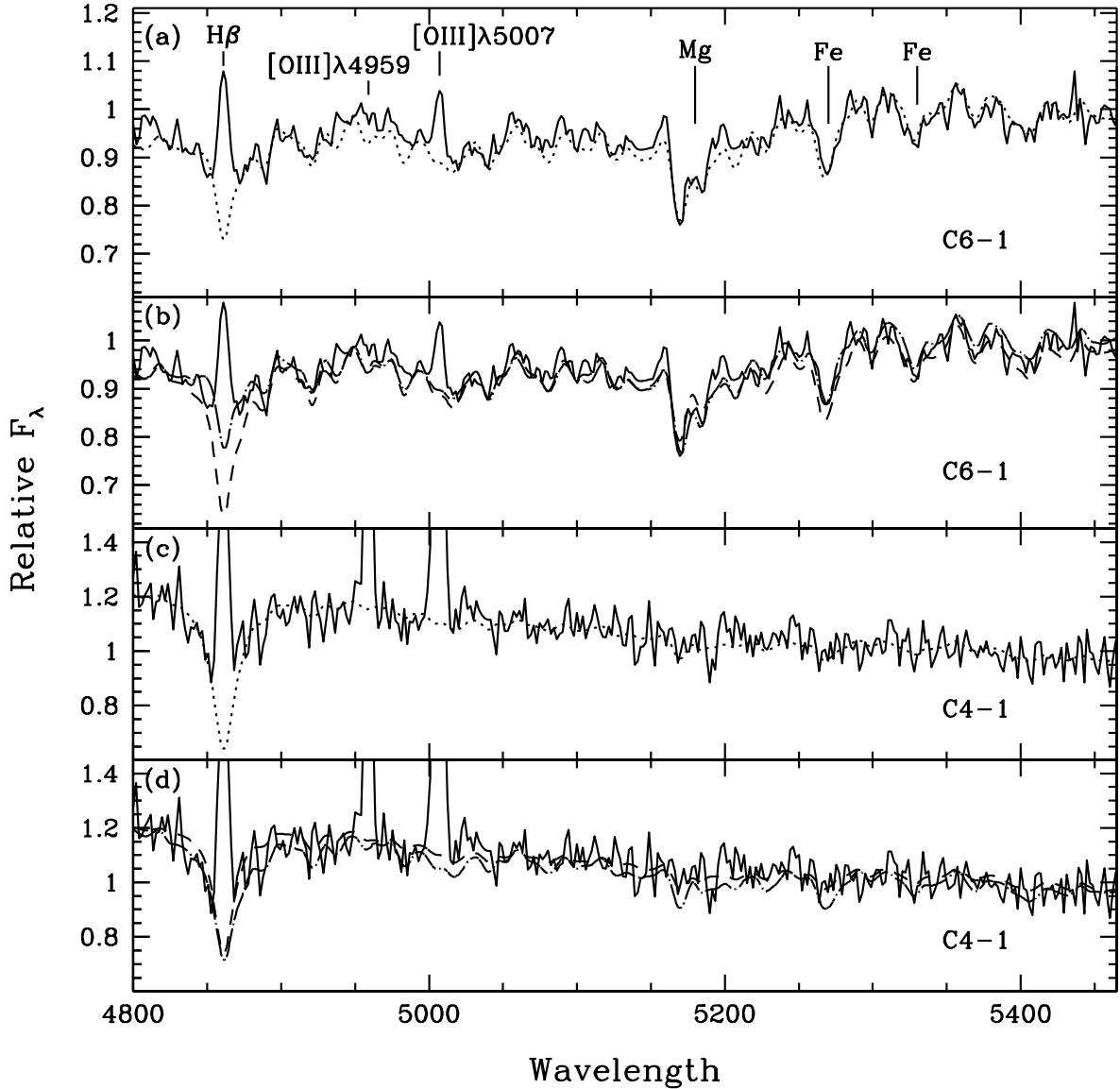,width=6.5in}}
  \caption{ Correction of the H$\beta$ emission line strengths for underlying 
H$\beta$ absorption.
Solid lines -- observed spectra; dotted lines -- best fitting model;  
dashed lines -- youngest acceptable model; dot-dashed lines -- oldest acceptable model.
The strongest emission and absorption features in this wavelength region are 
labelled in panel (a).
(a) C6-1 with the best fitting model, age=3.2 Gyr, \MET=-0.4.
(b) C6-1 with oldest (age=13.2 Gyr, \MET=-0.7) and youngest (
  age=1.0 Gyr, \MET=0.0) acceptable models. 
(c) C4-1 with the best fitting model, age=250 Myr, \MET=0.0.
(d) C4-1 with oldest (age=630 Myr, \MET=0.0) and 
  youngest (age=100 Myr, \MET=0.0) acceptable models.  Note that the depth of
the H$\beta$ absorption reaches a maximum for an SSP of 200 Myr, and so
appears weaker in both the older and younger models. }
  \label{emisfit}
\end{figure*}

\subsubsection{Emission line measurements\label{emlinemeasure}}

Emission line strengths are measured using the IRAF task {\tt fitprofs}.
Nearby lines are measured simultaneously, using the deblend option, and 
constraining the fit to have a single width for all the nearby lines. 
We use the variance spectra to estimate the ``inverse gain'' parameter, used by 
{\tt fitprofs} in its Monte Carlo estimation of the uncertainties. 

	In many cases, the measurement of the H$\beta$ emission strength is
clearly affected by underlying stellar absorption.  We correct this individually
for each galaxy.  The adopted correction procedure is as follows.
The wavelength coverage of the emission line is defined at its base, the widest 
point.  The continuum level is defined with a straight line between
the two sides of the base.  The emission line flux and equivalent width are
then measured above this continuum level.  The V1999 single burst stellar 
population model spectra, convolved to our instrumental 
resolution, are then used to estimate the effect of the underlying stellar absorption
profile on the emission measurements.  For each galaxy, we find the model which
best fits the absorption lines in the restframe wavelength range 
$4800 \mbox{\AA} < \lambda < 5450 \mbox{\AA}$.
We also determine what range of model ages and metallicities brackets a
reasonable fit to the spectrum.  The fits are all done by eye.  
We define a continuum level on the models,
by drawing a straight line between the model values at either side of the 
emission line wavelength interval, and measure the equivalent width of
absorption below this level.  We then combine the equivalent width of 
absorption seen in the models with the equivalent width of emission measured
above the interpolated continuum in the data to get a corrected H$\beta$
equivalent width measurement.  Finally, we scale the H$\beta$ flux measurement
by the ratio of the corrected to uncorrected equivalent widths to determine
the total emission line flux corrected for underlying absorption.
Examples of this technique applied to two of the galaxies are shown in 
Figure \ref{emisfit}.
Table \ref{hbetatable} lists the measured uncorrected H$\beta$ equivalent widths 
as well as the parameters of the best fitting, youngest, and oldest models, and their
corresponding absorption equivalent width corrections.

\begin{deluxetable}{lcccccccccc}
\tabletypesize{\scriptsize}
\tablecaption{H$\beta$ equivalent width corrections \label{hbetatable}}
\tablewidth{0pt}
\tablehead{
\vspace{-0.10in} 
& & \multicolumn{3}{c}{best-fit model\tablenotemark{a}} & \multicolumn{3}{c}{max EW(H$\beta$) model\tablenotemark{b}} &
\multicolumn{3}{c}{min EW(H$\beta$) model\tablenotemark{c}} \\
& & \multicolumn{3}{c}{\hrulefill} & \multicolumn{3}{c}{\hrulefill} &
\multicolumn{3}{c}{\hrulefill} \\ 
\colhead{Name} & \colhead{EW(H$\beta$)$_{\rm em}$} &
\colhead{age} & \colhead{[Fe/H]} & \colhead{EW(H$\beta$)$_{\rm abs}$} & 
\colhead{age} & \colhead{[Fe/H]} & \colhead{EW(H$\beta$)$_{\rm abs}$} & 
\colhead{age} & \colhead{[Fe/H]} & \colhead{EW(H$\beta$)$_{\rm abs}$}  \\
& (\AA) & (Gyr) & & (\AA) & (Gyr) & & (\AA) & (Gyr) & & (\AA) 
}
\startdata
C3-2 &  2.0$\pm$0.1 & 1.58&  0.2 & -1.3& 0.63 & 0.0 & -1.0& 11.0 &-0.7 & -1.9 \\
C4-1 &  7.5$\pm$0.6 & 0.25&  0.0 & -2.2& 0.10 & 0.0 & -1.8& 0.63 & 0.0 & -2.2  \\
C4-5 &  0.9$\pm$0.3 & 0.79&  0.0 & -1.2& 0.40 & 0.0 & -1.0& 1.58 & 0.0 & -1.5  \\
C6-1 &  2.7$\pm$0.3 & 3.16& -0.4 & -1.6& 1.00 & 0.0 & -1.3& 13.2 &-0.7 & -2.4 \\
C8-3 &  3.4$\pm$0.4 & 0.50&  0.0 & -1.1& 0.10 & 0.0 & -0.9& 0.79 & 0.0 & -1.2 \\
N10-2&  1.3$\pm$0.1 & 7.94& -0.4 & -1.0& 1.00 & 0.2 & -0.8& 17.4 &-0.7 & -1.4 \\
N9-1 &  11.5$\pm$0.3& 0.40&  0.0 & -2.6& 0.03 & 0.0 & -1.7& 1.00 & 0.2 & -2.9 \\
N9-2 &  27.7$\pm$0.9& 0.25&  0.0 & -4.4& 0.10 & 0.0 & -3.4& 0.63 & 0.0 & -4.5 \\
P1-3 &  7.2$\pm$0.4 & 1.00&  0.0 & -1.3& 0.40 & 0.0 & -1.0& 6.31 &-0.7 & -1.9  \\
P6-1 &  1.3$\pm$0.2 & 3.16& -0.4 & -1.0& 1.00 & 0.2 & -0.8& 11.0 &-0.7 & -1.4 \\
U1-4 &  2.6$\pm$0.1 & 0.63&  0.0 & -2.0& 0.50 & 0.0 & -1.9& 0.79 & 0.0 & -2.3 \\
\enddata
\tablenotetext{a}{Parameters for the ``best-fit'' stellar population model and 
the equivalent width of H$\beta$ absorption over the narrow wavelength region
for which we measure emission.  The ``best fit'' as well as the determination of
the maximum and minimum reasonable corrections were done by eye.  See 
Section \ref{emlinemeasure} for details.}
\tablenotetext{b}{The model which requires the largest correction to the H$\beta$
equivalent width, but is still a reasonable fit to the other absorption features in
the spectrum.  This is typically the youngest, highest metallicity, marginally acceptable model.}
\tablenotetext{c}{The model which requires the smallest correction to the H$\beta$
equivalent width, but is still a reasonable fit to the other absorption features in
the spectrum. This is typically the oldest, lowest metallicity, marginally acceptable model.}
\end{deluxetable}

We list two sets of uncertainties for the H$\beta$ fluxes in Table \ref{specdatatable}. The first set are the random uncertainties derived from the fit to the
emission line flux, the same as for the other emission lines.  Additionally,
we include a set of uncertainties showing the range of the absorption 
corrections consistent with the reasonable fitting models.

The V1999 models, which use an empirical spectral library, are limited in
their ability to model young ages at sub-solar metallicities.  We also 
investigated using the Starburst99 models \citep{star99}, which cover
very young ages at both solar and 1/20 solar metallicity.  However, the
narrow wavelength coverage of their models does not permit us to fit the 
metallic absorption features at longer wavelength, which provide the most
constraint on the range of reasonable models.

	Telluric absorption in the atmospheric B band ($\lambda \simeq$ 6840 \AA)
affected our line strength 
measurements for two galaxies.  The [\ion{N}{2}]$\lambda$6584 line was absorbed in the
galaxy C5-3.  Nebular physics predicts [\ion{N}{2}]$\lambda$6584 = 2.95 $\cdot$ [\ion{N}{2}]$\lambda$6548 \citep[p. 61]{ost89}.
For the 9 LSB galaxies with emission and where strong H$\alpha$ does not make
it impossible to measure 
the [\ion{N}{2}]$\lambda$6548 line, the mean [\ion{N}{2}]$\lambda$6584 / [\ion{N}{2}]$\lambda$6548 
ratio is 2.9$\pm$0.9. We use the measured value of [\ion{N}{2}]$\lambda$6548 in C5-3 and
the theoretical ratio to calculate the non-absorbed [\ion{N}{2}]$\lambda$6584 strength
(0.27$\pm$0.08 relative to H$\beta$). 
For the galaxy N9-1,
[\ion{N}{2}]$\lambda\lambda$6548,6584, and H$\alpha$  all suffered some
absorption.  For this galaxy, we estimate the additional uncertainty on the 
[\ion{N}{2}]$\lambda$6584/H$\alpha$ ratio 
to be 10\%, and we also likely underestimate the H$\alpha$/H$\beta$ ratio by 10\%. 
The uncertainties on the relative emission line fluxes
listed for N9-1 in Table \ref{specdatatable} have been adjusted to account for this.
In the following the [\ion{N}{2}]$\lambda$6584/H$\alpha$ ratio is simply called [\ion{N}{2}]/H$\alpha$.  

We correct the relative emission line strengths for galactic reddening using 
the Cardelli extinction law (eqs. 3a and 3b in 
\citet{car89}), with the galactic reddening measurements from \citet{sch98} as listed in Table \ref{littable}.
R$_V$=3.1 is assumed.  The amount of galactic reddening is variable across the field of NGC 1569.  We follow \citet{kob97} and adopt
the value of 0.5 mag \citep{bur82}, so that we may make a direct comparison to their results (see section \ref{litcomp}).
We have not corrected the emission line strengths for dust reddening internal to
the galaxies.  We discuss the dust content of these galaxies in Section \ref{dust}.  The emission
line ratios used for ionization determination and abundance analysis (e.g. \OFIVE/\HB\ and \NHA) utilize 
lines with small wavelength separations, and these ratios are not affected by reddening at the level seen
in these galaxies.
The H$\alpha$ equivalent widths are also corrected for
the effects of cosmological expansion, 
EW($\lambda _{\rm rest}$)=EW($\lambda _{\rm obs}) \cdot (1+z)$ 
\citep[see][page 156]{pet97}.   

\begin{deluxetable}{ccrccccclcccccc}
\rotate
\setlength{\tabcolsep}{0.05in}
\tablecaption{Spectroscopic Data\label{specdatatable}}
\tablewidth{589pt}
\tabletypesize{\tiny}
\tablehead{
\vspace{-0.15in}
 & & & & \multicolumn{3}{c}{Absorption Line Indices} & 
\multicolumn{8}{c}{Emission Line Fluxes} \\
\\
 & & & & \multicolumn{3}{c}{\hrulefill} & 
\multicolumn{8}{c}{\hrulefill} \\
\colhead{Name} & \colhead{S/N\tablenotemark{a}} & 
\colhead{cz$_{hel}$\tablenotemark{b}} & \colhead{M$_{\rm B}$\tablenotemark{c}}& 
\colhead{H$\beta$} & \colhead{Mgb} & \colhead{$<{\rm Fe}>$} &
\colhead {EW(H$\alpha$)} & \colhead{H$\beta$\tablenotemark{d}} &
\colhead{[OIII]$\lambda$5007} &\colhead{[NII]$\lambda$6548} &  \colhead{H$\alpha$} &  \colhead{[NII]$\lambda$6584} & 
\colhead{[SII]$\lambda$6717} &  \colhead{[SII]$\lambda$6731} 
}
\startdata
 P1-3  &19& 17505&-19.67&   \nodata   &0.9$\pm$0.4  &1.4$\pm$0.4 &31$\pm$1 & 1.00$\pm$0.05$^{+0.07}_{-0.04}$     &  1.48$\pm$0.05 & 0.20$\pm$0.04  &  3.44$\pm$0.07 & 0.70$\pm$0.06 & 0.89$\pm$0.05 & 0.65$\pm$0.05  \\
 P6-1  &34& 10891&-18.53&   \nodata   &3.1$\pm$0.2  &2.1$\pm$0.2 & 8.0$\pm$0.2  & 1.00$\pm$0.17$^{+0.15}_{-0.09}$& \nodata  & 0.75$\pm$0.07 & 5.0$\pm$0.1  & 2.3$\pm$0.1   &1.70$\pm$0.09  & 1.13$\pm$0.08  \\  
 P9-4  & 7&  4108&-14.67&  \nodata    &\nodata      &\nodata     &73$\pm$7    & 1.00$\pm$0.06                  &  3.1$\pm$0.1 & 0.04$\pm$0.04 & 3.76$\pm$0.0.09 & 0.30$\pm$0.05  & 0.56$\pm$0.06 & 0.36$\pm$0.06  \\  
 C1-2  &19&  4615&-16.26&2.0$\pm$0.4  &3.0$\pm$0.4  &2.2$\pm$0.3 &  \nodata   &    \multicolumn{1}{c}{\nodata}                & \nodata  &\nodata    & \nodata     & \nodata & \nodata      &  \nodata    \\ 
 C1-4  &70&  3883&-17.35&1.9$\pm$0.1  &3.5$\pm$0.1  &2.50$\pm$0.09 &  \nodata   &    \multicolumn{1}{c}{\nodata }               & \nodata  &\nodata    & \nodata     & \nodata & \nodata      &  \nodata     \\
 C3-2  &41&  4528&-17.63&   \nodata   &2.4$\pm$0.2  &2.0$\pm$0.2 &11.0$\pm$0.1  & 1.00$\pm$0.06$^{+0.18}_{-0.09}$& \nodata & 0.33$\pm$0.04 & 3.84$\pm$0.05  & 1.49$\pm$0.04   &0.82$\pm$0.05& 0.76$\pm$0.06  \\ 
 C4-1  &14&  4927&-16.48&   \nodata   &1.0$\pm$0.5  &0.8$\pm$0.5 &30$\pm$2    & 1.00$\pm$0.07$^{+0.00}_{-0.04}$&  1.53$\pm$0.07 & 0.09$\pm$0.04 & 2.64$\pm$0.09  & 0.25$\pm$0.05 & 0.56$\pm$0.06 & 0.39$\pm$0.05  \\
 C4-5  &23&  4633&-16.01&   \nodata   &1.1$\pm$0.3  &1.7$\pm$0.3 & 6.3$\pm$0.3  & 1.00$\pm$0.25$^{+0.12}_{-0.13}$&  0.5$\pm$0.1 & 0.47$\pm$0.13 & 2.9$\pm$0.1  & 0.9$\pm$0.1  & 1.3$\pm$0.1   & 0.9$\pm$0.1  \\    
 C5-3  &12& 12944&-17.21&   \nodata   &0.9$\pm$0.6  &1.0$\pm$0.5 &63$\pm$1  & 1.00$\pm$0.06     &  3.78$\pm$0.09 & 0.09$\pm$0.03 & 3.53$\pm$0.08  &0.27$\pm$0.08\tablenotemark{e} & 0.62$\pm$0.05 & 0.41$\pm$0.04  \\ 
 C5-5  & 5&  5473&-16.49&   \nodata   &\nodata      &\nodata     &25$\pm$2    & 1.00$\pm$0.28                  &  4.3$\pm$0.4   & \nodata    &   \nodata  & 5.8$\pm$0.4 & \nodata  &  \nodata    \\
 C6-1  &22& 25035&-20.13&   \nodata   &3.1$\pm$0.3  &2.2$\pm$0.3 &14.4$\pm$0.6  & 1.00$\pm$0.12$^{+0.17}_{-0.08}$& \nodata   & 0.65$\pm$0.06 & 4.3$\pm$0.1   & 1.96$\pm$0.09  &   \nodata &  \nodata    \\
 C8-3  &17&  3648&-15.91&   \nodata   &1.1$\pm$0.4  &1.0$\pm$0.4 &16.3$\pm$0.6  & 1.00$\pm$0.12$^{+0.03}_{-0.03}$&  3.2$\pm$0.2 & 0.25$\pm$0.06& 3.2$\pm$0.1   & 0.31$\pm$0.07 & 0.74$\pm$0.09 & 0.42$\pm$0.07  \\  
 N3-1  &72& 25222&-19.67&1.7$\pm$0.1  &4.7$\pm$0.1 &3.03$\pm$0.09&  \nodata   &   \multicolumn{1}{c}{ \nodata}                & \nodata  & \nodata    & \nodata     & \nodata & \nodata      &  \nodata     \\ 
 N9-1  &38& 15128&-19.38&   \nodata   &1.6$\pm$0.2  &\nodata     &41$\pm$1  & 1.00$\pm$0.02$^{+0.02}_{-0.07}$    &  0.71$\pm$0.02 & 0.12$\pm$0.01 & 2.88$\pm$0.02  & 0.80$\pm$0.01  & 0.81$\pm$0.02 & 0.57$\pm$0.02  \\
 N9-2  &13& 73629&-19.84&   \nodata   &1.3$\pm$0.6  &1.3$\pm$0.5 &   \nodata  & 1.00$\pm$0.03$^{+0.00}_{-0.04}$&  1.09$\pm$0.03  &  \nodata & \nodata  & \nodata  & \nodata      &  \nodata     \\ 
 N10-2 &38& 20668&-20.69&   \nodata   &2.8$\pm$0.2  &2.2$\pm$0.2 & 8.1$\pm$0.2  & 1.00$\pm$0.11$^{+0.17}_{-0.07}$&  0.33$\pm$0.08 & 0.70$\pm$0.09 & 3.80$\pm$0.08  & 1.90$\pm$0.06  & 0.57$^{+1.1\tablenotemark{f}}_{\pm0.07}$ & 0.32$^{+0.9\tablenotemark{f}}_{\pm0.07}$ \\
 N10-4 & 8&  7447&-16.20&   \nodata   &\nodata     &\nodata      & 7$\pm$1  &   \multicolumn{1}{c}{ \nodata}                     & \nodata & \nodata     & \nodata     & \nodata & \nodata      &  \nodata   \\
 U1-4  &53&  3416&-16.86&   \nodata   &1.0$\pm$0.2 &0.8$\pm$0.1 &11.9$\pm$0.2  & 1.00$\pm$0.05$^{+0.06}_{-0.03}$&  1.13$\pm$0.04 & 0.11$\pm$0.02 & 2.18$\pm$0.03 & 0.32$\pm$0.03  & 0.57$\pm$0.02 & 0.43$\pm$0.02  \\ 
 U1-8  &35& 32123&-20.27&2.0$\pm$0.3  &3.0$\pm$0.2  &2.2$\pm$0.2 &  \nodata   &   \multicolumn{1}{c}{ \nodata}               & \nodata  & \nodata     & \nodata     & \nodata & \nodata      &  \nodata     \\ 
\\
 NGC3872&153&3114&-21.83\tablenotemark{g}&1.47$\pm$0.05&5.04$\pm$0.05&2.83$\pm$0.04&  \nodata   &   \multicolumn{1}{c}{ \nodata}  & \nodata  & \nodata & \nodata    & \nodata & \nodata      &  \nodata     \\* 
 UGC3844&38& 3192&-20.77\tablenotemark{g}&1.5$\pm$0.2  &4.6$\pm$0.2  &3.5$\pm$0.2 &   \nodata   &   \multicolumn{1}{c}{ \nodata}  & \nodata  & \nodata & \nodata    & \nodata & \nodata      &  \nodata     \\* 
 NGC1569\tablenotemark{h}&120& -96& -17.23\tablenotemark{g}& \nodata     &0.68$\pm$0.07&0.71$\pm$0.06&95.6$\pm$0.6&  1.00$\pm$0.00 &  5.96$\pm$0.01& 0.06$\pm$0.00 &  3.92$\pm$0.00 & 0.14$\pm$0.00   & 0.20$\pm$0.00 & 0.14$\pm$0.00  \\* 
\enddata
\tablenotetext{a}{Signal-to-noise per \AA \ of the continuum, measured between 5050\AA \ and 5150\AA.}
\tablenotetext{b}{Heliocentric radial velocity: Based on emission lines, if present.  Otherwise based on cross-correlation with absorption line template.}
\tablenotetext{c}{Total absolute magnitude: Based on our photometry, if available, otherwise from OBC97. Assuming H$_0$=75 kms$^{-1}$Mpc$^{-1}$, and distances
based on redshifts in the CMB reference frame.}
\tablenotetext{d}{Emission line fluxes are normalized to f(\HB)=1. The first
set of uncertainties on \HB \ are the random uncertainties and the second set are
the range of absorption corrections consistent with the stellar population models.  If no second set is listed, the correction is negligible compared to the uncertainty due to random error.}
\tablenotetext{e}{Calculated from [NII]$\lambda$6548 $\times$ 2.95}
\tablenotetext{f}{Lower value is the intrinsic uncertainty, upper value is a systematic uncertainty, caused by poor flux calibration at the extreme red end
of the spectrum.  See Figure \ref{AGNdiagnostic}.}
\tablenotetext{g}{Data from \anchor{http://leda.univ-lyon1.fr}{the LEDA database}
(\url{http://leda.univ-lyon1.fr}), and adjusted to H$_0$=75 kms$^{-1}$Mpc$^{-1}$.}
\tablenotetext{h}{Uncertainties listed as 0.00 are less than 0.005.}
\end{deluxetable}

We present the measured emission line strengths as well as the equivalent
width of H$\alpha$ in Table \ref{specdatatable}.  
These data do not have an absolute flux calibration.  There are several reasons
for this.  As objects are tracked across the sky by moving the HET prime focus instrument platform, the
region of the primary mirror seen by the instrument changes.  Furthermore, not all of
the primary mirror is seen all the time, and thus the telescope throughput varies during an object track.
Additionally, the PSF at the spectrograph entrance slit changed (sometimes dramatically increasing in
size) during exposures, and some of the observations were taken through cirrus clouds.
Consequently, we tabulate only relative fluxes, and for convenience we have
scaled them to an H$\beta$ flux of 1.
For Table \ref{specdatatable}, we list both the random and systematic 
uncertainties on the scaled \HB \ flux,
rather than incorporating these into the uncertainties of the other flux
measurements and listing line ratios.  The uncertainties on H$\beta$ are 
incorporated into the \OFIVE/\HB\ ratios plotted in Figure \ref{AGNdiagnostic} and the 
H$\alpha$/H$\beta$ ratios shown in Figure \ref{dustfig}.

\subsubsection{Absorption line indices\label{absindices}}

We have measured the strength of the stellar absorption lines H$\beta$, Mgb,
Fe5270, and Fe5335 using the line
index definitions of the Lick/IDS system \citep{wor94}.  Several calibrations
were made to match our observations to the characteristics of the IDS 
spectrograph and put our measurements onto the Lick system.  Before measuring the
indices, it was necessary to match our spectra to the instrumental 
resolution of the IDS spectrograph.  We convolved our spectra with a wavelength
dependant Gaussian to accomplish this.  Additionally, stellar features in
galaxy spectra are broadened by the internal velocity dispersion of the
galaxies, and this broadening will have an effect on the measured strength of
the line indices. To facilitate comparison with stellar population models, we
correct our observations to the equivalent strength for a system with zero
velocity dispersion, following the method of \citet*{dav93}:
we determined the correction by taking a spectrum of a K giant star, already 
convolved to the IDS resolution, and 
convolving it with Gaussians of increasing velocity width, in 25 km/s 
increments, up to 400 km/s.  We measured the index strengths on this suite of
stellar spectra, and used the offsets between each velocity width and the 
unbroadened spectrum to correct the galaxy measurements.  Since we do not
actually resolve the velocity dispersion of the LSB galaxies, we assume
a value of 75 km/s.  The correction factors by which the measurements are
multiplied are as follows: H$\beta$ (1.004); Mgb (1.02); Fe5270 (1.02); 
Fe5335 (1.04).   The correction is such that for galaxies whose true velocity 
dispersion is larger than we have assumed (75 km/s), we will derive index
values which are weaker than the actual values for that stellar population if it
were observed with zero intrinsic velocity dispersion.  If the actual velocity dispersion
of the galaxy is 150 km/s instead of 75 km/s, then we will systematically underestimate
the H$\beta$ strength by 1\%, the Mgb strength by 3\%, the Fe5270 strength by 5\%,
and the Fe5335 strength by 8\%.  
For N3-1 and the two elliptical galaxies we correct for a velocity 
dispersion of 250 km/s.  NGC 3872 has a velocity dispersion of $\sigma$=243km/s
\citep{dav87}, while we simply assume 250 km/s for UGC 3844 and N3-1.
The correction factors are H$\beta$ (1.02); Mgb (1.17); Fe5270 (1.16);  
Fe5335 (1.36).

Uncertainties for the index strengths are derived from the variance spectra.
Essentially, we use poisson statistics on the flux in both the line and
continuum bandpasses, additionally accounting for the contribution caused
by the readout noise, and flatfielding, relative flux calibration, and sky
subtraction uncertainties.

In several cases, residuals from imperfect subtraction of strong skylines 
(O$\lambda$5577, NaD) fall in the index bandpasses.  When the residuals lie in 
the index continuum bandpasses,  we correct the residuals by interpolating
over the affected region, using values determined from comparison with the 
V1999 stellar population models.  Observations corrected in  
this way are N10-2 (Mgb red continuum, Fe5270 blue continuum), N3-1 (Mgb blue 
continuum), U1-8 (Fe5270 red continuum), and C6-1 (Mgb blue continuum).  In some
cases, the sky residuals lie on the absorption line itself, and this cannot
be corrected.  Both Fe5270 and Fe5335 are affected in N9-1.  
The Fe5335 line is affected for C5-3 and U1-8 while the Fe5270 line is affected
for P1-3.
There are strong correlations between Fe5270, Fe5335, and $<\!{\rm Fe}\!>$ 
(defined as $<\!{\rm Fe}\!>$ = (Fe5270 + Fe5335) / 2). Using the LSB
galaxy data not affected by the large sky subtraction uncertainties, we derived 
the relation Fe5335 $= (1.14\pm0.08) \cdot$ Fe5270 $- (0.74\pm0.24)$.  We use this
relation to correct the affected indices.

Intrinsic emission from these galaxies could affect the line indices Mgb and 
H$\beta$.
\citet{gou96} have found that the [NI] emission doublet at 5199 \AA \ can
affect the strength of the Mgb index in early-type galaxies which have ionized
gas emission.
Many of these LSB galaxies have emission lines, so this is cause for concern.  
However, the
[NI] doublet is strong in LINERs, not \HII\ regions,  and none of these galaxies 
show line ratios consistent with this type of nuclear activity (see Section
\ref{galactivity}).  Additionally,
since these are integrated spectra, rather than nuclear spectra, 
we do not expect the Mgb index to be contaminated.  The \HB \ index, on the
other hand, is certainly affected by emission in all but three of the galaxies
in our sample.  For one more galaxy (U1-8), the \HB\ emission must be small.
The observed wavelength range for U1-8 does not include either \HA\ or \OTHREE.  While
the spectrum is dominated by absorption features, there is a small amount of emission
seen at \OFIVE\, which suggests that the \HB\ absorption index will also be affected
by emission.  The equivalent width of \OFIVE\ emission is 1.2$\pm0.4$ \AA.  Trager et al. (1999)
determine the relation between \OFIVE\ and \HB\ emission strength in elliptical galaxies
to be  EW(H$\beta$) $\simeq 0.6 \cdot$EW([\ion{O}{3}]$\lambda 5007$).  
This ratio is metallicity sensitive, and
will tend to decrease with decreasing metallicity.  The integrated galaxy spectra
studied by \citet*{kob99} show ratios as low as  EW(H$\beta$) $\simeq 0.25 
\cdot$EW([\ion{O}{3}]$\lambda 5007$) at the metallicity appropriate for 
U1-8 ($\simeq 0.4$ Z$_\odot$; see Figure \ref{hbplots}).  
We therefore apply a correction to the \HB\ absorption index for U1-8 of $0.5\pm0.3$ \AA.

For the 15 LSB galaxies with strong emission lines, a correction of the \HB\ absorption index 
for the \HB\ emission (and its 
uncertainty) would be a significant fraction of the \HB\ absorption strength.  
Consequently, we do not present this absorption index for any of the other LSB 
galaxies.

The absorption line strengths of \HB \ , Mgb, and \meanfe\ are presented 
in Table \ref{specdatatable}.

\subsubsection{Comparison to literature data\label{litcomp}}

Our sample has a very
wide range in radial velocities, from 3000 km/s to more than 
60,000 km/s (z=0.2).  These same galaxies have been observed in the \HI\ survey 
of \citet{one00}, who give detections for twelve of them.
We find several discrepancies between our redshift measurements and the 
redshifts reported in that survey.  The \HI\ survey only covered
the velocity space out to 13,000 km/s so more distant objects would not have 
been detected.  However, for the galaxies P1-3, C6-1, and N9-2, lower \HI\ redshifts 
must have been attributed incorrectly to these higher redshift galaxies.  
The \HI\ detections for P1-3 and C6-1 can be explained by beam confusion.
At 21cm, the beam-size of the Arecibo 305m radio telescope is almost 3\arcmin, and both of
these galaxies appear in the sky at a projected distance of less than 3\arcmin 
from a large spiral galaxy at lower
redshift.  The \HI\ detection ascribed to P1-3 is probably NGC 7631 
(cz$_{\rm HI}=3746$ km/s; cz$_{\rm opt}=3741$ km/s 
\citetext{\citealp{fal99}; hereafter UZC}, 
and the detection ascribed to C6-1 is probably NGC 2595 
(cz$_{\rm HI}=4322$ km/s; cz$_{\rm opt}=4320$ km/s, UZC).   
The mis-identification also explains
why these galaxies appeared to have extreme M$_{\rm HI}$/L ratios, and 
lie more than $5\sigma$ from the I band Tully-Fisher relation \citep{one00}.  
Similar cases of beam confusion explain
discrepant measurements for the lower redshift LSB galaxies as well.  The
detection ascribed to C8-3 is probably UGC 4308 (cz$_{\rm HI}=3570$ km/s; 
cz$_{\rm opt}=3566$ km/s, UZC), and the detection of C5-5 is more likely
UGC 4416 (cz$_{\rm HI}=5524$ km/s; cz$_{\rm opt}=5555\pm45$ km/s, UZC).

There are also cases where the optical and \HI\ redshifts are discrepant, but
beam confusion is not obviously the reason:  N9-2 (listed above),
C4-1, C1-2, and possibly P9-4 (where the optical and \HI\ velocities differ
by $3\sigma$).  C1-2 in particular is a hard case to explain the \HI\ detection.
The difference between optical and \HI\ velocities is $\sim$3900 km/s so the
\HI\ detection is clearly a different object.  However, there are no obvious
candidates within 15\arcmin\ of the pointing, which lie at the redshift of the \HI\
detection, and very few bright objects in that area for which optical redshifts
were not obtained by the redshift survey of \citet{zab00}.  We are forced to
conclude that either the \HI\ observation was errant, due to either noise
or incorrect pointing,  or that
there are very large \HI\ clouds with little associated optical light.
Two starless objects have been found by the HIPASS survey \citep{kil00,ryd01} 
but the first one had only a 4 km/s velocity width, and the second was associated
with the galaxy NGC 2442, so neither one is a good analogue.

The remaining four galaxies, C5-3, P6-1, N10-4, and U1-4 are in good
agreement, less than $1\sigma$ separating the optical and \HI\ measurements.  
 
Recently, \citet{chu02} used the VLA to map the \HI\ around the galaxies P1-2, 
P1-3, C6-1, C4-1, and C4-2 from the OBC97 sample.  They also 
discovered the beam contamination issues affecting the \citet{one00}
Arecibo 305m single-dish results.  The maps
for P1-2, P1-3 and C6-1 show clearly that the \HI\ is all at the position of 
neighboring spiral galaxies.  The \HI\ detection for C4-2 was confirmed, but the
map of C4-1 did not detect any \HI\ even though the VLA observations were 
sufficiently deep to detect the \HI, if it were present at the level measured at
Arecibo.  This result suggests that the N9-2 and C1-2 \HI\ detections may also
be noise artifacts, and not starless gas clouds.

We compare our emission line strength measurements for the post-starburst dwarf
galaxy NGC 1569 with the results of \citet{kob97}.  Our aperture lies perpendicular 
to the C and D slit placements of their work and near the two super-star clusters.  
\citeauthor{kob97} found that the combination of galactic and intrinsic reddening is both 
large and variable over the face of this galaxy.
In accordance with their methods, we choose a value 
for the total extinction (galactic plus intrinsic) which brings our measured 
H$\alpha$/H$\beta$ ratio to 2.83 after applying the reddening correction.
We measure a total E(B$-$V) of 0.83, which is within the range of values found by
\citeauthor{kob97} (following \citeauthor{kob97} and \citet{bur82}, we assume that
0.5 mag of the reddening originates in our galaxy).
Once we apply
the extinction correction to all the emission lines, our relative flux
measurements for the lines H$\gamma$ (0.45$\pm$0.01), [\ion{O}{3}]$\lambda\lambda$4959,5007 (1.94$\pm$0.01, 5.72$\pm$0.01, respectively), 
[\ion{S}{3}]$\lambda$6312 (0.008$\pm$0.001), [\ion{N}{2}]$\lambda\lambda$6548,6584 (0.042$\pm$0.001, 0.098$\pm$0.001, respectively), 
HeI$\lambda$6678 (0.023$\pm$0.001), 
and [\ion{S}{2}]$\lambda\lambda$6717,6731 (0.137$\pm$0.002, 0.100$\pm$0.001, respectively) all lie within the range covered by 
the \citeauthor{kob97} measurements for this part of the galaxy.  The only discrepant line
is [\ion{O}{3}]$\lambda$4363 (0.124$\pm$0.007), which lies on the blue end of our spectrum where the 
flux calibration is less secure.  Over the wavelength range 4800 \AA \ to 
7000 \AA, our emission line measurements are fully consistent with the results
of \citeauthor{kob97}.

The absorption line index strengths for NGC 3872 are compared with the
results of \citet{tra98}.   We extract the spectra using a spatial aperture
equivalent to the Lick/IDS aperture, and make velocity dispersion corrections
appropriate for $\sigma_{\rm NGC3872}=250$ km/s.
\citet{tra98} measure the following values:
H$\beta$:  1.28$\pm$0.22; Mgb: 4.91$\pm$0.30; \meanfe: 2.22$\pm$0.22.
Our measurements for H$\beta$ and Mgb are within 1$\sigma$ of their values.
For \meanfe, our measurement is about 2.5$\sigma$ higher than theirs.

\section{Presentation of the spectra\label{datapresentation}}
\begin{figure*}[th!]
 \label{lsbplots}
  \figurenum{3a}
  \centerline{\psfig{figure=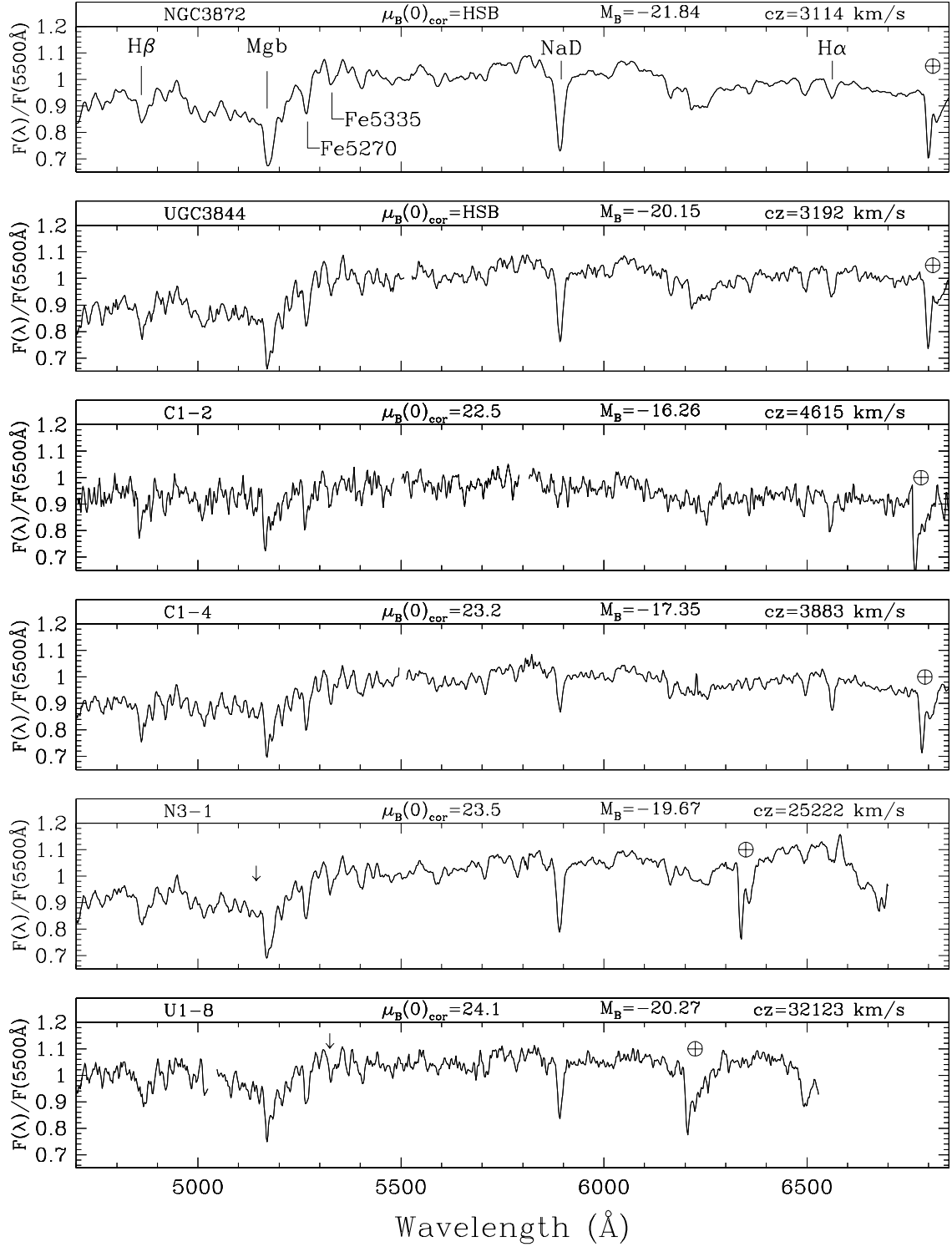,width=6.5in}}
  \caption{ 
Presentation of spectra for galaxies at low redshift (z $<$ 0.11).  The strongest
absorption features are labelled in (a), and the strongest emission lines
are labelled in (c).
(a) Objects with absorption dominated spectra.  The top two galaxies are bright 
ellipticals for comparison purposes, while the bottom four spectra are LSB galaxies.  
Gaps in the data mark regions affected by large sky subtraction residuals. 
The location of the atmospheric B band absorption ($\simeq 6840$ \AA\ in the observed frame) 
is marked with $\oplus$.
Vertical arrows denote the regions affected by sky subtraction residuals which were corrected via
interpolation using the V1999 stellar population model spectra (see Section
\ref{absindices}).
Note that
with the high S/N of the observations for NGC 3872, C1-4, and N3-1, the majority of
the bumps and wiggles are real features, and not noise.  }
\end{figure*}

\renewcommand{\textfraction}{0}
We present the spectra in Figures~\ref{lsbplots}a-e and \ref{highzspec}.
The spectra for these 19 LSB galaxies qualitatively resemble 
the spectra of HSB galaxies over the full range of spectra seen for all 
Hubble types, 
from ellipticals all the way through late-type spirals and 
starbursting irregular galaxies.  For comparison, we show our
spectra of the two elliptical galaxies, NGC 3872 and UGC 3844, in 
Figure \ref{lsbplots}a, and the spectrum of the post-starburst irregular 
galaxy NGC 1569 in Figure \ref{lsbplots}d.

\begin{figure*}[th!]
  \figurenum{3b}
  \centerline{\psfig{figure=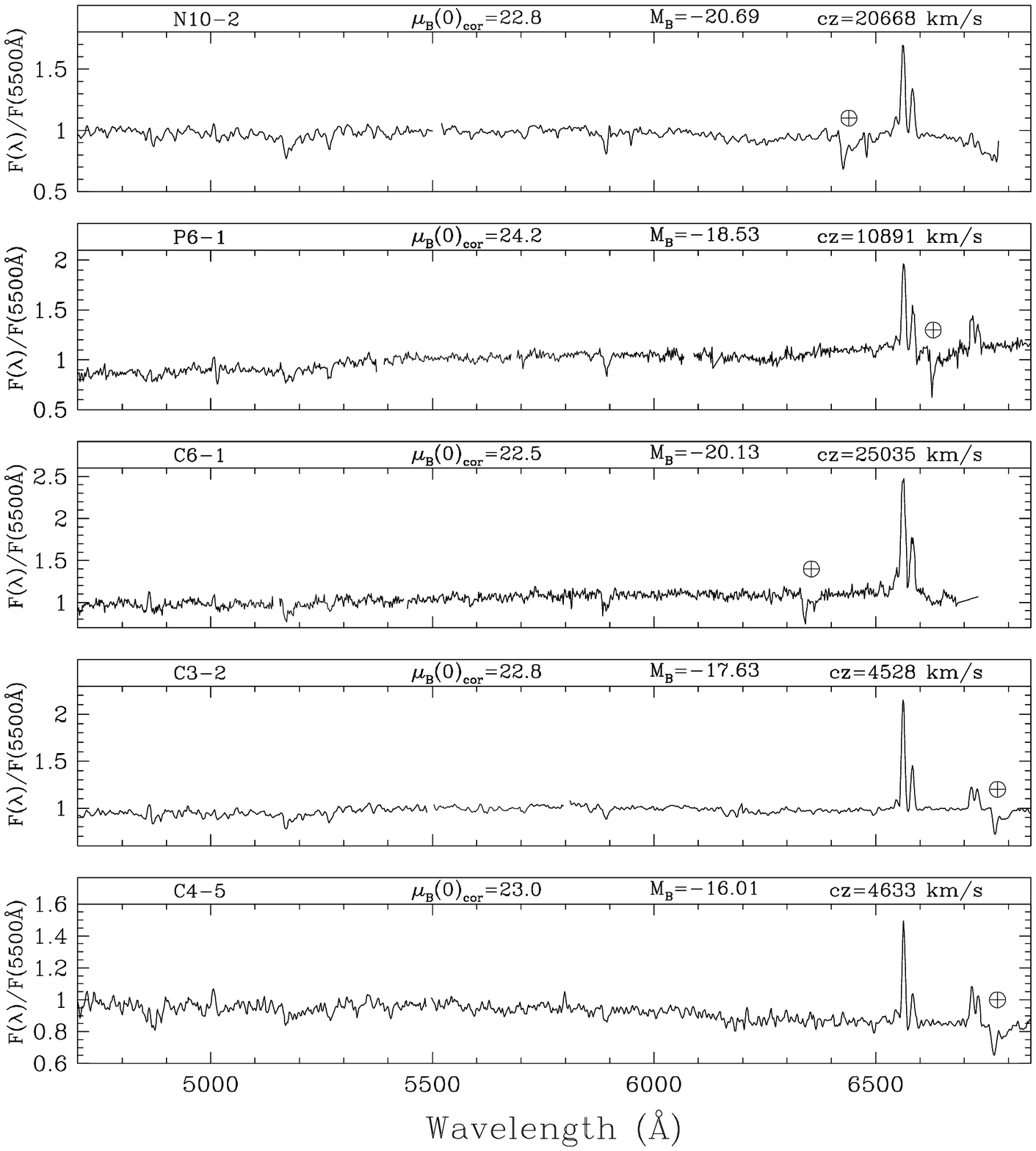,width=6.5in}}
  \caption{  
Spectra for LSB galaxies with emission lines, ordered by the strength of
[\ion{N}{2}]$\lambda$6584 relative to \HA. }
\end{figure*}

\begin{figure*}[th!]
  \figurenum{3c}
  \centerline{\psfig{figure=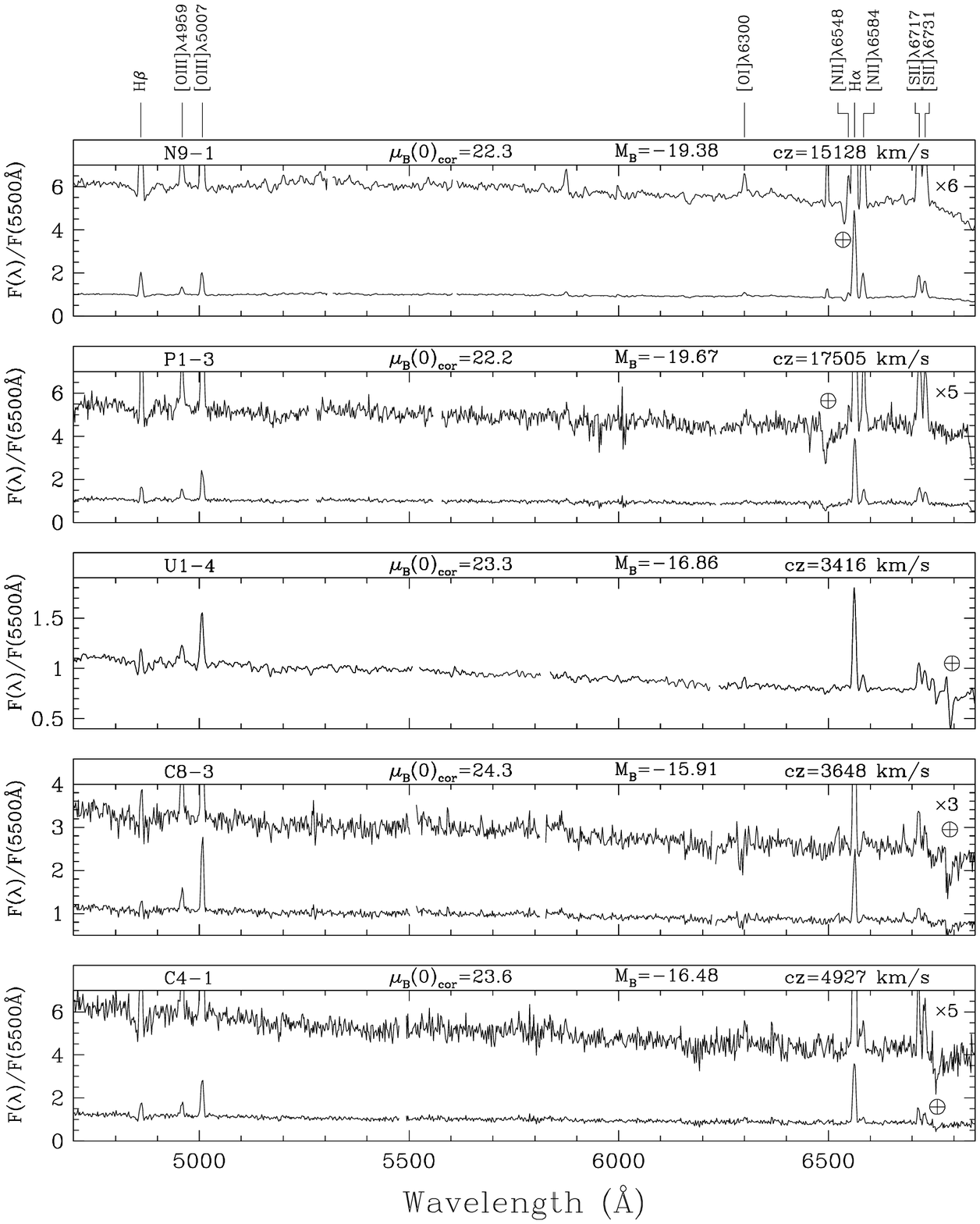,width=6.5in}}
  \caption{  
Spectra for LSB galaxies with emission lines, ordered by the strength of
[\ion{N}{2}]$\lambda$6584 relative to \HA. }
\end{figure*}

\begin{figure*}[th!]
  \figurenum{3d}
  \centerline{\psfig{figure=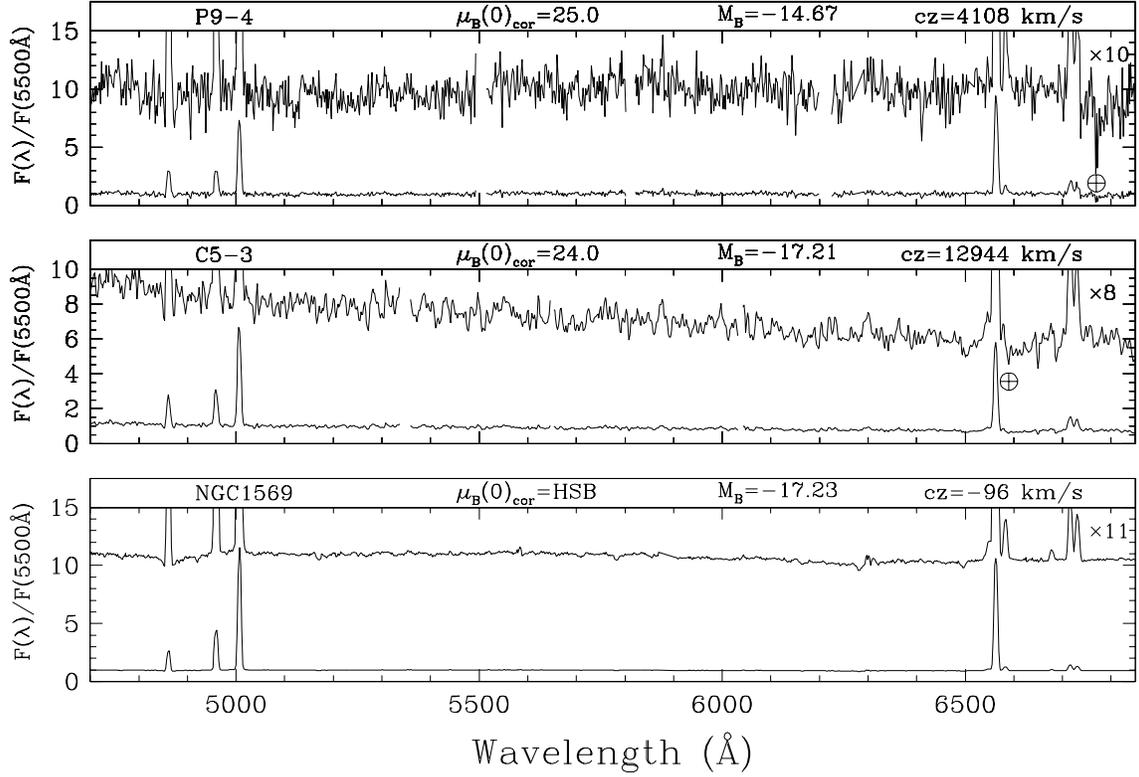,width=6.5in}}
  \caption{  
Spectra for galaxies with very weak [\ion{N}{2}]$\lambda$6584 and large
EW(\HA).  NGC 1569 is a high surface brightness post-starburst galaxy.  }
\end{figure*}
\renewcommand{\textfraction}{0.2}

\begin{figure*}[th!]
  \figurenum{3e}
  \centerline{\psfig{figure=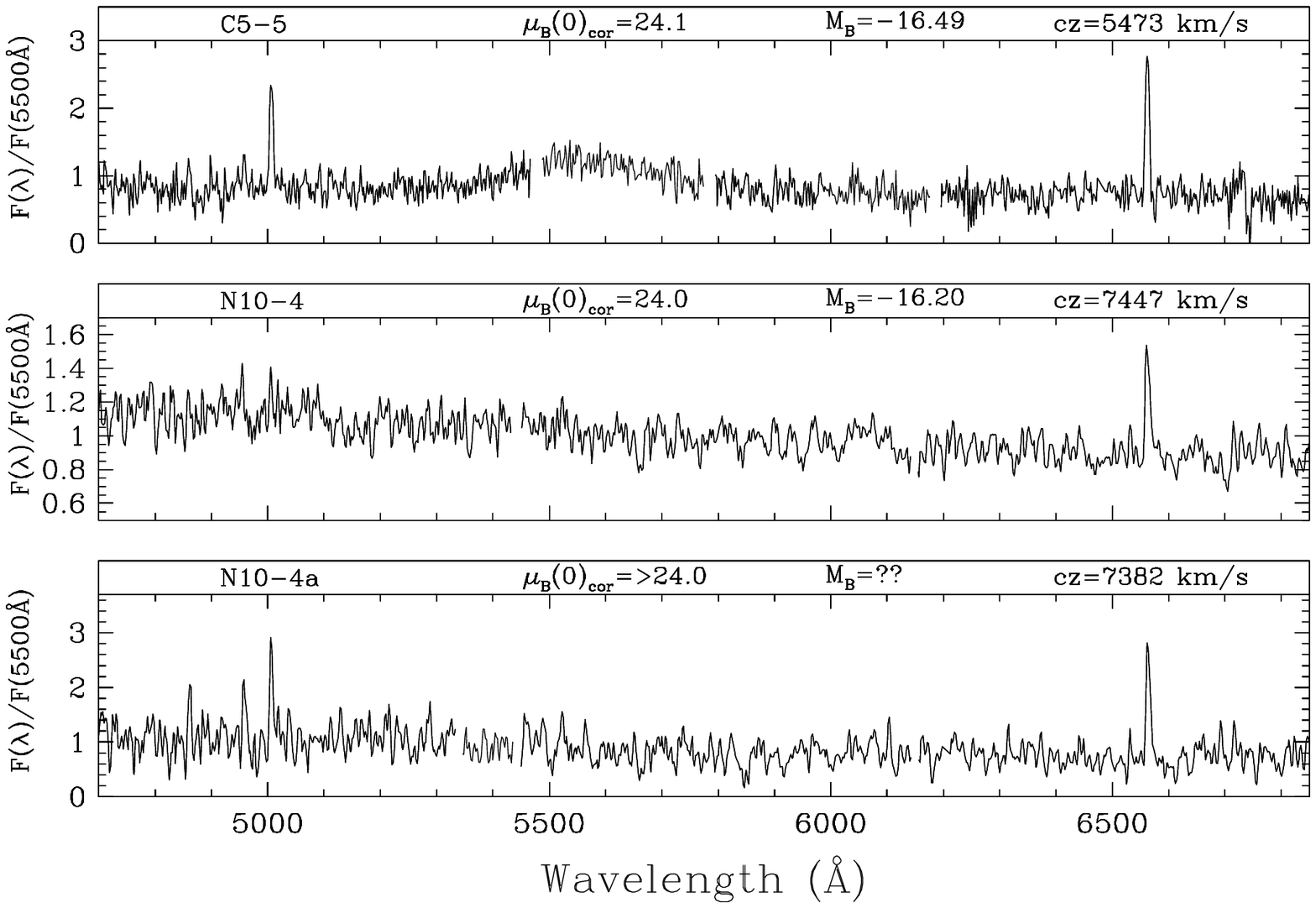,width=6.5in}}
  \caption{ 
The S/N of the spectra for N10-4 and C5-5 are too low to estimate the 
galaxies' metallicities, but the spectra are suitable for determining redshifts.
See the text for a description of N10-4a.  }
\end{figure*}

\setcounter{figure}{3}

The emission line galaxies have been ordered by the strength of their 
[\ion{N}{2}]$\lambda$6584 emission relative to H$\alpha$, from strongest to weakest.  
This is essentially a sequence of decreasing metallicity as we will show  
in Section \ref{metindicator}. 
The strength of the [\ion{O}{3}]$\lambda$5007 emission line increases along the 
sequence, due to a combination of the decreasing metallicity and the associated 
decreasing amount of dust reddening in the emission regions. 
This will be discussed in Section \ref{dust}.

\begin{figure*}[th!] 
  \centerline{\psfig{figure=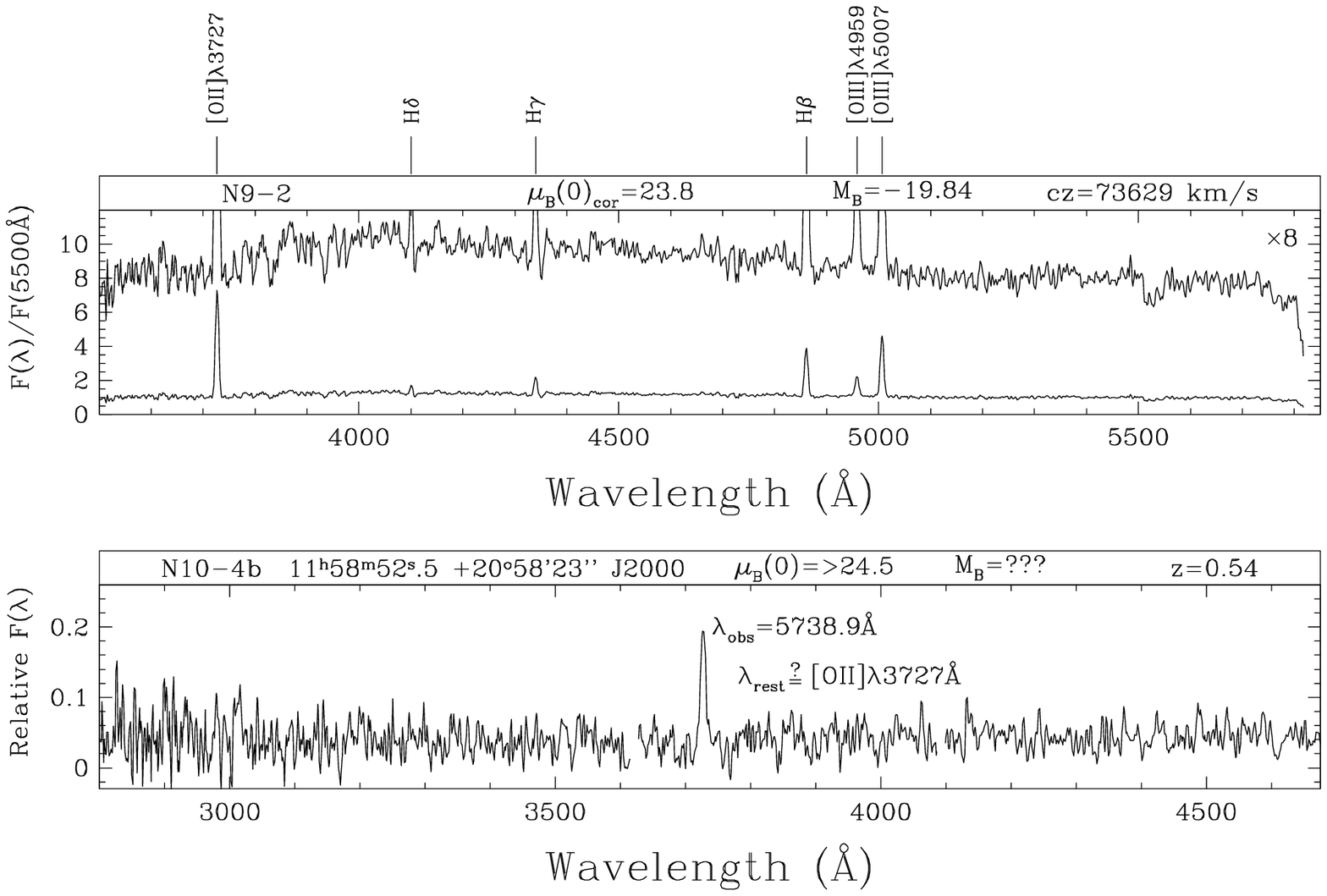,width=6.5in}}
\caption{Spectra for galaxies with z$ > 0.2$: a) N9-2 is at 
z$\simeq$0.245, and is most likely a normal high surface brightness spiral 
galaxy affected by cosmological surface brightness dimming. b) N10-4b is 
detected as a single emission line at $\lambda = 5738.9$\AA, which we identify
as [\ion{O}{2}]$\lambda3727$ emission at z=0.54.  No object is seen 
in the broadband imaging with the HET or PFC at the location of this
object, suggesting it could be a high redshift LSB galaxy.}
\label{highzspec}
\end{figure*}

\begin{figure}[tb!]
  \centerline{\hspace{0.3in}\psfig{figure=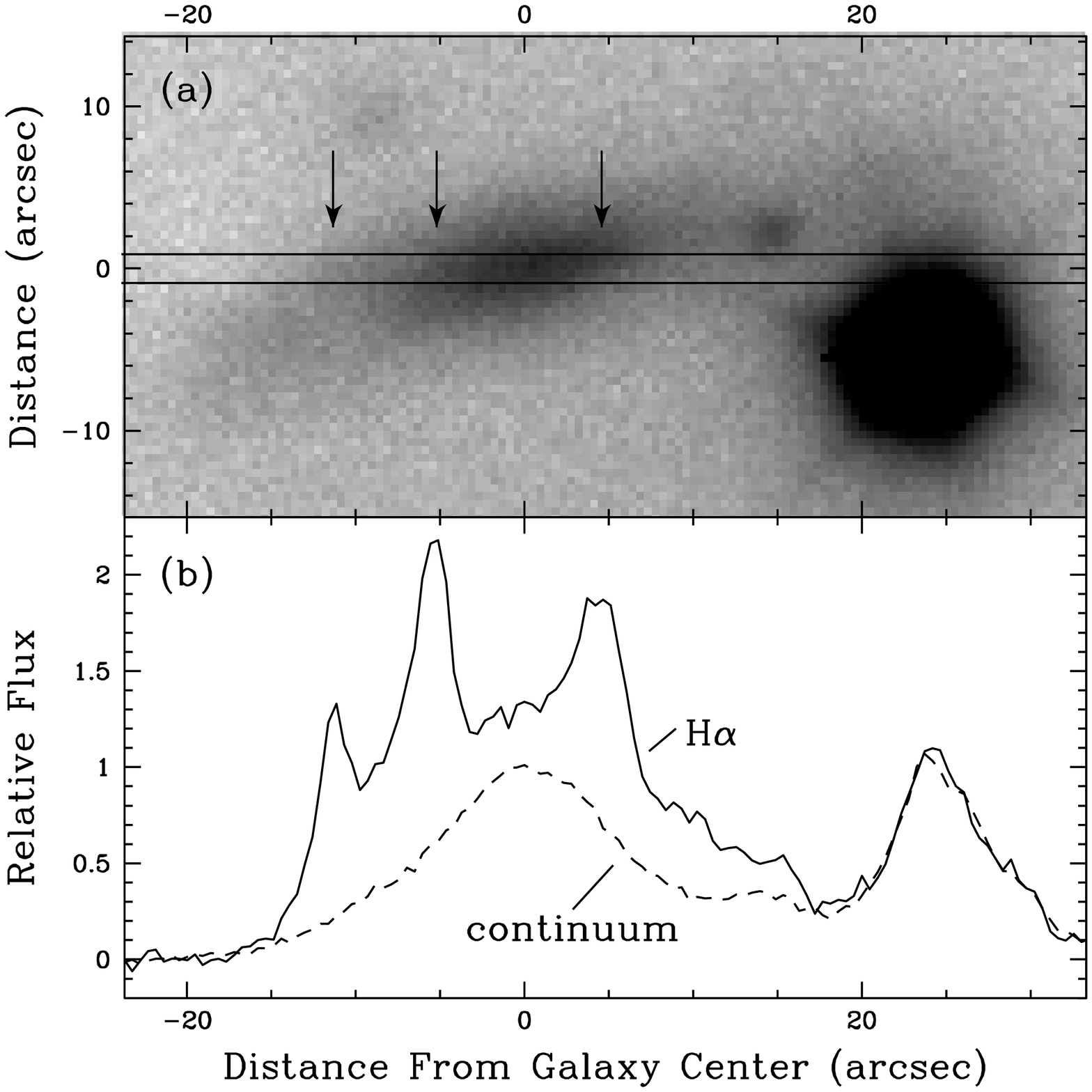,width=4.0in}}
\caption{
The galaxy U1-4: (a) image of U1-4 taken with the HET during set-up for the
spectroscopic observation.  The object at the lower right is a bright star.  The three
arrows correspond to the regions of the slit where emission peaks are seen.  (b) spatial cut
across the 2D spectrum.  Solid line - flux of the H$\alpha$ line (average over 6569.0\AA\ - 6565.9\AA\ 
in the U1-4 restframe); Dashed line - continuum flux blueward of H$\alpha$ (average 
over 6495.3\AA\ - 6540.8\AA\ in the U1-4 restframe).  The H$\alpha$ emission is more spatially extended
than the stellar continuum light, and three peaks are seen in the emission (spatial) profile, at
-12'', -5'' and +5'' from the galaxy center.  We don't see the emission regions in (a), which is a 
broadband image.  The spatial profile of the star in (b) shows the image PSF for this observation
(the wavelength regions in the restframe of the star are 6633.7\AA\ - 6640.7\AA\ and 
6569.3\AA\ - 6615\AA).}
\label{u14plot}
\end{figure}

The spectral extraction was done using global apertures, which blend together 
stellar light and line emission from both \HII\ regions and diffuse emission regions.
While we do not generally have sufficient spatial resolution to observe gradients in the
galaxy properties, we note that the galaxy U1-4 has a clear inner and outer region.
The inner region of radius 3\arcsec\ is dominated by a bright continuum with stellar
absorption features.  Outside this region there is a peak in the emission on both
sides of the nucleus at about 5\arcsec.  Further out, both the emission and 
continuum fall off.  This is illustrated in Figure \ref{u14plot}.  
The spectroscopic aperture we use for this galaxy is 
2\arcsec\ $\times$ 32\arcsec\ and includes light from both the inner and outer regions.

Analysis of the galaxies in Figure~\ref{lsbplots}(a-d) will be presented in
the following sections.  Here we discuss in detail the galaxies presented in Figures
\ref{lsbplots}e and \ref{highzspec}.

The galaxy N9-2 (Figure \ref{highzspec}) has
a redshift of 0.246 (cz=73629 km s$^{-1}$).  The restframe wavelength coverage for
this galaxy is significantly bluer than the rest of the sample, and we do not have 
observations of the wavelength region near H$\alpha$.  After applying the (1+z)$^4$ 
cosmological
surface brightness correction and a $k_{\rm B}$ correction appropriate for an Scd galaxy 
near z=0.2 of 0.45 mag \citep{fre94}, N9-2 has $\mu_B(0)=22.1$ well within the range of 
high surface brightness galaxies.  N9-2 is just a normal HSB spiral galaxy, seen at
moderate redshift, and so we remove it from the rest of the analysis.

The galaxies N10-4 and C5-5 (Figure \ref{lsbplots}e) both have $\mu_B(0)=24$ and are among the faintest in our
sample. The S/N of the observations are sufficient to determine redshifts for these two
galaxies, but insufficient for a more detailed analysis.  N10-4 is particularly interesting,
however, because of its companions.  There were two other objects which serendiptously
fell onto the slit during the observations. We refer to these as N10-4a and N10-4b.
N10-4a (11$^h$58$^m$52$^s$.5, 20$^\circ$58\arcmin23\arcsec\ J2000)  
lies 4h$^{-1}$ kpc away (12\arcsec) from the center 
of N10-4.  We
detect emission lines of \HA, \HB, and [\ion{O}{3}]$\lambda\lambda$4959,5007 in the companion,
shifted 65 km s$^{-1}$ blueward (cz$_{hel}$=7382 km s$^{-1}$).  We do not detect anything
in the broadband setup image at this position, making it difficult to determine whether
this is a separate system, or a knot of emission in the outer disk of N10-4.
Additionally, there is another companion 16\arcsec .5 in the opposite direction along the
slit (N10-4b, 11$^h$58$^m$50$^s$.8, 20$^\circ$58\arcmin33\arcsec.5 J2000). 
This object is also not seen in the setup 
image, or in images from the 
digitized sky survey. 
We do not detect any continuum from this object and
only one emission line is seen.  This line is detected in all four observations of N10-4.
The spectrum is presented in Figure \ref{highzspec}.
The observed wavelength of this emission line is 5738.9 \AA, which translates to 5600 \AA\ in
the restframe of N10-4.  This wavelength does not match any commonly observed galaxy emission lines.
Instead, we suggest that this object is a background galaxy, and the single line
observed at 5738.9 \AA\ corresponds to [\ion{O}{2}]$\lambda$3727 seen at a redshift of 0.54.
At this redshift, our
restframe wavelength coverage is 2790 \AA \ to 4700 \AA, where we do not expect to see any other 
strong emission lines.  

The SED of C5-5 (Figure \ref{lsbplots}e) has a bump in it, centered at 5600 \AA\ (5670 \AA\ in the observed frame)
which we do not have an explanation for.  The bright spiral galaxy UGC 4416 fills the
edge of the slit, and some of its light might be scattering in. There is also a spatial 
offset between the emission line peak and the peak of the continuum for this galaxy, 
though the emission is still consistent with coming from C5-5.  Due to the low S/N of
these observations, we do not include either N10-4 or C5-5 in the remainder of the 
analysis.

\section{Emission Line Source: \HII\ regions vs. AGN\label{galactivity}}

\begin{figure*}[bth!]
  \centerline{\psfig{figure=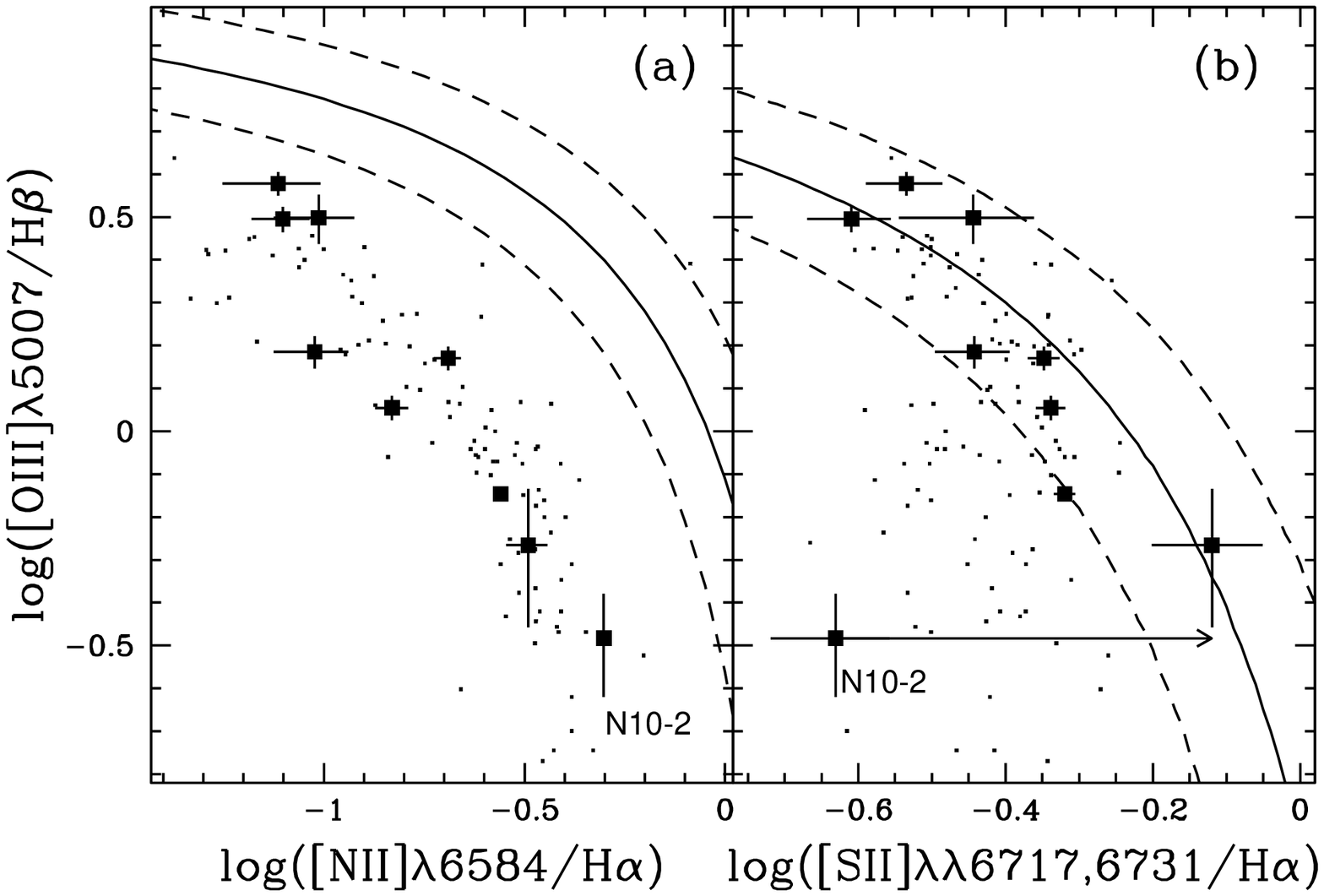,width=6.5in}}
\caption{Ionization diagnostic diagrams: large boxes -- LSB galaxies; 
small boxes -- galaxies in the NFGS sample \citep{jan00}; lines -- theoretical
classification line and $\pm$0.1 dex error range separating AGN from starburst (\HII) galaxies, 
from \citet{kew01b}.  The LSB galaxies are all well within the \HII\ region
for the [\ion{N}{2}]/H$\alpha$ vs [\ion{O}{3}]/H$\beta$ diagnostic.  There is a shift in
the data with respect to the theoretical separatrix in the 
[\ion{S}{2}]/H$\alpha$ vs [\ion{O}{3}]/H$\beta$ diagnostic.  However, the LSB galaxies are
all still consistent with starburst like spectra.  The shift of N10-2 between (a) and (b) is a result of
poor flux calibration at the extreme red end of the spectrum, which affects the [\ion{S}{2}] measurements.  
The arrow shows the uncertainty in [\ion{S}{2}]/H$\alpha$ for N10-2 due to this systematic effect.}
\label{AGNdiagnostic}
\end{figure*}

We use the diagnostic diagrams of \citet{vei87} to determine whether the emission lines
in the observed LSB galaxies come from active galactic nuclei (AGN) or \HII\ regions.
Figure \ref{AGNdiagnostic}a shows the [\ion{N}{2}]$\lambda$6584/H$\alpha$ vs. 
[\ion{O}{3}]$\lambda$5007/H$\beta$ diagnostic.
We overplot the theoretical separation line of \cite{kew01b} on the diagram.  
The LSB galaxies all clearly lie in the \HII\ side of the diagram.  They also lie
in the same area of the diagram as the HSB galaxies from the Nearby Field Galaxy
Survey \citep[NFGS;][]{jan00}.   Figure \ref{AGNdiagnostic}b shows the 
[\ion{S}{2}]$\lambda$6717$+\lambda$6731/H$\alpha$ vs. [\ion{O}{3}]$\lambda$5007/H$\beta$ 
emission line diagnostic.  The interpretation here
is less clear.  The LSB galaxies all lie on the separatrix, and there is considerably
more scatter amongst the NFGS galaxies as well.  \citet{kew01a} note that the
[\ion{S}{2}]$\lambda$6717$+\lambda$6731/H$\alpha$ ratio is sensitive to the local density, 
and therefore might 
be a less reliable diagnostic than the [\ion{N}{2}]$\lambda$6584/H$\alpha$ ratio.  
N10-2 shifts dramatically between the two diagnostics, relative to the rest of the
LSB galaxies.  This is an instrumental effect, as the redshifted [\ion{S}{2}] lines fall just at
the red end of our spectral coverage, where the flux calibration is less secure.
All the galaxies with emission lines are consistent with star formation as the ionizing
source, rather than AGN.
 
\section{Mean Age and Metallicity of the Stars\label{empirical}}
\begin{figure*}[bht!]
  \centerline{\psfig{figure=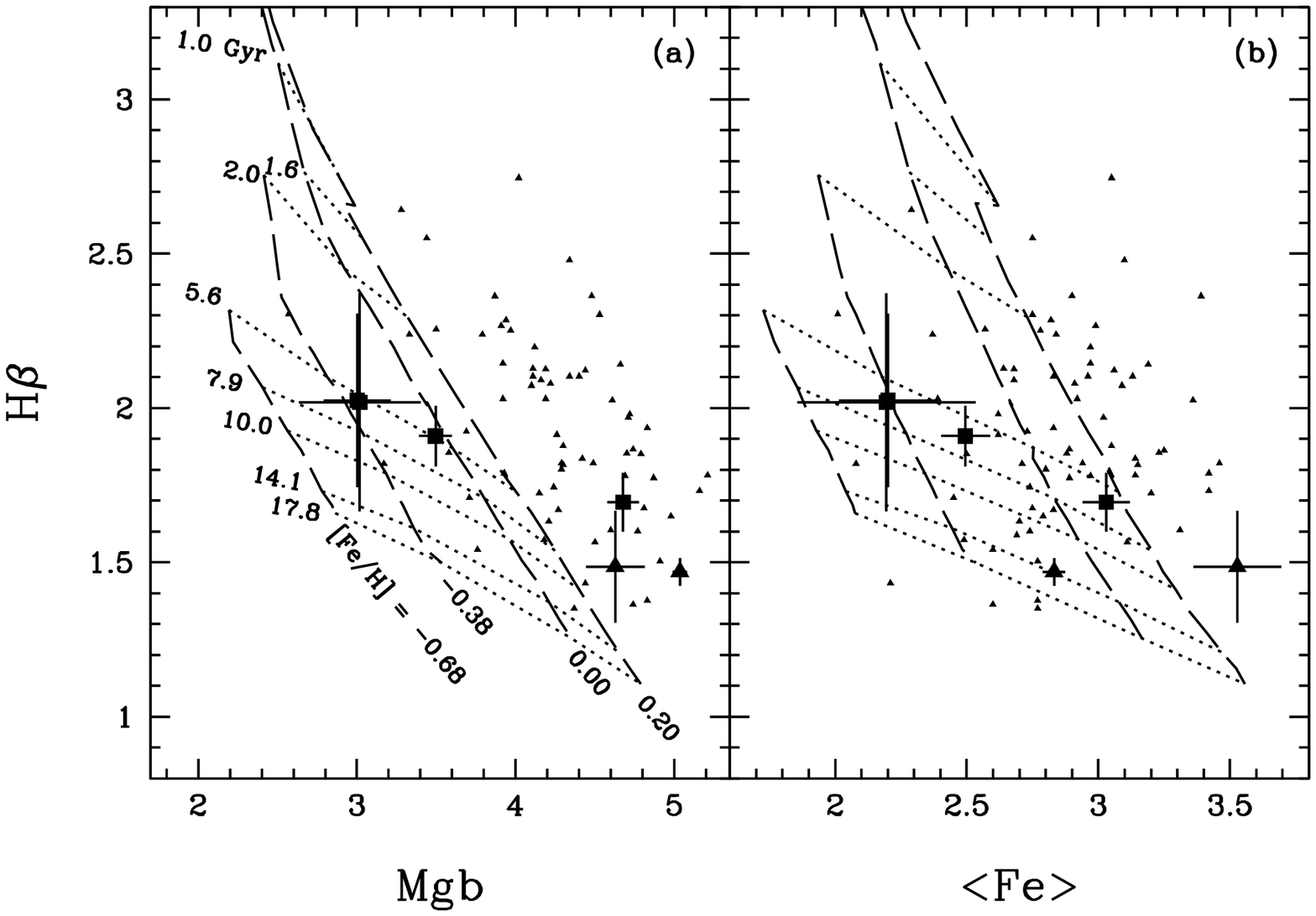,width=6.5in}}
\caption{H$\beta$ absorption line index vs. the Mgb index (left) and the 
\meanfe\ index (right). Boxes: LSB galaxies; Large triangles: our data of HSB 
ellipticals; small triangles: Coma cluster elliptical and S0 galaxies from \citet{jor99}.  The model 
grid shows lines of constant age (dotted) and constant metallicity (long 
dash), from V2000 as labelled.  The same age and metallicity values
are used for the grid in both panels.  The four LSB galaxies plotted are
(in order of increasing Mgb, \meanfe\ strength) U1-8, C1-2, C1-4, and N3-1. 
Note that the points for C1-2 and U1-8
lie virtually on top of eachother.  The Coma cluster galaxies, 
N3-1, and NGC 3872, shift relative to the model grid in the two plots. This
is caused by the well known super-solar [Mg/Fe] ratio in luminous
elliptical and S0 galaxies, which is not reproduced by the models.  }
\label{hbplots}
\end{figure*}

The degeneracy between the effects of age and metallicity on broadband optical colors 
and individual spectral features has been well documented \citep{oco76,wor94a,ros94}.  
This degeneracy can be broken by using combinations of two or more absorption features, 
with different sensitivity to age and metallicity \citep{wor94,jor97,vaz99,kun01}.  
We use the line index H$\beta$ which is more age sensitive, in conjunction with the
more metal sensitive indices Mgb and \meanfe.
In Figure \ref{hbplots} we plot grids of age and metallicity for single stellar
populations derived from the V2000 models, on the axes of
\HB\ vs. Mgb and \HB\ vs. \meanfe.  
Within this grid, we determine the mean luminosity weighted age and metallicity for
the stellar populations of the four absorption line dominated LSB galaxies, C1-2, C1-4,
N3-1, and U1-8.
We also plot the two elliptical galaxies which we observed, 
as well as the locus of points
covered by the Coma cluster elliptical and S0 galaxies from \citet{jor99}.

The three LSB galaxies C1-2, C1-4 and U1-8 all lie together on the plots, in the
low metallicity regime.  They all have mean ages around 5.5 Gyr.
C1-4 has near solar metallicity, while C1-2 and U1-8 have about half solar metallicity.

In contrast, N3-1 has a much higher metallicity, appearing at 1.5 times solar metallicity
in the \HB\ vs. \meanfe\ plot, and even higher in the \HB\ vs. Mgb plot.  
This is the regime shared by
luminous elliptical and S0 galaxies.  Furthermore, note how N3-1 as well as the elliptical and S0
galaxies shift with respect to the model grid between the two plots.  This is the
signature of super-solar [Mg/Fe] abundance ratios.  This enhancement effect 
is well known for luminous galaxies, both spirals and ellipticals,  
and the strength of the enhancement is correlated
with galaxy mass and velocity dispersion \citep{wor98,jor99,kun01}.
The other
three LSB galaxies do not move relative to the model grid between the two plots; these
galaxies all have solar [Mg/Fe] abundance ratios.

\begin{figure}[bht!]
  \centerline{\hspace{0.8in}\psfig{figure=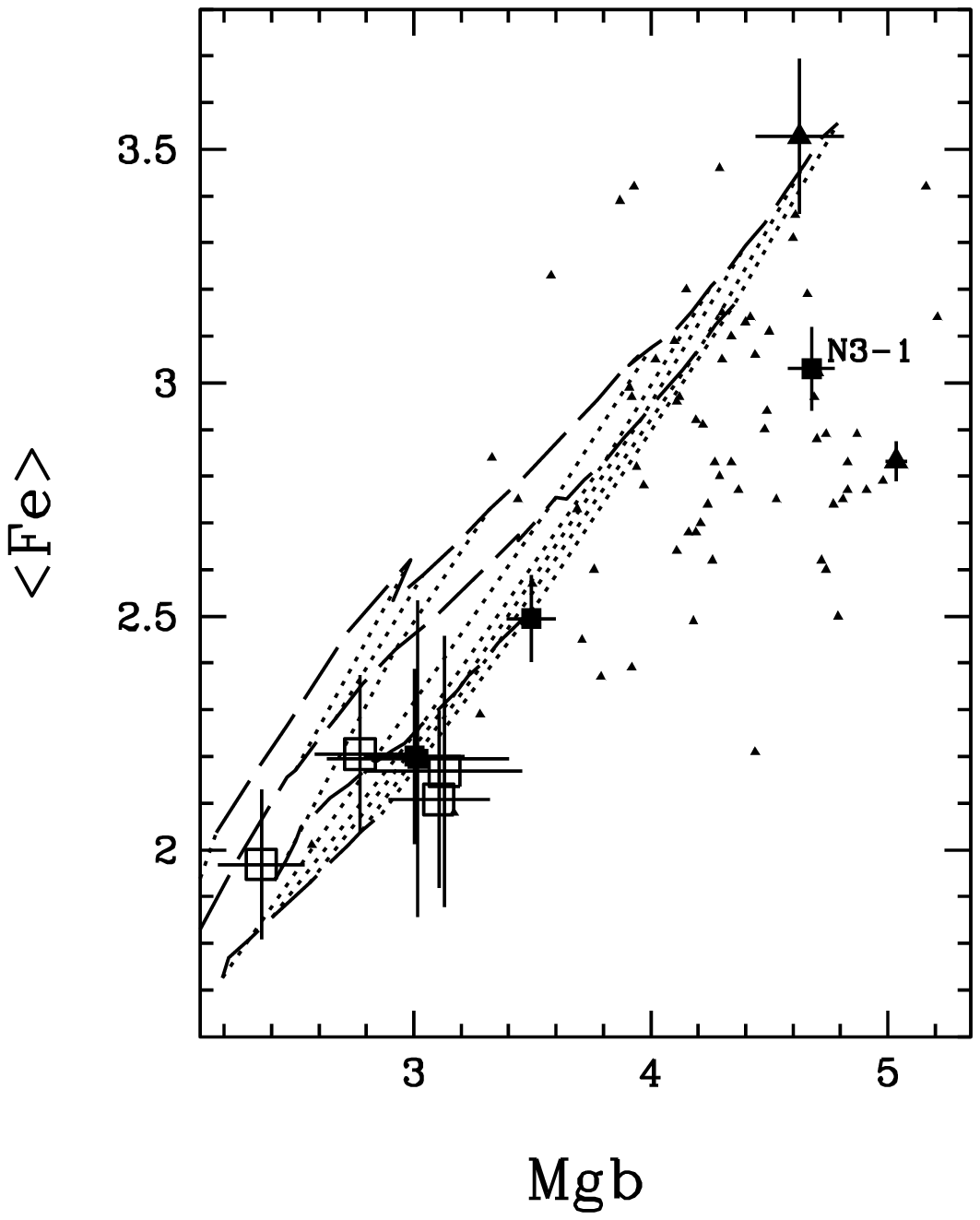,width=4.0in}}
\caption{The \meanfe\ index vs. the Mgb index.  Filled boxes: LSB galaxies with no H$\beta$ emission;
Open boxes: LSB galaxies with measured H$\beta$ emission; Large triangles: our data of HSB
ellipticals; small triangles: Coma cluster elliptical and S0 galaxies from \citet{jor99}.  The same models used
for the grid in Figure \ref{hbplots} are used here.
The Mgb and \meanfe\ indices have similar sensitivity to
age and metallicity, so the model grid is very narrow.  N3-1, like many of
elliptical galaxies, lies off the model grid, suggesting super-solar [Mg/Fe]
abundance.  See the text for details.}
\label{mgfeplot}
\end{figure}

Figure \ref{mgfeplot} more clearly illustrates the [Mg/Fe] issue.  This plot shows
the \meanfe\ index plotted against the Mgb index.  Since these two indices have similar
metallicity and age sensitivity, the model grid for differing ages and metallicities 
is not well separated in this plane.  Galaxies with solar [Mg/Fe] 
should fall onto the narrow region covered by the model grid, as indeed most of
the LSB galaxies do. On this Figure we also show the LSB galaxies with emission lines 
which have reliable Mgb and \meanfe\ measurements.  The bulk of the elliptical 
and S0 galaxies 
as well as N3-1 fall to the right of the model grid, a failure of the models to account
for the [Mg/Fe] appropriate to these galaxies.

There is another caveat with these models, especially important when dealing with low
metallicity stellar populations.  These models do not include blue horizontal branch or
post-AGB stars, which will have an effect on measured Balmer line strengths \citep{mar00}.
However, at the lowest metallicities found here (a tenth solar to a fifth
solar), galactic globular clusters do not show a blue horizontal branch, and thus we
do not expect contamination by these blue stars to be a significant effect.  Furthermore,
the effect of these stars on integrated spectra is to increase the strength of the 
\HB\ absorption, leading to younger apparent ages. The main conclusion of this
section is that some LSB galaxies do in fact have old mean ages; a correction for these
blue stars would only strengthen our conclusion.


%
%

\section{Gas Phase Metallicity and Age Indicators\label{metindicator}}

\begin{figure}[tb!]
  \centerline{\hspace{0.1in}\psfig{figure=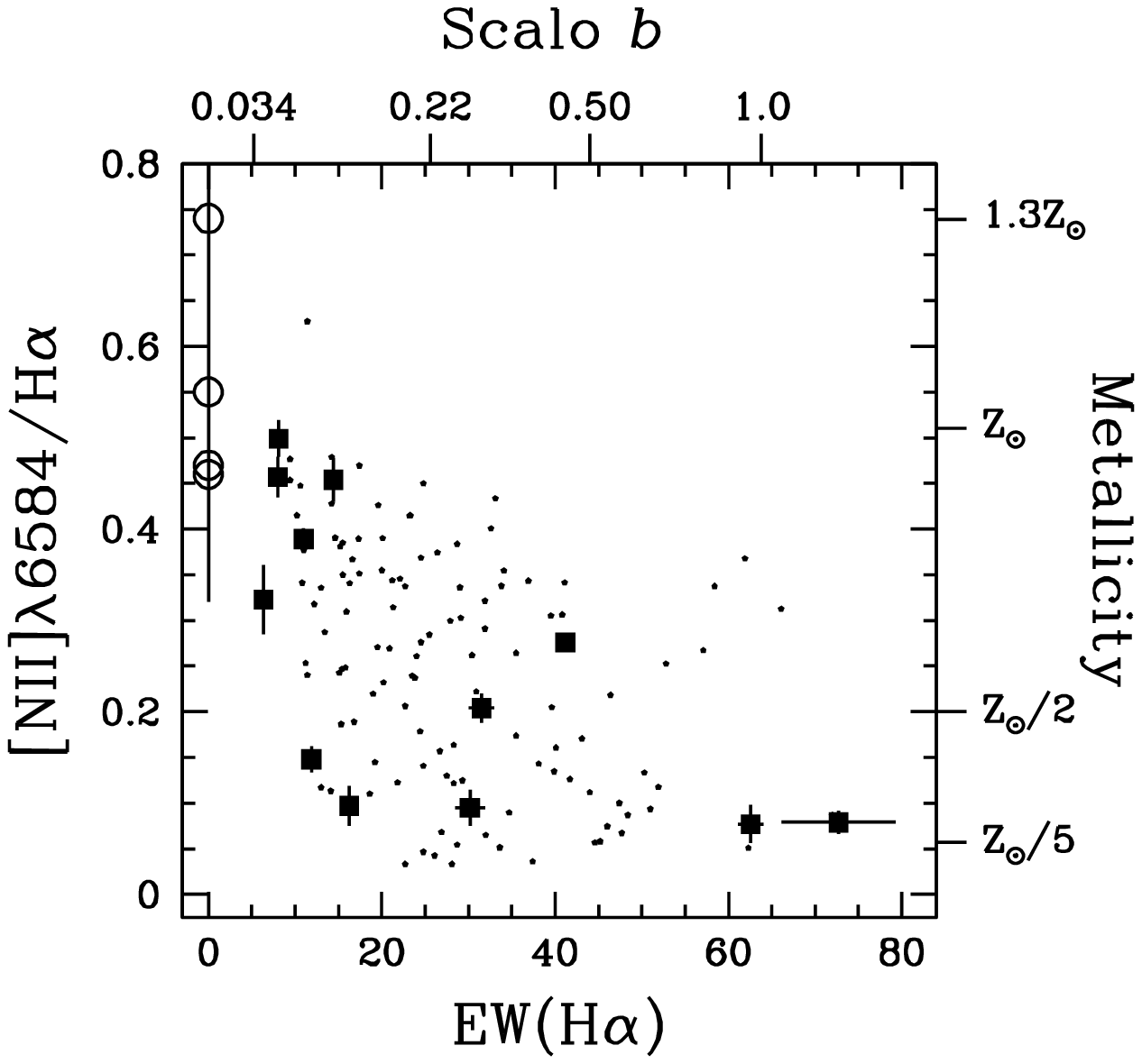,width=4.0in}}
\caption{The metallicity indicator [\ion{N}{2}]/H$\alpha$ plotted against 
EW(\HA), a measure of the birthrate parameter.  Large
boxes: LSB galaxies with measured emission lines; circles: LSB galaxies with no
measurable emission lines whose positions are based on the extrapolation 
described in Section \ref{metindicator}; small boxes: HSB galaxies from the 
NFGS \citep{jan00}.  The right axis shows the [\ion{N}{2}]/H$\alpha$ 
metallicity calibration of \citet{den01}, appropriate for \HII\ regions.  The
top axis shows the EW(H$\alpha$) calibration of the Scalo birthrate parameter for
one star formation model detailed in \citet{ken94}.  See text for details.}
\label{met_bplot}
\end{figure}

Much as we used the stellar absorption indices to study the older stellar populations, we would
like to use emission line strengths and ratios to characterize the metallicity
of the gas and the properties of the young stellar populations.  
The equivalent width of \HA, EW(\HA), compares the flux from hot young stars, 
capable of ionizing nebulae, to the red flux from the stellar population, 
which is dominated by long-lived lower mass stars for a standard 
\citep[e.g.][]{sal55} initial mass function (IMF).  
Consequently, the EW(\HA) can be used as a mean age indicator for the integrated 
stellar population.  The conversion between EW(\HA) and the \citet{sca86}
birthrate parameter, defined as {\it b}~$\equiv{\rm\frac{SFR}{<SFR>_{past}}}$, 
is sensitive to both the form of the star formation history (SFH) and the IMF, 
and hence is quite model dependant.
A single burst of star formation will only produce
\HA\ emission for $\sim 30$ Myr after which the massive stars 
(M$_\star > 15$ M$_\odot$) which produce the ionizing flux have all died off.  
Models with continuous star formation and multiple burst models that produce
an underlying old population plus a new burst will both produce \HA\ emission at later times.
\citet*{ken94} use stellar population and nebular models
to derive a relation between EW(\HA) and {\it b}
appropriate for galaxies with a constant or 
exponentially decreasing SFR.  The relation they derive, based on a \citet{sal55} IMF with
upper and lower mass limits of 100 and 0.1 M$_\odot$, 
a 10 Gyr age and varying exponential decay times,  shows a monotonic
increase of {\it b} with EW(\HA), and has {\it b}=1 at EW(\HA)=64 \AA.   
\citet{vdh00} found that chemo-evolutionary models with an exponentially decreasing SFR 
were sufficient to describe the LSB galaxies in their sample with (B$-$V)$>$0.4 but that
additional star formation bursts were needed to explain bluer galaxies.
The measured EW(\HA) for an old-plus-burst stellar population will of course 
depend on both the age of the burst, and the stellar mass fraction that it embodies.
In Figures \ref{met_bplot} and \ref{allcomp} we show the {\it b} calibration of \citet{ken94} on the right
vertical axis, but give the caveat that this calibration is inappropriate for an 
old-plus-burst SFH.


Traditional methods of studying gas phase chemical abundances in \HII\ regions require 
measurements of several H or He recombination lines, along with collisionally excited
lines from multiple ionization states of a heavier metallic species.  \citet{ost89} 
details the method for studying individual \HII\ regions.  \citet{kob99} extend this
work and look at the
effect on the derived abundances of smearing together the light from multiple \HII\ 
regions in a single galaxy into a single spectroscopic aperture, using abundance analysis
methods such as the semi-empirical strong-line R$_{23}$\footnote{log(R$_{23}$)=
log($\frac{{\rm [O II]}\lambda3727 + {\rm [O III]}\lambda\lambda4959,5007}{H\beta}$)}
method \citep{pag79,ost89,mcg91,mcg94}.
The wavelength range covered by our spectra rarely includes the [\ion{O}{2}]$\lambda$3727 \AA\
line used in the R$_{23}$ method.
Instead, we choose to use the metallicity indicator [\ion{N}{2}]$\lambda$6584/H$\alpha$ 
(hereafter [\ion{N}{2}]/H$\alpha$) proposed by
\citet{sto94}.  This measurement
has the advantage of being monotonic (as opposed to the double valued nature of the
R$_{23}$ method) and is essentially reddening independant by virtue of the small 
wavelength split between the two lines.  
Empirical calibrations between [\ion{N}{2}]/H$\alpha$\ and log(O/H) (determined using the R$_{23}$ method) have been derived from 
observations of individual \HII\ regions \citep[e.g.][]{van98} and for galaxies \citep*{den01}.  
The drawback to using [\ion{N}{2}]/H$\alpha$ is that this ratio 
is sensitive to the ionization state, which will vary between \HII\ regions and
the diffuse ISM, and is difficult to determine for globally averaged spectra.
\citet{kew02} provide a theoretical calibration of [\ion{N}{2}]/H$\alpha$ vs. 
metallicity for starburst galaxies, including the effects of the ionization parameter.
\citet{mcg94} find for
\HII\ regions in LSB galaxies that the ionization parameter log$<$U$>$ varies between
$-2$ and $-3.5$, while \citeauthor{den01} find that log$<$U$>$=$-2.0$ is most representative
for their sample of galaxies.  However, a change in log$<$U$>$ of +0.5 dex or $-1.0$ dex
leads to an uncertainty in the metallicity zeropoint of $\pm 0.3$ dex. 
There was no correlation seen between metallicity and log$<$U$>$ in the 
\citeauthor{mcg94} dataset. 
In the plots that follow, we label one axis with the \citeauthor{den01} empirical
calibration of [\ion{N}{2}]/H$\alpha$, but we give the caveat that the zeropoint 
of this calibration may be high by $\simeq$0.3 dex and that any systematic 
changes of the ionization parameter with other parameters of these galaxies such as age
may lead to a misinterpretation of the [\ion{N}{2}]/H$\alpha$ ranking as a 
strict metallicity scale.  

\begin{figure*}[bht!]
  \centerline{\psfig{figure=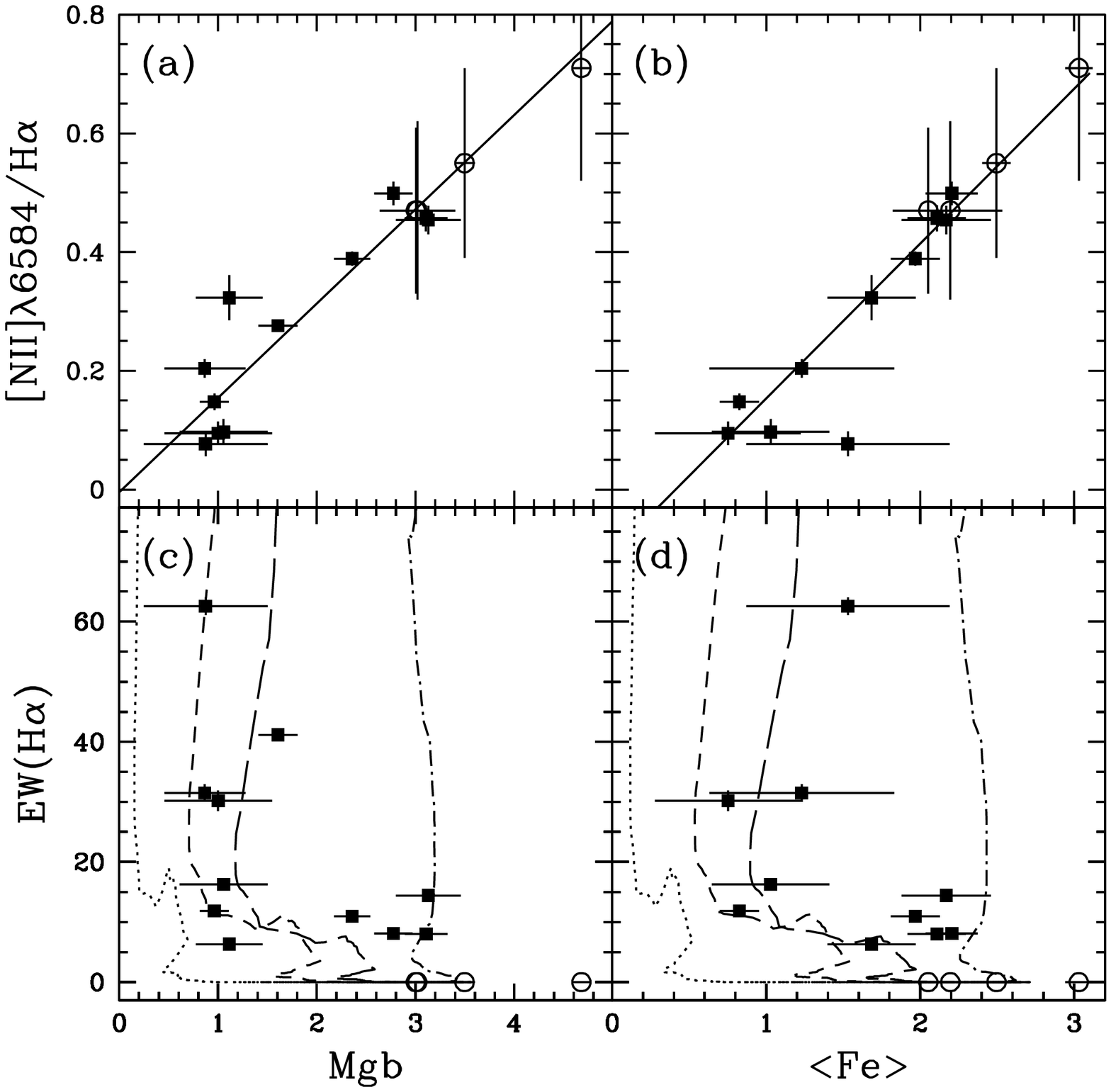,width=6.5in}}
\caption{
Comparison of gas phase indicators and stellar absorption line indices.
The gas phase [\ion{N}{2}]/H$\alpha$ ratio plotted against (a) the stellar 
Mgb line index and (b) the \meanfe\ line index.  The best fitting relations
are shown as solid lines.
(c) The EW(\HA) vs. the Mgb index.  (d) EW(\HA) vs. the \meanfe\ index.
Boxes: LSB galaxies with measured emission lines; circles: LSB galaxies with no measured
emission, see Section \ref{metindicator} for details.  The lines in panels (c) and (d) are evolutionary tracks for an old-plus-burst
stellar population, with a 5 Gyr old population, and bursts which comprise (from left
to right) of 10\% (dotted), 2\% (short-dash), 1\%(long-dash) and 0.1\% (dot-dash) of the
total stellar mass.  The kink in these evolutionary tracks occurs at $\sim 7.5$ Myr after
the burst, and the EW(\HA) drops below 1 \AA\ within 20 Myr of the burst for all the
tracks.  The tracks all end at Mgb = 3.6 and \meanfe\ = 2.7, EW(\HA) $\simeq$ 0, which are
the values of the underlying population.}
\label{met_stelmetplotb}
\end{figure*}

We compare the gas phase metallicity indicator [\ion{N}{2}]/H$\alpha$ with the stellar
absorption indices Mgb and \meanfe\ in Figure \ref{met_stelmetplotb}.  There are
clear correlations with each index.  A Kendall's $\tau$ test gives a probability
of 1.6$\%$ that [\ion{N}{2}]/H$\alpha$ and Mgb are not correlated, and a probability of 0.08$\%$
that [\ion{N}{2}]/H$\alpha$ and \meanfe\ are not correlated.
The best fit relations are: 
\begin{equation}
{\rm [N II]/H\alpha}= -0.11(\pm0.13) + 0.26(\pm0.08)\ \cdot <\!{\rm Fe}\!>
\label{N2meanfe}\end{equation}
\begin{equation}
{\rm [N II]/H\alpha}= -0.01(\pm0.09) + 0.16(\pm0.05) \cdot {\rm Mgb}~~~
\label{N2mgb}\end{equation}
It is perhaps surprising that the correlations are so strong, given the 
age-metallicity degeneracy 
of the stellar indices and the variation of
the [\ion{N}{2}]/H$\alpha$\ calibration with ionization parameter.  
Panels c and d in Figure \ref{met_stelmetplotb} show the age indicator EW(\HA) plotted against
the absorption line indices.  At low Mgb strength, we see a wide range of EW(\HA)
strengths, which diminishes at higher Mgb strengths.  In models with continuous, exponentially
decreasing SFR, the metallicity increases as the gas fraction decreases. 
Simultaneously, the SFR drops and the fractional galaxy mass made up by low mass
stars increases.  This will result in a decrease of EW(\HA) and simultaneous increase of Mgb 
over time, leading to an upper envelope in the EW(\HA) vs Mgb
plane, much as we see here.  For ``open box'' chemical evolution models (models in which enriched gas is
allowed to escape due to stellar winds and supernovae),  the low EW(\HA) and low Mgb region 
will get filled in by galaxies with low total mass, where the star formation rate drops
once the gas is expelled, and before metallicity builds up.  In any case, once star formation
stops completely, the EW(\HA) will drop to zero, and the Mgb index will slowly strengthen as
the stellar population ages (at constant metallicity).   This picture holds equally 
well for the \meanfe\ strength as for Mgb.
A quantitative analysis of where models with continuous star formation 
would lie on the EW(\HA) vs. metal absorption index plots requires models that
contain both nebular emission spectra,
and stellar absorption line strengths.  Unfortunately, there are not yet any such
models in the literature. 
Within the picture of continuous star formation, we can also give a qualitative explanation for the strong correlation 
seen between [\ion{N}{2}]/H$\alpha$\ and Mgb or \meanfe.  [\ion{N}{2}]/H$\alpha$ reflects the 
current metallicity of the gas. The ionization parameter is a small source of scatter, but 
is not (apparently) significant.  The Mgb and \meanfe\ strengths are responsive to both
the metallicity and age of the stellar population, but since these are linked by
the gas consumption and enrichment history, a correlation exists between the current metallicity
and the age and enrichment history of the stellar population.  

An alternative explanation can be given by a model
with an old stellar population underlying a young burst of star formation.
In this case, the mass fraction of the burst and metallicity of the old population
will determine the evolutionary track of a galaxy through the EW(\HA) vs. Mgb (or \meanfe) 
plane.  The mass fraction determines how much blue continuum light from the young stellar 
population there will be to veil the strength of the underlying metal line absorption.  The
\HA\ flux is driven by very short-lived stars,  while the continuum veiling effect
evolves over longer timescales ($\simeq$ 0.5 Gyr vs. $\simeq$ 30 Myr for the \HA).  
Consequently, the tracks will move vertically down the diagram, and then to the right
at late times.  The track will end at the absorption line strength of the underlying
population for small bursts, and the offset between where the track descends vertically and 
where it ends depends on the mass fraction of the burst.
To illustrate this, we use the Starburst99 models \citep{star99} for the evolution 
of an instantaneous burst population, along with the V2000 stellar population models for 
the underlying population.  In Figure \ref{met_stelmetplotb}(c-d) we plot the data for the LSB galaxies together with
model tracks for galaxies with a 5 Gyr old, solar metallicity underlying population, and a 
10\%, 2\%, 1\%, and 0.1\% (by mass) instantaneous burst.  It is clear that
by varying the age and metallicity of the underlying population, or by varying the burst
mass fraction, it is possible to fill in every part of the plot.  While we do not find LSB
galaxies in all parts of the plot, we have a small sample and do not strongly constrain
against the possibility of galaxies in the high EW(\HA) and high Mgb (or \meanfe) region of 
the diagrams.

It is possible to fully explain the EW(\HA), Mgb, and \meanfe\ data with an old ($>$ 5 Gyr) 
solar metallicity population and varying burst mass fractions.  
However, with this model the \NHA\ vs. Mgb correlation becomes difficult to explain.
The \NHA\ value should reflect the metallicity of the burst, which is presumably greater
than or equal to the metallicity of the underlying stellar population.  If the underlying stellar population is
always metal rich and the burst mass fractions were random, pushing the absorption line
indices to lower levels, there would not be a correlation between  \NHA\ and the absorption
indices.  Consequently, it appears that old-plus-burst models are only viable if the old populations
have a range of metallicities, and if any one burst constitutes only a small fraction of the 
stellar mass (less than $\sim 0.3\%$) such that the absorption indices are not perturbed from
that of the underlying population.

Our understanding of these correlations is hampered by the small size
of our sample, and discontinuity between models which predict nebular
emission strengths from young stellar populations, and models which predict
the stellar metal absorption line indices for older, cooler populations.
Ideally, we could gleen a better understanding of these correlations 
from a larger sample of LSB or HSB galaxies.  However, we have been unable to 
find published studies of HSB galaxies which include data on both the 
stellar absorption indices and emission line strengths.  The NFGS and
Sloan Digital Sky Survey (though the latter is prone to aperture effects) 
represent two datasets upon which it would be enlightening to make this study.

We cannot measure gas phase abundances for the four absorption line dominated galaxies,
but we use the correlations in Equations (\ref{N2meanfe}) and (\ref{N2mgb}) 
to assign values of [\ion{N}{2}]/H$\alpha$\ to these galaxies.
We do this primarily so that we can include these four galaxies on the remainder of
the diagnostic diagrams, where we use [\ion{N}{2}]/H$\alpha$ as a metallicity indicator.
We interpolate the relations get values for 
C1-2 and U1-8, and extrapolate to get values for C1-4 and N3-1.  The values determined
from each of the two correlations are averaged, to reduce the uncertainties. The assigned 
[\ion{N}{2}]/H$\alpha$ values are as follows:  0.47$\pm$0.15 for C1-2; 0.47$\pm$0.14 for U1-8;
0.55$\pm$0.16 for C1-4; 0.71$\pm$0.19 for N3-1.  These four galaxies are shown as
open circles on Figure \ref{met_stelmetplotb}. 
The absorption line dominated galaxies seem to lie at one end of a continuous sequence
of spectral properties, as illustrated by Figs. \ref{lsbplots}a - \ref{lsbplots}d.  This
suggests that the extrapolation of these correlations is reasonable.  However, we do
note that stellar populations at constant metallicity will evolve to higher Mgb
and \meanfe\ with age once the star formation stops.  
Though we do not see a large
age spread among the four absorption dominated systems, age will affect the extrapolation
of these relations to higher index strengths.  
By including both the absorption and emission line galaxies in Figures \ref{met_bplot} 
and \ref{allcomp} we attempt to show these galaxies potentially have very similar
evolutionary histories (though they are perhaps at different stages of that evolution).
The emission associated with a small burst of star formation is very short-lived, and 
though the emission line flux can be significant, the associated effect on the chemical
evolution and continuum luminosity could be slight, and so understanding these
two "classes" of LSB galaxies, absorption and emission line galaxies, is necessarily 
intertwined.

In Figure \ref{met_bplot} we plot [\ion{N}{2}]/H$\alpha$ against
the EW(\HA) for the LSB galaxies.  
The metallicity and birthrate calibrations discussed at the beginning of this section 
are included on the axes opposing the measured parameters.
We also plot the HSB galaxies from the NFGS \citep{jan00}, which satisfy the
condition that EW(\HA)$>$9 \AA\ (i.e. those which have a small measurement 
uncertainty for the  [\ion{N}{2}]/H$\alpha$ line ratio).  The LSB galaxies fill
the same region of the [\ion{N}{2}]/H$\alpha$ vs. EW(\HA) parameter space, suggesting
a similar star formation and chemical enrichment history to that of HSB galaxies.
At low metallicities there is 
a wide range of mean ages, however the relative current SFR decreases as metallicity 
increases.
This makes sense for continuous star formation, as the bulk of the gas in these 
galaxies must have been previously 
processed into stars to raise the mean metallicity, and consequently there is less
gas around to make new stars.  

\section{Dust Content\label{dust}}
\begin{figure*}[bt!]
  \centerline{\psfig{figure=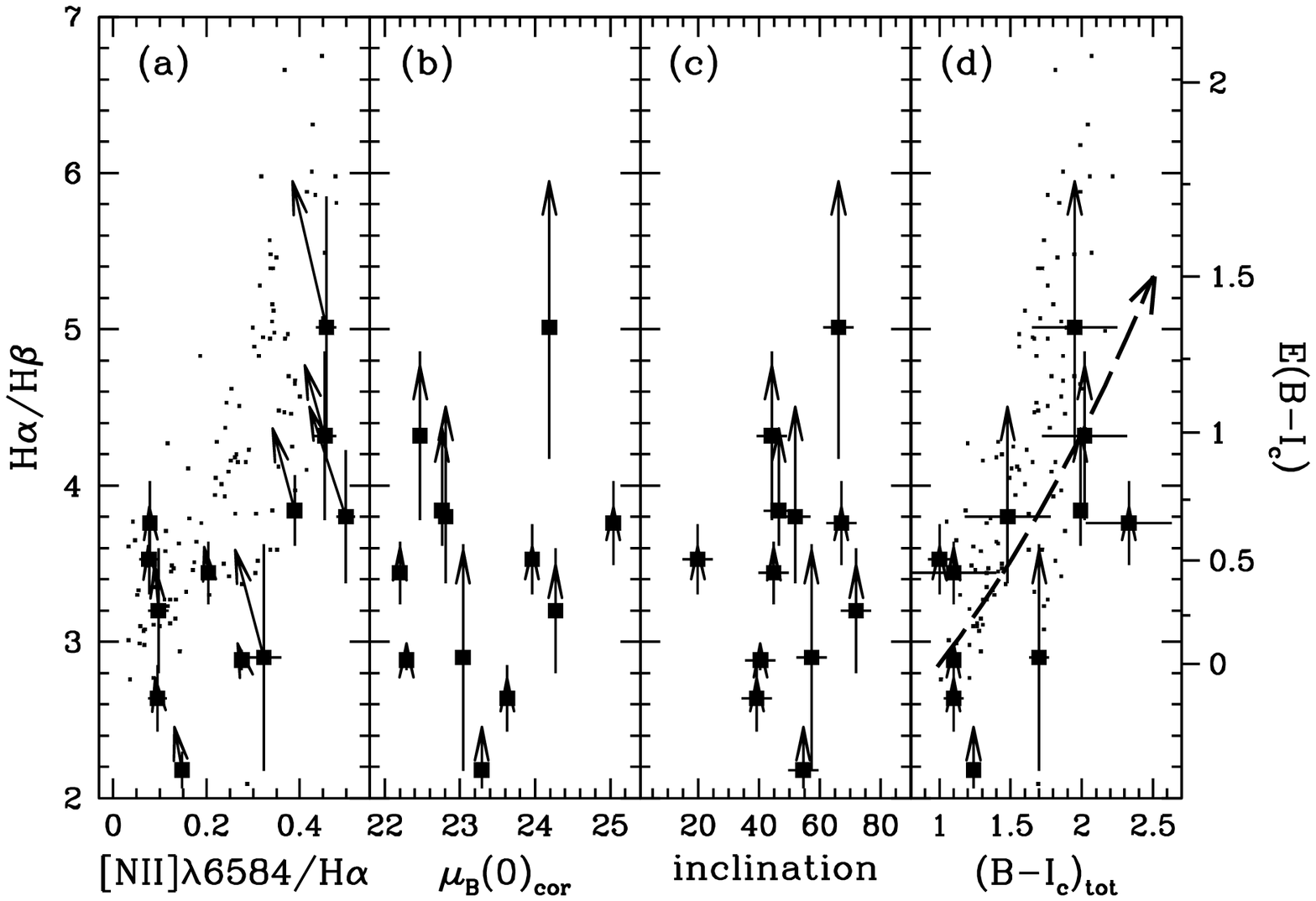,width=7.0in}}
\caption{The Balmer ratio, H$\alpha/$H$\beta$, plotted against (a) the 
metallicity indicator [\ion{N}{2}]$\lambda6584$/H$\alpha$, (b) the central surface brightness
$\mu_{\rm B}(0)_{\rm cor}$, (c) galaxy inclination (edge-on = $90^{\rm o}$), 
and (d) The total (B$-$I$_c$) color.  Large boxes -- LSB galaxies;
small points -- galaxies from the NFGS \citep{jan00}.  The right side y-axis 
lists the E(B$-$I$_{\rm c}$) values derived from equation \ref{dusteq}. 
The vertical arrows show where the LSB galaxies would move if we made a correction of 
1.5 \AA\ to the H$\alpha$ equivalent width to account for underlying stellar 
absorption.  The dashed arrow in panel (d) shows the locus of reddening vectors for the LSB galaxies.
The reddening vector for the NFGS galaxies (whose colors have been transformed 
from observed (B$-$R$_{\rm c}$) colors) would be $\simeq$10\% steeper.
The uncertainties shown for the Balmer ratios do not include 
those caused by uncertainties in the absorption correction to H$\beta$. 
The NFGS data has not been corrected for underlying stellar absorption at H$\alpha$  
or H$\beta$.  Applying an arbitrary correction of 1.5 \AA\ to the equivalent width of
both H$\alpha$ and H$\beta$ does not significantly shift the location of the locus of 
HSB galaxy points.}
\label{dustfig}
\end{figure*}

The amount of internal reddening in LSB galaxies is not well known, but of
great importance to attempts at modelling their formation and evolution.
In Figure \ref{dustfig}a we plot the Balmer ratio,  H$\alpha/$H$\beta$, versus
the metallicity indicator [\ion{N}{2}]/H$\alpha$.  The Balmer ratio can be used to
derive the internal reddening of the emission line regions.  
Using the Cardelli extinction law \citep{car89}, with
R$_{\rm V}$=3.1,  we derive the following relation:
\begin{equation}
{\rm E(B-I_c)} =2.375\cdot {\rm E(B-V)} = \frac{{\rm log}(\frac{f(H_\alpha)}{f(H_\beta)})}{0.172} - 2.65
\label{dusteq}
\end{equation}
This relation is used to place an E(B$-$I$_{\rm c}$) scale on the right axis of 
Fig.\ref{dustfig}.
We use the NFGS \citep{jan00} as a comparison sample of
HSB emission line galaxies.  The NFGS galaxies were not observed in the I-band. 
We convert their (B$-$R$_{\rm c}$) colors to (B$-$I$_{\rm c}$) colors using the relation
(B$-$I$_{\rm c}$) = $1.34~\cdot$~(B$-$R$_{\rm c}$)~+~0.18, derived from the mean galaxy colors of \citet{fuk95}. 
Central surface brightnesses were not derived for the NFGS galaxies.
In the LSB data, we have not corrected the measured H$\alpha$ 
fluxes and equivalent widths for underlying stellar absorption.  
This correction will move points upwards and to the left in the
H$\alpha$/H$\beta$ vs. [\ion{N}{2}]/H$\alpha$ plane.  While the uncorrected data show
a trend for LSB galaxies to have lower extinction than HSB galaxies at the same
metallicity, an H$\alpha$ equivalent width correction of 1.5 \AA\ would move most
of the LSB galaxies onto the locus of HSB galaxy points.  

\citet{mcg94}, in his study of LSB galaxy \HII\ regions, found evidence for dust
in some galaxies, with internal extinctions as high as E(B$-$V) $\sim$ 0.5 (corresponding
to E(B$-$I$_{\rm c}$) $\sim$ 1.3).
The galaxies in that study were similar to the lower metallicity galaxies 
([\ion{N}{2}]/H$\alpha < 0.2$) in Figure \ref{dustfig}a 
for which we find a similar range of reddenings.  His
study did not include any of the red or giant LSB galaxies for which we find
generally larger reddening values.

The dust distribution in galaxies tends to be clumped, and these reddening 
values apply only to the regions where the emission lines are produced.  
The emission line regions and light from the stellar disk are not spatially 
resolved in our observations and both contribute to the extracted spectra.
Consequently, we cannot constrain the reddening affecting the 
general stellar disk, which dominates the continuum luminosity of the galaxy,
and thus we do not make a correction to the global photometry for internal
extinction.

Figure \ref{dustfig}a shows a trend for the Balmer ratio (and presumed 
reddening) to increase with increasing gas-phase metallicity.  
In Figure \ref{dustfig}(b-d) we plot the Balmer ratio versus
photometric parameters dominated by the stellar disk.  There is no correlation at
all between H$\alpha$/H$\beta$ and central surface brightness over the
range 22$\leq \mu_{\rm B}$(0)$_{\rm cor} \leq 25$,
nor is there a correlation between H$\alpha$/H$\beta$ and galaxy inclination.
The correlation between (B$-$I$_{\rm c})_{\rm tot}$ color and the Balmer ratio is shown for the NFGS
galaxies in Figure \ref{dustfig}d.  The LSB galaxies fall onto the same locus of points, 
especially after applying a correction for underlying H$\alpha$ absorption of 1.5 \AA.
The LSB galaxies appear to lie along the reddening vector in Figure \ref{dustfig}d.
This might imply that the dust is not local to the emission region, but is instead 
mixed throughout the galaxy.
Alternatively, it is possible that the colors are not caused by dust reddening of 
the stellar light, and the relation between H$\alpha$/H$\beta$ and
(B$-$I$_c$)$_{\rm tot}$ is not the fundamental one.  Rather, the amount of dust 
increases with increasing metallicity (Fig. \ref{dustfig}a), and the stellar populations 
also become redder with increasing metallicity (cf. the tight correlation between 
[\ion{N}{2}]/H$\alpha$ and (B$-$I$_{\rm c}$)$_{\rm tot}$ shown on Fig.~\ref{allcomp}c, 
and the corresponding discussion in Section \ref{globcomp}), which in turn results in the 
correlation seen here between H$\alpha/$H$\beta$ and (B$-$I$_c$)$_{\rm tot}$.
If the colors are driven by dust reddening over the whole galaxy, then we would
expect to also see a correlation between galaxy inclination and reddening, with
higher reddening measurements for galaxies closer to edge-on.  We do not detect
any such correlation, and thus conclude that the correlation between H$\alpha/$H$\beta$
and (B$-$I$_c$)$_{\rm tot}$ is secondary, and both parameters are driven primarily
by the metallicity.

Regardless of the interpretation of the galaxy colors, it is clear that dust is present
in abundance in some LSB galaxies, particularly the red and high metallicity ones, but
the absence of a correlation between dust and $\mu_{\rm B}$(0)$_{\rm cor}$ suggests that dust is  
not the cause of the low surface brightness for LSB galaxies of any color. 

The relative emission line strengths presented in Table \ref{specdatatable} have not been 
corrected for internal extinction within these galaxies.  The line ratios we use in our
analysis (e.g. [\ion{N}{2}]/H$\alpha$) all have narrow wavelength spreads, and are not
significantly affected by dust in the amount we detect.  The uncertainty of a correction for intrinsic reddening, however,
would be significantly larger than the correction itself (which is negligible), and so we do not apply any correction.  

\section{Emission Line Properties Versus Global Photometric Properties\label{globcomp}}

\begin{figure*}[tb!]
  \centerline{\psfig{figure=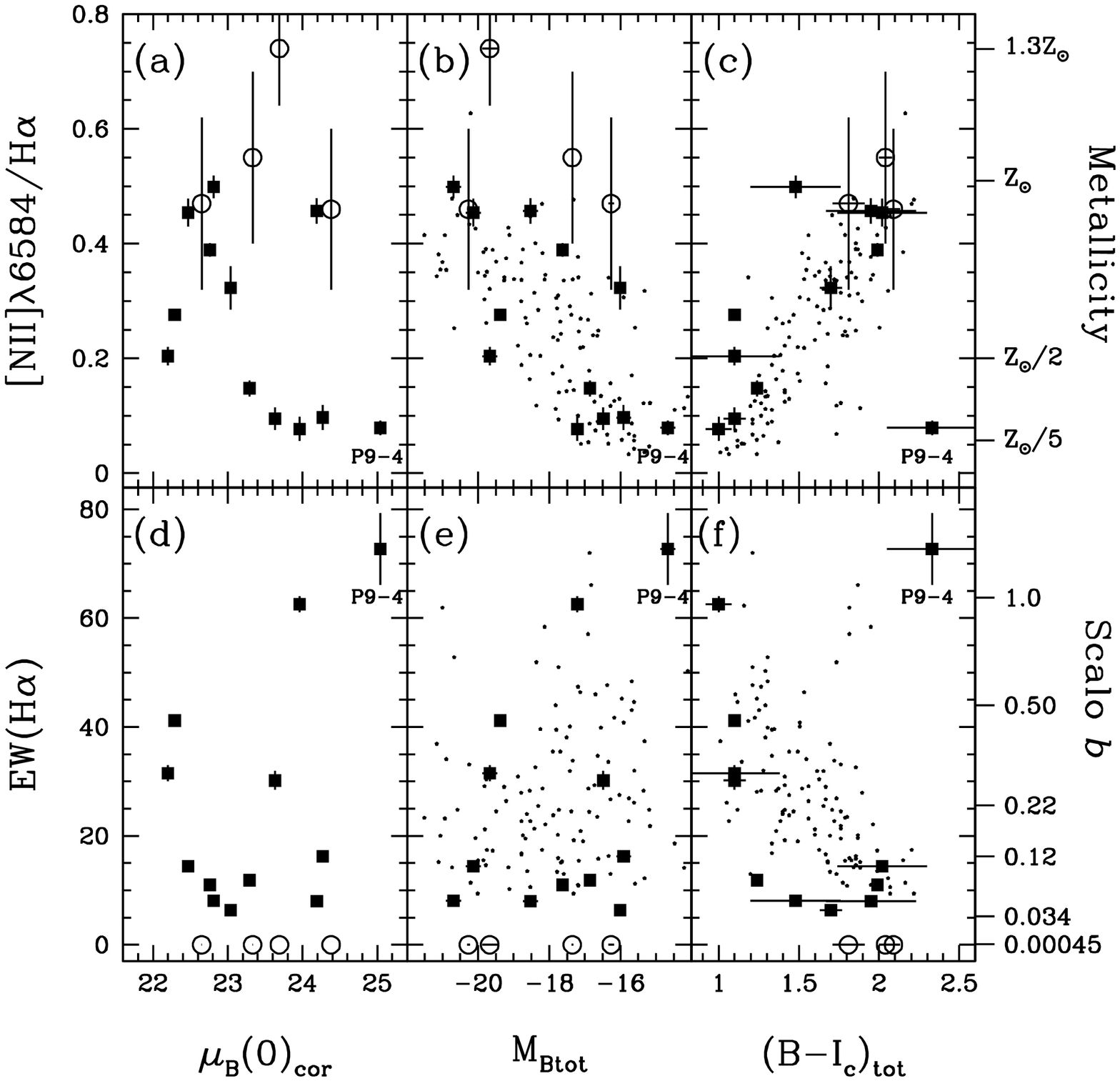,width=7.0in}}
\caption{
The metallicity indicator [\ion{N}{2}]/H$\alpha$ and the age indicator EW(H$\alpha$)
plotted against central surface brightness, $\mu_{\rm B}$(0)$_{\rm cor}$, absolute
magnitude, M$_{\rm Btot}$, and (B$-$I$_{\rm c}$) color.  Large boxes: LSB galaxies with measured
emission lines; circles: LSB galaxies with no measured emission lines; small pentagons:
emission line galaxies in the NFGS \citep{jan00}.  On the right side axes are the 
metallicity and birthrate parameter calibrations described in Section 
\ref{metindicator}.   The correlation between metallicity and absolute magnitude found
for HSB emission-line galaxies is also seen in LSB galaxies.  We do not detect a significant 
correlation between metallicity and surface brightness in this sample.  
The strongest correlation is between color and [\ion{N}{2}]/H$\alpha$.
See text for comments on the published photometry for P9-4.}
\label{allcomp}
\end{figure*}

In Figure \ref{allcomp}, we show \NHA\ and EW(\HA) versus central surface
brightness, absolute magnitude, and (B$-$I$_{\rm c})_{\rm tot}$. 
We plot the absorption line galaxies in this figure as well, using the [\ion{N}{2}]/H$\alpha$ 
values derived and discussed in Section \ref{metindicator}. However, these four galaxies are
never included when we measure correlations between the parameters plotted in Figure \ref{allcomp}.

In Figures \ref{allcomp}c and \ref{allcomp}f, the outlier P9-4 at (B$-$I$_{\rm c}$)=2.33 (OBSCI97) 
is purportedly the reddest object in the sample, despite its high EW(\HA) and metal poor nature.
The continuum slope in our HET spectrum of this object is very flat.  In comparison,
the galaxies at (B$-$I$_{\rm c}$)=1.3 have bluer continua, but those at (B$-$I$_{\rm c}$)=1.8 have redder continua.
The continuum slope of N10-2 (B$-$I$_{\rm c}$)=1.48 is also very flat.  We suggest that there are 
errors in the published photometry for P9-4.  We label P9-4 in all panels of Figure \ref{allcomp} 
and we do not include the data for P9-4 in the determination of any of the correlations discussed below.

Figures \ref{allcomp}a and \ref{allcomp}d show that there is a broad range of 
both \NHA\ and EW(\HA) values for the whole range of surface brightnesses studied.  
A Kendall's $\tau$ test gives a probability of 54\% that there is no correlation
between EW(\HA) and $\mu_{\rm B}(0)_{\rm cor}$.  There is
a probability of 31\% that there is no correlation between \NHA\ and $\mu_{\rm B}(0)_{\rm cor}$.
In contrast, \citet{bel00} found correlations both between mean age and K-band 
central surface brightness, and between mean metallicity and K-band central 
surface brightness.  Their modelling involved an assumption of exponentially 
decreasing star formation rates all beginning at the same time in the past (12 Gyr), 
with different exponential timescales, $\tau$, which thus produced stellar populations with 
different luminosity-weighted mean ages at the current epoch.  They varied $\tau$ and 
metallicity in their models to produce a grid in the (R$_{\rm c} - $K) vs. (B$-{\rm I_c}$) 
plane, which was then used to interpret their data.  These models have a very similar SFH
to the models of \citet{ken94}, which we use for the right side scale on Figure \ref{allcomp}(d-f)
and which predict a monotonic relationship between EW(\HA) 
and $\tau$, and consequently mean age.   However, as mentioned above, we do not see
a correlation between EW(\HA) and B-band central surface brightness.   
\citeauthor{bel00} make an unweighted least-squares fit to their data which 
produces a change in mean age of $0.84\pm0.08$ Gyr for a change of two magnitudes
in central surface brightness, equivalent to the range covered by our sample, with 
lower luminosity-weighted mean ages occuring at lower central surface brightness.  Their fit to the 
metallicity-surface brightness relation produced a change in the mean metallicity of
0.26 dex for the same two magnitude change in central surface brightness, with higher 
metallicity at higher surface brightness.  Our sample size, though
similar to the LSB galaxy study of \citet{bel00b}, is a small fraction of the combined
HSB-plus-LSB galaxy sample of \citeauthor{bel00}.  It is quite possible that our sample
selection, which deliberately covers a wide range of galaxy colors and surface
brightnesses, has led to more scatter than would be present in a volume-limited sample
of galaxies, and that therefore we might not be sensitive to these predicted modest 
changes in mean age and metallicity.

The line ratio \NHA\ versus the absolute magnitude is shown in Figure
\ref{allcomp}b. 
This plot makes clear the point that LSB galaxies are not synonomous with dwarf
galaxies, but cover a range of absolute magnitudes comparable to that of HSB galaxies.
The emission line LSB galaxies show a correlation between \NHA\ and absolute magnitude.
A Kendall's $\tau$ test gives a probability of 5\% that no correlation is present. 
The relation for LSB galaxies matches that of the NFGS galaxies.  The 
LSB galaxies without emission also follow the correlation, but form a ridge on the faint, high metallicity
edge of the galaxy distribution.  This may be reflecting a higher B-band 
mass-to-light ratio in these older objects, if the \NHA-M$_{\rm Btot}$ relation is a 
result of a more fundamental metallicity-mass relation.
Alternatively, it could reflect a systematic error in the 
interpolation/extrapolation of the \NHA-Mgb and \NHA-\meanfe\ relations used for the  non-star-forming
galaxies.

The distribution of the LSB galaxies is a good match to 
that of the NFGS galaxies in Figure \ref{allcomp}e.
We do not detect any correlation between EW(\HA) and M$_{\rm Btot}$.  A
Kendall's $\tau$ test gives a probability of 63\% that there is no correlation for
the emission line LSB galaxies.
A rough scaling of the absolute SFR can be determined by multiplying
the EW(\HA) by the total R-band luminosity, which includes the continuum near \HA.
Within our sample, the galaxies with the 
largest (EW(\HA) $\cdot$ L$_{\rm B}$) are the ones at the higher surface brightness end of
our range, with $\mu_{\rm B}(0)_{\rm cor} \lesssim 23.0$ \msa.
Galaxies with lower total SFR will be found in the lower right of the EW(\HA) vs. M$_{\rm Btot}$ diagram.
Our sample selection is neither complete nor representative, so we cannot comment on how the absolute star-formation
rates of LSB and HSB galaxies compare.  
\citet{vdh00} found that the mean present-day SFR for a sample of late-type LSB galaxies was
about one-tenth the SFR of similar HSB galaxies.  

The relations between \NHA\ and EW(\HA) with (B$-$I$_{\rm c}$) color, shown in Figs.~\ref{allcomp}c and
\ref{allcomp}f, are comparable in strength to those between \NHA\ and the stellar absorption
indices Mgb and \meanfe.  A Kendall's $\tau$ test applied to the emission line LSB galaxies
gives a probability of 2\% that there is no correlation between \NHA\ and (B$-$I$_{\rm c}$)$_{\rm tot}$.
Similarly, there is a probability of only 0.7\% that EW(\HA) and (B$-$I$_{\rm c}$)$_{\rm tot}$
are not correlated for the emission line LSB galaxies.
(B$-$I$_{\rm c}$) color is sensitive to changes in age, metallicity, and dust reddening, and
all three may play some role in explaining the correlations.  
If we use the \citet{den01} calibration for \NHA, the resulting
metallicity range spanned by the data is less than one dex. 
At any constant age less than 5 Gyr, 
the V2000 stellar population models predict a maximum (B$-$I$_{\rm c}$) 
color range of only 0.6 mag for a range in [Fe/H] from -1.0 dex to +0.2 dex. 
The ${\rm (B-I_c)_{tot}}$ range spanned by our data is more than one magnitude, so 
varying only the metallicity of the stellar population is not a sufficient explanation
for the colors.  Stellar populations with mean ages older than 5 Gyr will show a wider color range 
over that range of metallicity, but are not supported by the ages determined 
in Section \ref{empirical} for the four absorption line dominated LSB galaxies.

Including age in the interpretation of the colors requires us to make assumptions about the SFH in
these galaxies.  In Section \ref{metindicator}, we discussed three plausible star
formation scenarios: (1) continuous, but decreasing star formation rate with a build-up
of metallicity, (2) old-plus-burst models with an old stellar population underlying a
significant ($> 1\%$ by mass) burst of star formation, and (3) old-plus-burst SFH models
where the burst was small ($< 0.3\%$ by mass) and produced \HA\ emission, but had little
effect on the global photometric properties.  Scenario (2) was rejected because it would
not maintain the strong correlation between \NHA\ and Mgb or \meanfe.  Scenario (3) was
acceptable simply because the blue flux from the young population did not significantly
veil the strength of the underlying metal absorption strength, and the \NHA\ - metal index
correlations were unperturbed.  However, this assumption also means that the global colors
will not be significantly changed by the young burst.  Consequently, for this SFH we must
invoke a screen of dust as the explanation for the large range of (B$-$I$_{\rm c}$) colors.
Scenario (1) predicts a correlation between age and metallicity, which then produces a tight
correlation between metallicity and the Mgb, \meanfe\ strengths for the stellar population.
By invoking a correlated evolution of age and metallicity, we avoid adding scatter to 
parameters affected by both, including color.  Furthermore, the range of (B$-$I$_{\rm c}$) colors
in this picture is at least as great as the range seen in our data.  The V2000 models for 
a 1 Gyr population with one-fifth solar metallicity predict (B$-$I$_{\rm c}$)=1.57, while a
4 Gyr solar metallicity stellar population would have (B$-$I$_{\rm c}$)=2.55.  Mean ages less than 
1 Gyr yield colors bluer than (B$-$I$_{\rm c}$)=1.5, and while the zeropoint of the model colors is
poorly calibrated, the range is more accurate.  Dust, while certainly present in the
emission line regions, is not necessary to explain the global colors in this scenario.

\section{Discussion\label{discussion}}
The previous sections show that LSB galaxies cover a wide range of metallicities, and 
as a class they may contain both old and young stellar populations.  Dust is present 
in some LSB galaxies, and may affect their global colors.  Star formation occurs 
continuously in these galaxies, but with global rates which have varied exponential decay 
timescales.  The metallicities of
the galaxies are coupled to their ages, such that galaxies with older mean ages have
reached a higher metallicity.   Bursts that occur after
long periods without star formation must make up a very small fraction of the total galaxy
mass.  

The detailed study of LSB galaxies is still a young field, partly because the observations
are very difficult to make.  The study of LSB galaxies originated with the realization that selection
effects in imaging surveys had an impact on our understanding of the galaxy population,
and galaxy formation and evolution in general.  While we are now more conscious of some
of the biases, samples chosen for follow-up study are still subject to selection effects. 
It is interesting to look at the studies in the literature in the context of our
results, and determine what effects selection or observational biases have had on both our and the
published conclusions.

Previous spectroscopic studies of LSB galaxy chemical abundances \citep{mcg94,ron95,deb98} 
were done on galaxies selected to have at least one bright \HII\ region.  These studies
concluded that LSB galaxies were all metal-poor, with gas phase metal abundance less than
one-tenth to one-fifth solar.  These galaxies were selected from photographic surveys which
were much more sensitive to blue galaxies than red galaxies at the surface brightness limit.
Furthermore, the \HII\ regions selected were the brightest \HII\ regions in the galaxy, and 
lower metallicity \HII\ regions will have higher ionization parameters and be brighter than
high metallicity \HII\ regions with similar numbers and ages of stars.  
Compared to our sample, and based on the implicit
color criterion imposed by the photographic surveys, these studies were limited to having
galaxies as blue or bluer than the five bluest galaxies in our survey.  The five bluest
galaxies in our study (C5-3, N9-1, U1-4, C4-1, and P1-3) all have active star formation, 
and define the low metallicity end of the sample. Though they do overlap the range of surface brightnesses
and total absolute magnitudes covered by the rest of our sample, they are certainly not representative
of the full continuum of LSB galaxy properties found for the sample.

\citet{ger99} made an N-body simulation of the SFH for one of the blue 
late-type LSB galaxies from \citet{deb98}, and found that they required both low mass surface density and 
low metallicity to explain the properties of the galaxy.  The SFH they derived has a slowly
decreasing SFR when averaged over timescales of order 100 Myr, but has significant 
(factors of 2) deviations from
the mean SFR on timescales of order 20 Myr. It is these deviations from the mean, they say,
which explains the color range seen for LSB galaxies in general.  Their models do not predict 
absorption line strengths, but if the colors are significantly affected by the 
SFR deviations, then the Mgb and \meanfe\ absorption line indices must also be
affected.   Their simulation cannot be used to explain the full range of LSB galaxy parameters
that we see.  First of all, despite some uncertainty in the metallicity zeropoints of the
\NHA, Mgb and \meanfe\ indicators, we find LSB galaxies with metallicities near solar, and 
certainly at least half-solar.  Their models require metallicities less then about one-fifth
solar to maintain a low SFR.  Furthermore, the range of color (and we infer also Mgb, \meanfe)
is decoupled from increases in metallicity, in conflict with the \NHA-Mgb, \NHA-\meanfe, and 
\NHA-(B$-$I$_{\rm c}$) correlations.   Determining whether the range of colors (and thus Mgb and \meanfe)
they predict at any given metallicity could be hidden within the scatter of these correlations
requires more quantitative modelling than we have done.  The \citeauthor{ger99} models predict that $\sim 20\%$ of
the late-type LSB galaxy population will have red colors, (B-V) $\sim$ 1.  However, those
galaxies should be gas rich and metal-poor, and while the gas fractions of the red LSB galaxies in
our sample remain unconstrained, their metallicities are higher than predicted by \citeauthor{ger99}.
In addition, they predict that these red galaxies will have  $\mu_{\rm B}(0) \simeq 24.5$ \msa, 
which is fainter than the galaxies selected for our sample.  This surface brightness is well below
the completeness limit of the OBC97 survey, and no other large surveys for red LSB galaxies
have yet been published.  

\citet{vdh00} modeled the SFH of the galaxies studied by \citet{deb98} using a 
galactic chemical and photometric evolution model.  They were able to explain
the colors, magnitudes,  and gas phase metallicities of most of the blue LSB galaxies with
a SFH that had a continous, exponentially decreasing SFR over a period of 14 Gyr.
For the bluest galaxies in the sample (with (B$-$V) $<$ 0.45) an additional burst of 
star formation was needed.  While our analysis rejects the possibility for significant
late-time, i.e.~high metallicity, bursts, none of the galaxies in our sample are that blue.  

 \citet{bel00b} used near-infrared and optical colors to study the
stellar populations of blue, red, and giant LSB galaxies.  They did not require that the
galaxies have \HII\ regions, nor that they had previous detections in \HI.  However, their
sample did have a higher average surface brightness than the spectroscopic studies, so 
that they could get sufficient S/N in the K-band imaging.  Their sample has a similar central
surface brightness distribution to ours.  Like us, they found that LSB galaxies cover a range
of metallicities, with some having metallicities near solar.   They also found that some red LSB galaxies had 
rather old luminosity-weighted mean ages.  They only had five red LSB galaxies in their
survey and they determined that those with old ages and solar metallicities were
all at the bright end of the LSB surface brightness range, $\mu_{\rm B}(0) \sim 22.5$ \msa.  
We find old, metal rich
stellar populations in some red galaxies with $\mu_{\rm B}(0) \simeq 24.0$ \msa.
By combining the \citeauthor{bel00b} LSB sample with
a sample of HSB disk galaxies, \citet{bel00} found a clear correlation between mean age
and surface brightness, which we do not detect.  The discrepancies in both cases can be
explained by small sample sizes, especially of the reddest LSB galaxies in the sample.

\citet{imp01} obtained low resolution ($\cong$ 20\AA), low S/N optical spectra for 250 galaxies selected
from the APM survey of \citet{imp96}, 93 of which have central surface
brightnesses fainter than $\mu_{\rm B} = 22$\msa (their LSB cutoff).  They also obtained HI spectra for
145 LSB galaxies.  \citet{bur01} analyze these data, and measure Balmer decrements (H$\alpha$/H$\beta$) and 
[O/H] (using the R$_{23}$ method) when the optical spectra are of sufficient quality (43 HSB and 17 LSB galaxies have measured [O/H]).  
They also measure the best-fit Tully-Fisher relation for the HI sample.  Their sample selection criteria were similar
to those of \citet{bel00b} and ourselves: no requirements for prior HI detections, or the presence of HII regions.
Though the quality of their data for individual galaxies may not be as good, they have a larger dataset to draw
conclusions from, and their conclusions match ours: there is evidence for old stellar populations in some LSB galaxies, and
the range of metallicities for LSB galaxies shows a large overlap with the range of metallicities seen in HSB galaxies.

A complete picture of the formation and evolution of LSB galaxies is still not in hand, as
many of the conclusions of these authors and ourselves are clearly still subject to
variance due to small and incomplete samples.  Future detailed studies should rely
on deep, complete, multi-color surveys with well determined detection criteria to 
gain a better overall picture of LSB galaxy formation and evolution.
The constraints on LSB SFH implied by the many correlations we detect are only
applicable to LSB galaxies with $1.0 < {\rm (B-I_c)} < 2.1$ and $\mu_{\rm B}(0) < 24.3 $\msa.

\section{Summary \& Conclusions \label{summary}}
We have used the Marcario Low Resolution Spectrograph on the 9.2m Hobby-Eberly Telescope
to obtain deep integrated spectra of the gaseous and stellar components of 19 LSB 
galaxies, covering a range of colors and surface brightnesses.  These spectra have sufficient
spectral resolution and S/N in the continuum to measure both emission line ratios and stellar
absorption line strengths.

The spectra of these galaxies qualitatively resemble the spectra of HSB galaxies covering
the full range of spectra seen in galaxies of Hubble types from E to Irr.  
In most of the parameters we measure, LSB
galaxy properties cover the same broad range as HSB galaxies.  Four galaxies have 
spectra whose features are dominated by old stellar populations, covering a range of abundances
from less than half solar to twice solar metallicity.  None of these four have been
detected in \HI.  The remainder of the galaxies show evidence for ongoing star formation,
in some cases at a significant rate compared to their average past SFR.
At least one massive red LSB galaxy shows evidence for super-solar [Mg/Fe] abundance.
This is in accord with similar findings for HSB galaxies, and suggests that the
mechanism responsible for the super-solar abundance ratios is effective in LSB galaxies as well.
All of the emission line galaxies are consistent with star formation, rather than AGN,
being the ionizing source.

There are tight correlations between the stellar absorption line indices and the gas
phase indicators \NHA\ and EW(\HA).  These have not been studied before, in LSB or HSB
galaxies.  We use these correlations to argue that the SFH
for these galaxies must be fairly smooth.  Large bursts of star formation, or
multiple small bursts separated by long quiescent periods, do not fit these relations.

Dust is present in some of these galaxies, with A$_V$ as large as 1.8 mags.  We
are unable to determine, though, whether the dust is isolated to the emission
line regions, or if it is mixed in with the the stellar population.

The redshift distribution for this galaxy sample is very broad, ranging from 3000 to 70000 km/s.
Additionally, we find several discrepancies between the published redshifts based on 
Arecibo 305m telescope \HI\ observations and our new optical observations.  In most cases, the discrepancy
can be explained by beam confusion affecting the Arecibo detection.  All of the galaxies
found by \citet{one00} to deviate significantly from the Tully-Fisher relation, we
show to suffer from significant redshift errors.   



\acknowledgments
\noindent{ACKNOWLEDGEMENTS}

\noindent We would like to thank the staff of the Hobby-Eberly telescope for obtaining the
data, and we heartily thank the many astronomers and engineers involved in building the telescope 
and its instruments.
We thank Greg Shields for useful discussions on 
nebular chemical abundances, and Karl Gebhardt for comments on an early draft 
of this paper.  We also thank the anonymous referee for comments which helped to improve
this paper. The Hobby-Eberly Telescope is operated by McDonald Observatory on behalf of 
The University of Texas at Austin, the Pennsylvania State University,
Stanford University, Ludwig-Maximilians-Universit\"at M\"unchen, 
and Georg-August-Universit\"at G\"ottingen. The Marcario Low Resolution Spectrograph 
is a joint project of the Hobby-Eberly Telescope partnership and the Instituto de 
Astronomía de la Universidad Nacional Autonoma de M\'exico.   
This material is based in part upon work supported by the Texas Advanced
Research Program under Grant No. 009658-0060-1977.
This research was supported in part by the Gemini Observatory, which is operated by the 
Association of Universities for Research in Astronomy, Inc., on behalf of the 
international Gemini partnership of Argentina, Australia, Brazil, Canada, Chile, 
the United Kingdom, and the United States of America.  This research has
made use of the NASA/IPAC Extragalactic Database (NED) which is operated by the 
Jet Propulsion Laboratory, California Institute of Technology, under contract 
with the National Aeronautics and Space Administration.  This research has made use of 
NASA's Astrophysics Data System Service.

\clearpage


\end{document}